%% file: ms.tex
\documentclass[twocolumn]{aastex62}
\usepackage{amssymb,amsmath,epsfig,graphicx, natbib}

\def\clm{}

\received{\today}
\shorttitle{Kinematics of Circumgalactic Gas}
\shortauthors{Martin \et}
\input mymac.tex

\begin{document}

%\title{The Angular Momentum of Circumgalactic Gas}
\title{Kinematics of Circumgalactic Gas:  Feeding Galaxies and Feedback}

\correspondingauthor{Crystal Martin}
\email{cmartin@physics.ucsb.edu}

\author{Crystal L. Martin}
\affiliation{Department of Physics, University of California, Santa Barbara,
Santa Barbara, CA 93106, USA}
\nocollaboration

\author{Stephanie H. Ho}
\affiliation{Department of Physics, University of California, Santa Barbara,
Santa Barbara, CA 93106, USA}
\nocollaboration

\author{Glenn G. Kacprzak} 
\affiliation{Center for Astrophysics and Supercomputing, Swinburne University
of Technology, Hawthorn, Victoria 3122, Australia}
\nocollaboration

\author{Christopher W. Churchill}
\affiliation{Department of Astronomy, New Mexico State University, Las Cruces, NM 88003, USA}
\nocollaboration



\begin{abstract}
We 
present observations of 50 pairs of redshift $z \approx 0.2$ star-forming galaxies 
and background quasars. These sightlines probe the circumgalactic medium (CGM) out 
to half the virial radius, and we describe the circumgalactic gas kinematics relative to the 
reference frame defined by the galactic disks.   We detect halo gas in \mgII\  absorption,
measure the equivalent-width-weighted Doppler shifts relative to each galaxy, and find 
that the CGM has a component of angular momentum that is aligned with the galactic disk.
No net counter-rotation of the CGM is detected within 45\deg\ of the major axis at any
impact parameter. The velocity offset of the circumgalactic gas correlates with the projected 
rotation speed in the disk plane out to disk radii of roughly $70$~kpc. We confirm previous 
claims that the \mgII\ absorption becomes stronger near the galactic minor axis and show 
that the equivalent width correlates with the velocity range of the absorption.
We cannot directly measure the location of any absorber along the sightline,
but we explore the hypothesis that individual velocity components can be
associated with gas orbiting in the disk plane or flowing radially outward
in a conical outflow. We conclude that centrifugal forces partially support the 
low-ionization gas and galactic outflows kinematically disturb the CGM  
producing excess absorption. Our results firmly rule out schema for the inner 
CGM that lack rotation and suggest that angular momentum as well as galactic winds 
should be included in any viable model for the low-redshift CGM.

%Galaxy inclinations, position angles, and rotation curves together 
%determine the projected velocity field of each galactic disk on the sky.
%We distinguish the near and far sides of these disks when our 
%high-resolution imaging resolves the winding direction of the spiral arms; 

%We determine the three-dimensional orientation of many galactic disks
% establishes
%the projected velocity field of any galactic outflows up to a scale factor.

\end{abstract}

%\keywords{editorials, notices --- miscellaneous --- catalogs --- surveys}


\section{Introduction} \label{sec:intro}

The growth of galactic disks requires a nearly continuous gas supply
which has been modified by feedback from massive stars. This accretion
dilutes the relative number of low metallicity stars \citep{vandenBergh1962,
Worthey1996,Woolf2012} and substantially lengthens the timescale for building disks
\citep{Kennicutt1983,Papovich2015}. How the sizes of disks grow as their stellar
mass increases depends primarily on the angular momentum of the accreted gas.
Observations of disks indicate inside-out growth \citep{Barden2005,
Gogarten2010,MunozMateos2011,GonzalezDelgado2014}, consistent
with the angular momentum of the accreted gas increasing with cosmic time.
The evolution of the stellar specific angular  momentum also
depends on the stellar feedback history \citep{Governato2007,Governato2010,Sijacki2012,Danovich2015}. 
Efforts to characterize the mass dependence of the stellar specific angular momentum 
may constrain the fraction of this angular momentum retained by galaxies \citep{Fall2018}.
Identifying the circumgalactic gas that feeds the inside-out growth of galactic disks, 
and its interaction with galactic winds, will be necessary to understand galaxy assembly.

Direct observations of disk accretion remain sparse.  
{\clm
Evidence for gas
accretion in redshift $z \approx 0.9 - 2.3$  galaxies has grown steadily in recent years,
aided by integral field spectroscopy \citep{Bouche2013,Bouche2016,Zabl2019}.
}
 The disk -- halo
interface at low redshift, however, has only been well described within a 
few kpc of the disks \citep{Sancisi2008}.  Even for the Milky Way galaxy, much of
the infalling gas may go unrecognized because only the highest velocity
clouds are easily distinguished from the gas disk \citep{Zheng2015}. Progress 
measuring physical conditions in the circumgalactic medium (CGM) over the past few years, however, 
make it timely to systematically study the kinematics of circumgalactic gas now.

These observations have identified a large reservoir of baryons surrounding 
galaxies. This circumgalactic 
{\clm
gas extends
}
to roughly the virial radius \citep{Shull2014}
and contains a substantial fraction of the baryons \citep{Werk2014,Bregman2018} and
metals \citep{Peeples2014} associated with the dark matter halo. The larger
column densities of the $O^{+5}$ ion in the CGM of star-forming galaxies compared
to passive galaxies \citep{Tumlinson2011} has generated great interest in the
CGM because the processes producing this dichotomy may explain why star formation 
is quenched in massive halos \citep{Kauffmann2003,Blanton2003,Schawinski2014}.
In cosmological hydrodynamical simulations, the \ovi\ absorbing gas lies behind the 
halo accretion shock and is maximal in \lstar\ galaxies because their virial 
temperature is close to the temperature $T \approx 10^{5.5}$~K where the $O^{+5}$ 
ionization fraction peaks \citep{Oppenheimer2016}. 
Feedback from supermassive black holes may suppress the
$O^{+5}$ fraction in the halos of red galaxies relative to the halos of
blue galaxies of similar stellar mass \citep{Nelson2018a}. 
The nucleus would not typically still be active by the time its outflow impacted
the gas properties at half the virial radius, so differentiating between AGN 
activity and halo mass is challenging observationally \citep{Berg2018}.
Simulations that zoom in on individual galaxies include more physics than cosmological 
simulations \citep{Hummels2013,Su2018}. They qualitatively agree that enhanced star formation 
feedback increases the strength of high-ionization absorption lines  as \cite{Heckman2017}
observed. Quantitatively, however, the star-formation feedback does not produce enough
\ovi\ absorption nor does it permanently quench star formation. A solution may require
a completely different schema for the CGM.  \cite{Stern2018} argue,
for example, that the \ovi\ absorption occurs beyond the accretion shock, 
where \ovi\ would be photoionized by the UV background, a low-pressure scenario.

The dichotomy between the CGM properties of blue and red galaxies extends to low-ionization
gas. These differences are less widely appreciated  because the halos of both star-forming
and passive galaxies produce low-ionization absorption-line systems. 
Large surveys routinely find an excess of strong \mgII\ absorbers around blue
galaxies relative to red galaxies \citep{Lan2014,Lan2018}. 
The large absorption strength requires a substantial velocity spread among 
velocity components or very large turbulent velocities. We call this the {\it kinematic
dichotomy} between the CGM of star-forming and passive galaxies.

Massive, star-forming galaxies generally have both stellar and gaseous galactic 
disks \citep{Schawinski2014}. Quasar sightlines near the position angle of the
minor axis of a disk detect larger \mgII\ absorption equivalent widths than 
do those near the major axis at similar impact parameter 
\citep{Bordoloi2011,Bouche2012,Kacprzak2012,Lan2014,Lan2018}. 
This minor-axis excess may be associated with galactic winds. Only 
minor-axis sightlines at very small impact parameters, however,  would
intersect the outflowing gas that is directly detected in galaxy spectra because
the higer covering fraction (and density) of gas at small radii dominates the outflow
absorption in galaxy spectra \citep{Martin2009}. It has been common to model the 
trajectory of cool clouds outward from the galaxy to 60~kpc or more \citep{Bouche2012,
Gauthier2012,Kacprzak2014,Schroetter2016}, but this interpretation is in direct conflict 
with hydrodynamical simulations. Individual clouds in the hot, supersonic wind are destroyed
over spatial scales of just a few kiloparsecs in hydrodynamical simulations  
\citep{Scannapieco2015,Bruggen2016,Schneider2017}. Other processes, including 
cooling hot winds \citep{Wang1995,Martin2015,Thompson2016} and the 
interaction of hot winds with the CGM \citep{Lochhaas2018,Su2018} for example, might
produce low-ionization gas that explains the excess minor-axis absorption. A comprehensive 
look at the relationship between galaxy properties and CGM kinematics would provide 
additional insight.

The kinematics of galactic disks are rarely 
measured for the hosts of quasar absorption-line systems. In their exploratory study,
\cite{Steidel2002} measured galactic rotation curves for a few galaxies at $z \approx 0.5$.
Remarkably, the Doppler shift of the galactic rotation often had the same sign as the velocity 
offset of the \mgII\ absorption. Corotation suggested an extended gas disk,
but simple disk models did not explain the large velocity width of the absorption systems \citep{Kacprzak2010,Kacprzak2011}.
This linewidth problem does not exclude an extended gas disk contributing a velocity component 
to a strong \mgII\ system, as seen for example in one low impact-parameter 
sightline passing through the halo of a Milky-Way-like galaxy at $z = 0.413$  
\citep{Diamond2016}.
Multiple quasar sightlines through the halo of a very nearby spiral galaxy also detect absorption
best described by a disk-like distribution of gas approximately planar to the observed
\ion{H}{1} disk yet draw attention to the noise from additional velocity components
\citep{Bowen2016}. Multiple sightlines through the same halo can rarely be observed
with current facilities however.

Our {\it Quasars Probing Galaxies} program targets quasars fortuitously located behind 
redshift $z \approx 0.2$ blue galaxies, detects low-ionization \mgII\ (and sometimes
\mgI) absorption near the galaxy redshift,  and characterizes the kinematics of the
circumgalactic gas relative to the galactic reference frame.
The galaxy sample is more homogeneous than those presented previously, allowing 
us to stack the galaxies and produce a multiple-sightline 
description of the average CGM kinematics. We presented 15 sightlines within 30\deg\ of 
the major axis in \cite{Ho2017}, finding no  \mgII\ systems with Doppler shifts in 
the opposite sense of the foreground galaxy's rotation. To better characterize the
rotation of the CGM, we have expanded the study to 50 sightlines which sample
the full range of azimuthal angle. We also measured rotation curves and obtained
high-resolution images to better describe the orientation of the galaxy.

In Section~\ref{sec:data} we describe the galaxy properties and the detection of associated
quasar absorption-line systems. 
The uniformity of the galaxy properties allows us
to discuss the geometry and kinematics of the average CGM with a large number of sightlines.
Section~\ref{sec:results} describes the absorption signature of low-ionization CGM
relative to the projected galactic disks.  Properties of the corotators
are compared to large, extended disks in Section~\ref{sec:thin_disk}, and we
discuss the constraints that minor-axis sightlines place  on galactic winds and outflows in Section~\ref{sec:winds}.
Section~\ref{sec:ovi} summarizes the empirical constraints on corotation.
Throughout the paper we assume a cosmology with $H_0 = 67.74$\kms\ Mpc$^{-1}$, $\Omega_0 = 0.3089$,
and $\Omega_{\Lambda} = 0.691$ \citep{Planck2015}.  The angular diameter distance at $z = 0.2$ corresponds
an angular scale of 3.30~kpc per arcsecond.

%%%%%%%%%%%%%%%%%%%%%%%%%%%%%%%%%%%%%%%%%%%%%%%%%%%%%%%%%%%%%%%%%%%%%%%%%%%%%%%%
\section{data} \label{sec:data}

We selected galaxy -- quasar pairs based on the properties of the 
foreground galaxies.  In contrast, traditional surveys detect
absorption-line systems in quasar spectra and then identify the 
galaxy halo.  By selecting the galaxies first, we can study
the halos of galaxies with very similar properties. It follows
that although a single sightline intersects each galaxy halo, the
galaxies can be {\it stacked} such that we sample the average CGM 
with 50 sightlines.  Registration of the pairs requires measurements
that establish the orientation of the galactic disk. In this section
we describe these selection criteria and observations, summarize the 
measured galaxy properties, and then present the absorption line systems 
associated with the galaxies.

\subsection{Selection of Galaxy -- Quasar Pairs}

The rest-frame UV bandpass contains many strong transitions from common ions,
but the number of quasars bright enough to observe in the UV from space is
small compared to the number accessible with ground-based telescopes in the visible.
We therefore restricted the sample to $z > 0.15$ galaxies in order to
redshift the \mgII\ $\lambda
\lambda 2796, 2803$ doublet longward of the atmospheric limit.  We also chose low
redshift  galaxies because we wanted to spatially resolve morphological features.

These star-forming galaxies have a median redshift of  $ z \approx 0.21$ so they
are fainter than the Sloan Digital Sky Survey DR9 (SDSS,\cite{Ahn2012}) spectroscopic sample. We used photometric
redshifts to select the galaxies but then measured spectroscopic redshifts from
our new spectra. These redshift revisions mean that our selection criteria are
not strict limits. 

Passive, red galaxies were excluded from the galaxy sample using a color cut
$M_u - M_r < 2.0$ \citep{Schawinski2014}.  We rejected galaxies with disks
observed face-on by requiring the SDSS $r$-band semi-minor axis to be less than 0.71 
times the length of the semi-major axis.  We later obtained higher resolution imaging 
and revised the disk inclination and position angle measurements for a few galaxies. 

Our target galaxies are the subset of this sample with a background quasar 
brighter than $u = 18.5$ and impact parameters $b <  100$~kpc, corresponding
to 10-50\% of the halo virial radius.
We inspected the SDSS DR9 spectra of these quasars, eliminating a few
Lyman-limit systems and misidentified stars, and obtained a parent sample
of galaxy -- quasar pairs.  Our observational campaign prioritized 
pairs with $ b < 60$~kpc.

\subsection{Observations}

%%%%%%% TABLE 1
%\input tables/table_obs.tex
\input table1.tex

Table~\ref{tb:obs} summarizes our new spectroscopic and imaging observations
of 50 galaxy -- quasar pairs. Measured properties of the galaxies and
associated quasar absorption-line systems follow in Sections~\ref{sec:galaxy_properties}
and \ref{sec:quasar_als}, respectively. Section~\ref{sec:environment} describes
the environments of the target galaxies.

\subsubsection{Spectroscopy}

We spectroscopically observed the quasars using the Low Resolution
Imaging Spectrograph (LRIS) at the W. M. Keck Observatory \citep{Oke1995,Steidel2004}. 
We designed custom slitmasks in order to simultaneously observe the quasar,
target galaxy, and several {\it filler galaxies} that had SDSS photometric redshifts
consistent with  the target galaxy, prioritizing the filler galaxies by their
luminosity and angular separation from the quasar. 
The mask position angles were not restricted to the parallactic angle because
LRIS has an atmospheric dispersion corrector \citep{Phillips2006}. 
We configured LRIS with the 1200 line mm$^{-1}$ blue grism blazed at 3400 \AA, 
D500 dichroic, and 900 line mm$^{-1}$ red grating blazed at 5500 \AA. The
1\farcs0 wide slitlets provided a resolution of 105-165\kms\ FWHM (full width
at half maximum intensity) with LRISb and 75-105\kms\ FWHM on the red side.

Following the procedure discussed in detail by \cite{Ho2017}, we removed fixed
pattern noise, rejected cosmic rays, and measured the shifts between 
individual frames. We computed a variance image when these frames were combined.
We wavelength calibrated the frames using arc lamp exposures, checked the 
dispersion solution using night-sky emission lines, and applied small ($< 25$\kms)
zeropoint corrections as warranted. We applied a heliocentric correction and extracted
one-dimensional spectra and error spectra. The RMS error in the dispersion solution
was $\sles\ 15$\kms\ over the region covering \mgII.

We measured \Ha\ Doppler shifts along the major axis of the target galaxy.
If the position angle of the LRIS slitlit was not within 30\deg\ of
the major axis, then we obtained longslit spectroscopy with the
Apache Point Observatory 3.5~m telescope.
We configured the blue and red sides of the Double Imaging Spectrograph with 
1200 lines mm$^{-1}$ gratings blazed at 4400 \AA\ and 7300 \AA, respectively.
We obtained a resolution of 50\kms\ with a 1\farcs5 slit width.
We integrated long enough
to detect \Ha\ across many spatial resolution elements. Table~\ref{tb:obs} 
provides the exposure times and observation dates. \cite{Ho2017} describe
the data reduction procedure.

The {\it Cosmic Origins Spectrograph} on HST \citep{Green2012} has observed two of the
50 quasars we observed with LRIS. We retrieved the G130M observations
from the HST archive for \j155504+362847 (GO 11598; PI Tumlinson) and 
\j165931+373528 (GO 14264; PI Kacprzak). All exposures covering
the \ovi\ doublet were compared.  Absorption in \ovi\ is detected near 
the redshift of \j155505+362848. The \ovi\ doublet is not clearly detected
at the redshift of \j165930+373527.

\subsubsection{High Resolution Imaging}

High resolution images were obtained using the quasar as the tip-tilt reference
for the Laser Guide Star Adaptive Optics system \citep{Wizinowich2006} on Keck II. We configured
the NIRC2 camera in the wide field mode providing an average plate scale of 0\farcs04~/ 
pix over the $ 1024 \times 1024$ InSb array. We observed with the $Ks$ filter
and 20~s integrations, nodding the telescope between coadditions.  We choose
a camera position angle and dither pattern that kept the galaxy -- quasar pair off 
the bad quadrant of the detector. The pixels undersample the diffraction limit 
at 2.146\um, so we also applied sub-pixel dithers at each nod position.

We removed fixed pattern noise and drizzled the frames using the pipeline 
provided by the UCLA/Galactic center group \citep{Ghez2008}.  The point-spread 
function (psf) estimated from the quasar profile has a median full width at
half maximum intensity of 0\farcs13. Data taken during periods of high winds  
has a psf elongated in the altitude direction which we attribute to wind shake.
We increased the number of coadditions under these challenging conditions, but
the low bandwidth wavefront sensor did not always close the loop. We flux
calibrated the images using a combination of
A0 star observations and the 2MASS Ks magnitudes of the quasars.
Table~\ref{tab:galaxy} provides the new Ks-band galaxy photometry.

We imaged 9 of the target galaxies with
the Hubble Space Telescope (GO14754). Short exposures
with WFC3/UV were obtained through the F390W and F814W filters.  The data
were reduced and calibrated with the WFC3 calibration pipeline and then drizzled.
Cosmic ray removal followed the steps described in Ho \et (2019, in prep), which
describes properties of the disks probed by major-axis sightlines.

%%%%%%% TABLE 2
%\input tables/galaxy_table.tex
\input table2.tex

\subsection{Galaxy Properties} \label{sec:galaxy_properties}

We measured galaxy redshifts from the bright optical emission lines in
the red LRIS spectrum. These spectroscopic redshifts supercede the SDSS
photometric redshifts, which are not accurate enough for describing the
velocity offsets of the circumgalactic gas. Three of the target galaxies 
had large redshift corrections; see notes to Table~\ref{tab:galaxy}.
These spectroscopic redshifts define the systemic velocity of the target galaxies.

To ensure consistency with the new redshifts, we revised the galaxy 
rest-frame colors and stellar masses via spectral energy distribution (SED)
fitting with FAST \citep{Kriek2009}. We added our new Ks-band photometry 
to the SDSS {\it ugriz} photometry obtained from the DR9 PhotoObjAll catalog. 
Including our {\it Ks} photometry lowered the stellar mass estimates significantly for 
five galaxies, effectively requiring younger SEDs. If the galaxy was not blended
with the quasar in archival GALEX images, the  FUV and NUV magnitudes
(often upper limits on the flux)  were included in the SED fit.
Adding this constraint on the UV slope lowered one of the mass estimates by a factor
of two, reddened J084723+254105, moving it into the green valley, but did not change
the  median  $M_u - M_r$ color of the sample.
The broad spectral coverage constrains the stellar masses well, although the meaning 
of the fitted parameters is different for three targets that turned out to be
blends of two galaxies. We note these exceptions in Table~\ref{tab:galaxy} where
we list the measured galaxy properties and define several derived properties including
masses.

\begin{figure} 
 \begin{center}
  \includegraphics[scale=0.65,angle=-90,trim = 20 20 80 0]{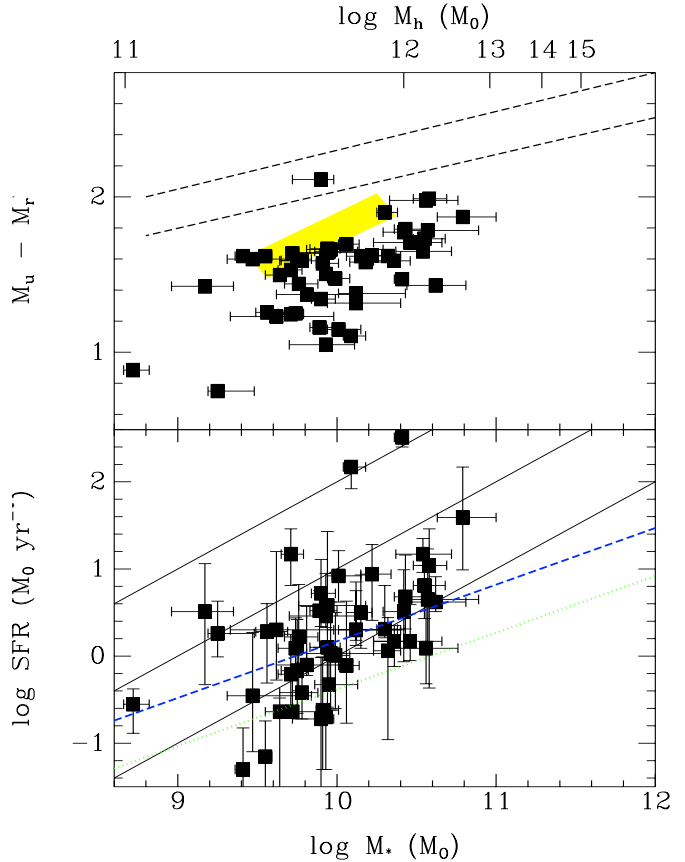}
   \end{center}
    \caption{Properties of galaxy sample.
      {\it Top:}
      Color -- mass diagram for the galaxy sample. The
      diagonal lines show the green valley defined by all morphological types, and
      the shaded region shows the peak density of late-type galaxies 
      \citep{Schawinski2014}.  The median color ($M_u - M_r = 1.59$)
      is 0.15~magnitude bluer than the typical late-type galaxy at
      the median stellar mass ($\log M_* = 9.94$) of our targets.
      {\it Bottom:}
      Solid lines represent  specific SFRs of $10^{-8}$, $10^{-9}$, and $10^{-10}$~yr$^{-1}$.
      The green dotted line denotes the division between star-forming and quiescent 
      galaxies at redshift $z = 0.2$ \citep{Moustakas2013}. The blue dashed line
      denotes the main sequence of star formation \citep{Salim2007}.
      }
     \label{fig:color_mass}
      \end{figure}

Figure~\ref{fig:color_mass} shows the resulting color -- magnitude and  color -- mass 
diagrams.  The stellar masses are typical for late-type galaxies at low redshift.
The color cut biases the sample  median  color 
bluer than the average late-type galaxy with $10^{10}$\msun\ of stars.
The median halo mass of a target galaxy is $10^{11.6}$\msun. The halo virial
radii have a median value of 170~kpc.

%LBRT 1

%LBRT
\begin{figure*} 
 \begin{center}
  \includegraphics[scale=0.74,angle=0,trim = 30 10 0 0]{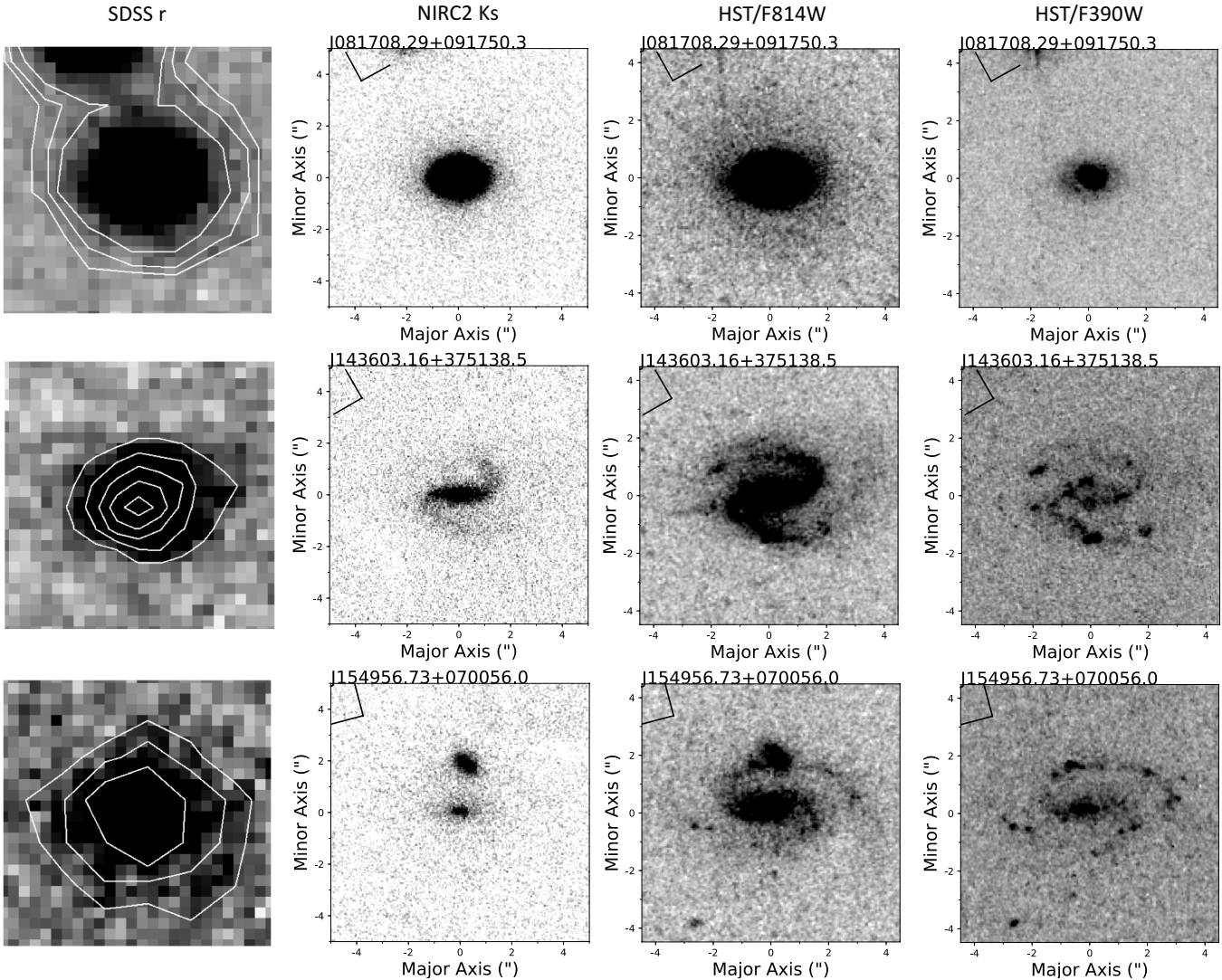}
   \end{center}
    \caption{Comparison of SDSS, NIRC2/LGSAO, and HST images.
      {\it Top:} 
      The position angle and axis ratio of a target galaxy 
      can often be accurately measured from SDSS images. 
      {\it Middle:}
      However non-axisymmetric structure can skew the 
      disk orientation measured from SDSS images.
      The NIRC2/LGSAO {\it Ks} images detect large-scale, 
      two-arm spiral patterns in roughly $ 10\%$ of the galaxies.
      Unsharp masking brings out spiral structure in additional
      10\% of the Ks images.
      {\it Bottom:} 
      High-resolution imaging also resolves close pairs of galaxies,
      allowing us to eliminate large orientation errors. 
      The bluer F390W images emphasize the youngest stellar population,
      and these images detect spiral structure in $\approx 90\% $ of
      the galaxies imaged.
      }
     \label{fig:nirc2_hst}
      \end{figure*}

The high-resolution images confirm that the  target galaxies have 
late-type morphologies as expected based on the color selection.
The SDSS images accurately determine the disk position angle and axis ratio
in many cases. An example is \j081708+081750, which is shown in the top 
panel of Figure~\ref{fig:nirc2_hst}. The bluest image in F390W resolves
flocculent spiral structure that is fairly axisymmetric. 

Young stellar populations provide the best contrast between the arms and
the underlying stellar disk, so it is surprising that some Ks images detect
spiral structure. The middle panel of Figure~\ref{fig:nirc2_hst}
shows an example, a two-arm spiral pattern in \j143606+375138.  The two-arm
pattern is typical of the arms resolved by our infrared imaging; these
arms may be driven by a recent interaction with another galaxy. 
Resolving spiral arms allows us to discuss the 3D orientation of the disks 
in Section~\ref{sec:discussion}. 

Upon close inspection the NIRC2 Ks images often reveal non-axisymmetric structure
that was not resolved by the SDSS imaging. These features include faint companions
as well as spiral arms. We ran GALFIT \citep{Peng2010} on the {\it Ks} images 
and found the disk position angles and inclinations typically require revisions
of only a few degrees. Table~\ref{tab:galaxy} lists the adopted disk position 
angles and inclinations. The seeing-limited resolution produced catastropic
systemic errors for the orientations of three pairs of blended galaxies.
For illustration, the bottom panel of Figure~\ref{fig:nirc2_hst} shows the
the NIRC2 Ks image of \j154956+070056. We adopted the fainter galaxy 
in the Ks image as our target because its coordinates were closer to the SDSS 
coordinates. Later, the HST images showed that the fainter galaxy in {\it Ks}
is the brighter member of the pair in F390W. Our target is a  well-defined disk; 
the companion galaxy is much redder, and its spectroscopic redshift has not been 
measured. The NIRC2 image also resolved \j142815+585442 into two galaxies. 
We have no high-resolution color data in this case, and we define the target 
galaxy as the brighter member of the pair.  We did not detect \mgII\ absorption 
associated with the third close pair, \j143512+360424. 

We defined a grid of rotation curves over a range of asymptotic velocities $\vc$
and  turnover radii, \rrc,  using the following parametric model: 
\begin{equation}
\vc (R) = (2 / \pi) \vrot \arctan (R/\rrc).
\end{equation}
For each target galaxy, we computed the projected velocities along the slit.
We convolved these position -- velocity curves
with a model point spread function (psf) and compared the result to
our \Ha\ position -- velocity measurements. We minimized the fit residual
to estimate the fitted asymptotic rotation speed, \vrot, and turnover 
radius, \rrc,  for each galaxy in Table~\ref{tab:galaxy}.  
We did not measure rotation curves \j142815+585422 and \j154956+070056 because
of the small separation between the paired galaxies.

\subsection{Galaxy Environment} \label{sec:environment}

Determining the environments of our target galaxies would require 
extensive wide-field spectroscopy. We flag potential group members
here using SDSS  photometric redshifts which are not precise enough to establish 
group membership. A subset of these group candidates were spectroscopically
observed as filler galaxies on the LRIS masks.

\subsubsection{Compact Groups} 

{\clm
Heavy elements have been shown to be more widely dispersed 
around group members than isolated galaxies \citep{Bordoloi2011,Johnson2015,Nielsen2018}. 
The groups in these studies  span a large range in mass and size. 
We will use the term {\it compact group} member to identify target galaxies
that may have another luminous galaxy nearby, specifically within one virial radius
in projection on the sky. Sightlines intersecting 
compact groups could show metal-line components from both halos, the
blended systems producing stronger than average absorption strength.
}

Using the same definition of a compact group, \cite{Ho2017} identified four major 
axis targets as potential group members: 
{\clm
\j084725+254104, \j091954+291345, \j102907+421737, and \j124601+ 173152. 
}
%\j084723+254105, \j091954+291408, \j102907+421752, and \j124601+173156. 
{\clm
We measure a redshift $z_{spec} = 0.18496$ for \j102906.98+421733.3, leaving only
\j102909.45+ 421721.4 and our target \j102907+421737 as a candidate group.
}
We add six more
potential group members -- \j081125+093616, \j081708+091751, \j092908+350942, 
\j101713+183232, \j104151+174603, and \j154956+070044.  The paired quasar spectra
did not detect \mgII\ absorption near the redshift of two of these candidates.

The richest environment is likely that of \j124601+173152. 
The environment of target \j124601+173152 includes three brighter galaxies 
with consistent photometric redshifts, but our target is the closet the quasar sightline.
Our target may be falling into the galaxy overdensity marked by
\j124559+173203 (40\farcs4 W, $\Delta V_{LOS} = 1010$\kms).  
\cite{Ho2017} flagged  \j124601+173152 because the host identification for
the \mgII\ absorption in the \j124601+173156 sightline was not unique.

We measure a spectroscopic redshift $z = 0.19013$ for the filler galaxy \j101712.54+183232.5,
which is just 119\kms\ redder than our target \j101711.80+183237.6.
The filler galaxy is closer to \j101713+183232 sightline than our target but fainter.

Our target \j104151.26+174558.7 lies just 4\farcs7 from the \j104151+174603
sightline, making it the most likely absorber host.  It is possible, however, that
a much more luminous galaxy, \j104152.39+174520.2 which is 45\farcs5 SE of the quasar,
contributes to the absorption system; its redshift $z_{spec} = 0.14189$ is  112\kms\ 
bluer than our target's redshift. 

The field surrounding galaxy \j152027+421519 includes
several objects whose photometric redshifts make them candidate group members. 
We measured a spectrocopic redshift $z_{fill} = 0.29908$ for the nearest one, 
\j152026.82+421529.4 (11\farcs0 NW), eliminating it as the source of the \mgII\ absorption.  

We conclude that most of the target galaxies do not
belong to a compact group containing a galaxy more luminous than our target galaxy.
The potential group members which we have flagged here identify sightlines where the
environment might contribute to the absorption signature.

\subsubsection{Interacting Galaxies}  \label{sec:interactions}

Galaxy interactions may be integral to feeding gas to low redshift
galactic disks. The galaxies identified above as potential group
members may have interacted recently with another massive galaxy.
Interactions with lower mass galaxies, including satellites, 
may also drive gas inflows. Here we  identify morphological features that
may mark disturbances caused by  galaxy flybys or mergers.

%     \label{fig:nirc2}
%LBRT
%\figsetstart
% \figsetnum{3}
%  \figsettitle{NIRC2/LGSAO {\it Ks}-band images.}

%   \figsetgrpstart
%    \figsetgrpnum{3.1}
\begin{figure}
 \begin{center}
  \includegraphics[scale=0.445,angle=0,trim = 20 160 0 0]{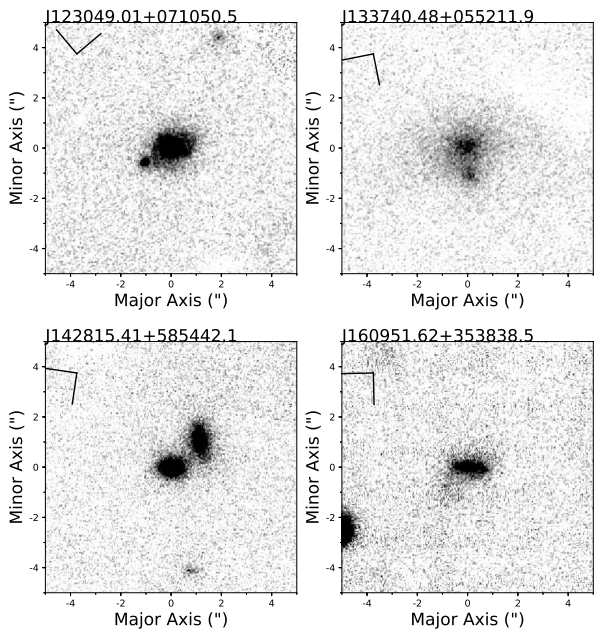}
   \caption{NIRC2/LGSAO {\it Ks}-band images.
     {\it (a)}
     Examples of  targets showing signs 
     of recent interactions. See Section~\ref{sec:interactions} for description.
     (Figure 3b and 3c are available online and show images for other target galaxies.)
     }
     \label{fig:nirc2}
     \end{center}
      \end{figure}

%      \figsetgrptitle{
%      }
%     \figsetplot{fig3a.pdf}

%      \end{figure}
% \figsetgrpend
%  \figsetend

We have already noted three examples of close pairs of galaxies.
The {\it Ks} images in Figure~3a
reveal smaller, fainter objects near our target galaxies.
The potential companions with the two smallest separations are
6.8~kpc SE of \j123049+071050 and 3.3~kpc E of \j133740+055211. 
We looked for nearby galaxies that might drive the two-arm infrared 
spiral patterns and found several candidates.
The spiral arms extending from \j091954.11+291345.3, shown in Ho \et (2019, in prep),
may be driven by an interaction 
with \j091954.44+291345.6 (red galaxy 4\farcs4 E of our target)
and/or \j091954.07+291336.6 (red galaxy 8\farcs7 SW). It is possible
that our target \j143603.16+375138.5 ($z_{spec}=0.31462$) recently interacted with 
\j143603.11+375153.2 (14\farcs7 N, $z_{ph} = 0.36 \pm 0.06$) and/or 
\j143601.36+375143.9 (21\farcs9 NW, $z_{ph} = 0.31 \pm 0.02$). 
The strongest absorber, \j160951+353843 ($W_r(\lambda)  = 2.2$\AA), shows 
a disturbance towards the east; we did not find a brighter galaxy that could have
interacted with the target, but an interaction with a less massive galaxy may explain
the irregular morphology.
One of the galaxies with spiral arms, \j145844.18+170522.2, 
has no neighbors. Figures 3b and 3c show the NIRC2 images of 
the targets not shown elsewhere in the manuscript; note, however,

We adopt a specific SFR of 1~Gyr$^{-1}$ as a definition for a recent starburst 
because it describes the dynamical timescale for galaxy -- galaxy interactions. Our 
sample includes just four galaxies with specific SFRs above this threshold, but the images
show morphological evidence for recent interactions in all of them. 
Two galaxies have specific SFR exceeding $ 10^{-8}$~yr$^{-1}$. One of these is the close
pair \j142815+585442; spatially resolved SEDs are needed to interpret the high specific SFR. The
companion near \j123049+071050 may have triggered the high  spcific SFR.  The other galaxy with
a similarly close companion, \j133740+055211, has the third highest specific SFR in the sample.
Our SED fitting is therefore consistent with the viewpoint that galaxy interactions
are required to generate starbursts at redshift $z \approx 0.2$. The galaxy \j150150+553227 
also satisfies this starburst criterion.\footnote{ 
           Its paired sightline  is the only one among these four that did not detect 
           \mgII\ absorption; see Section~\ref{sec:quasar_als}. No high-resolution images were obtained.}
The SDSS imaging indicates a more irregular morphology than is typical within our sample.
We conclude that interactions with low mass galaxies are common among the targets
with the highest specific SFRs, and we will discuss in Section~\ref{sec:discussion} whether evidence for
gas flows is more common in these systems.

\begin{figure} 
 \begin{center}
  \includegraphics[scale=0.50,angle=0,trim = 20 60 100 0]{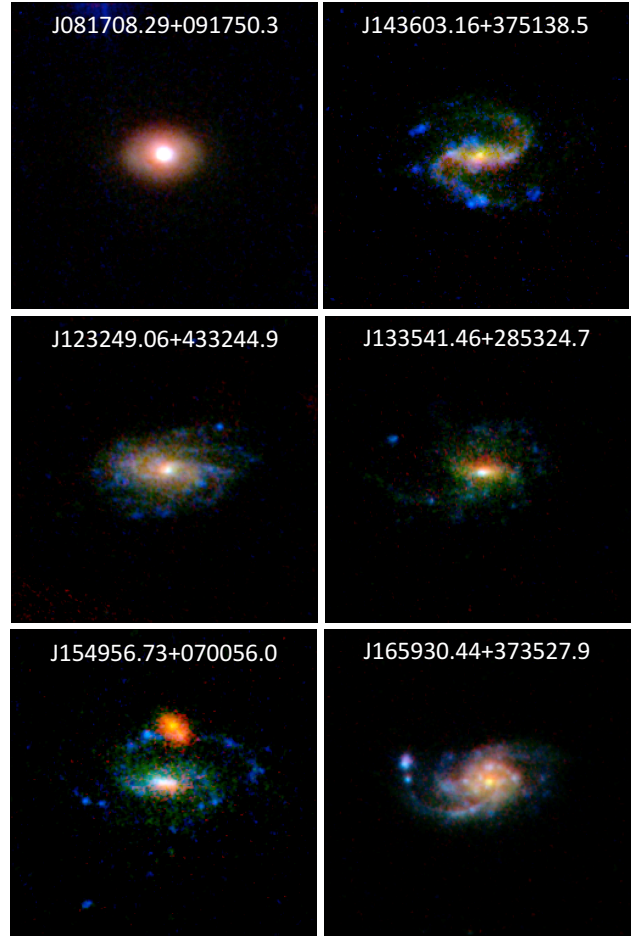}
   \end{center}
    \caption{Composite F390W (blue), F814W (green), and Ks (red) images. 
      The galaxies are ordered by specific SFR, increasing from left to right and
      from top to bottom. The cutouts are 10\arcsec\ by 10\arcsec.
      }
      \label{fig:color} 
       \end{figure}

Figure~\ref{fig:color} shows composite images ordered by specific SFR. 
\j081708+091750 is the only target
in our sample that may have an active nucleus, which we identify based
on the broad emission lines in our red LRIS spectrum.
The imaging of \j081708+091750 does not resolve spiral arms; the galaxy color
is redder than most, and the specific SFR is low.  The galaxy colors generally
become bluer toward larger radii, something most easily seen in the
\j143603+375138 image and the star-forming member of  \j154956+070056 pair.

%%LBRT
%\begin{figure*} 
%       \end{figure*}

\subsection{Quasar Sightlines \& Absorber Properties} \label{sec:quasar_als}

Measuring  galaxy orientation on the sky allows us to describe the properties
of the \mgII\ absorption relative to the projected plane of the galactic disk. 
Following the convention introduced in \cite{Bouche2012} (see their Figure~1), we
define the {\it azimuthal angle}  of a sightline as the angle $\alpha$ between the galaxy 
major axis and a line drawn from the galaxy's center to the quasar. By this definition,
$\alpha$ takes values from 0\deg\ to 90\deg;  Table~\ref{tab:quasar} provides a second
value that distinguishes among the four quadrants of the sky coordinates.
We calculate the impact parameter of the sightline  $b$ from the angular separation of the 
pair defines and angular diameter distance at the galaxy redshift.
Table~\ref{tab:quasar} lists $(b,\alpha)$ sky  coordinates for the quasars relative
to the target galaxy.

We continuum-normalized the quasar spectra and deredshifted each spectrum 
to the rest-frame of the target galaxy. We searched for \mgII\ absorption 
near the systemic velocity of the target galaxy. The doublet spacing 
robustly identified \mgII\ systems.\footnote{The \j142816+585432 spectrum cuts off
         between the 2796 and 2803 lines; detections of the 
         \feII\ 2586, 2600 determine \zabs.} 
We measured the absorption strength and Doppler shift of each transition
using custom software. We defined the integration limits by the intersection
of the spectrum (plus one-sigma errorbars) with the continuum level.
If this bandpass was narrower than the resolution element,
then the latter was adopted allowing us to compute limits on non-detections. 
We integrated the total equivalent width, $W_r$, and calculated the equivalent-width-weighted 
mean velocity, \vavg.  We repeated this procedure 1000 times, perturbing the 
flux in each pixel by a random deviate scaled by the uncertainty. The median
values and the one-sigma uncertainties computed from the resulting histograms
define the measured values. Table~\ref{tab:quasar} summarizes these 
absorption properties.

We detect \mgII\ absorption near the redshift of the target galaxy
in 33 of the 50 sightlines. The median equivalent width is 
$W_r(\lambda)  = 0.31$\AA. The optical depth  at line center would
be saturated for a single, thermal ($10^4$~K) component of this strength,
so the optically thin limit gives only a lower limit on 
{\clm
the 
}
column density 
$N(\mgII) > 7.3 \times 10^{12}$\col.

\begin{figure*} 
 \begin{center}
  \includegraphics[scale=0.75,angle=90,trim = 180 0 0 20]{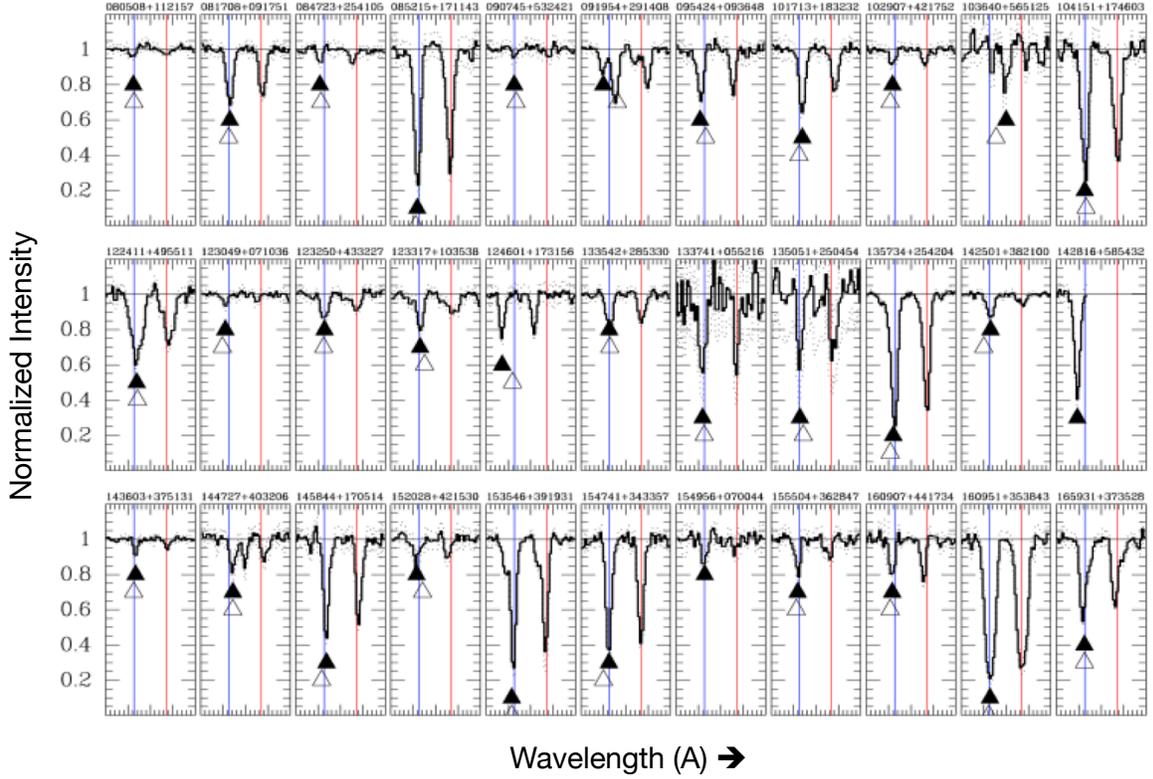}
   \end{center}
    \caption{Continuum normalized quasar spectra showing \mgII\ system
      near the target galaxy redshift.      Each spectrum
      has been deredshifted based on the emission-line redshift of the
      target galaxy. The vertical blue and red lines mark the vacuum wavelengths
      for \mgII\ 2796.352, 2803.531, respectively; tick marks are separated by 1~\AA.
      The equivalent-width-weighted velocity (solid triangles) is significantly
      Doppler shifted in 22 of these 33 detections. Open triangles mark the
      projected rotation speed of an extended galactic disk with a flat rotation
      curve.
      The dotted lines show the 1-sigma uncertainties on the normalized flux.  
      }
     \label{fig:mgII_profiles}
      \end{figure*}

Figure~\ref{fig:mgII_profiles} shows the line profiles. The \mgII\ $\lambda 2796$
line is always noticeably stronger than the {\clm $\lambda 2801$} line, so the \mgII\ $\lambda 2803$
transition is not completely saturated in these data.  Only the 
weakest systems, however, have doublet ratios consistent with the 2:1 
ratio of the optically thin limit. Some components of these systems must
have optically thick \mgII\ $\lambda 2796$. Their absorption troughs
are not black (zero intensity) because the spectral line-spread function
smooths individual components and blends velocity components together.

%LBRT

\begin{figure} 
 \begin{center}
  \includegraphics[scale=0.6,angle=0,trim = 0 0 0 0]{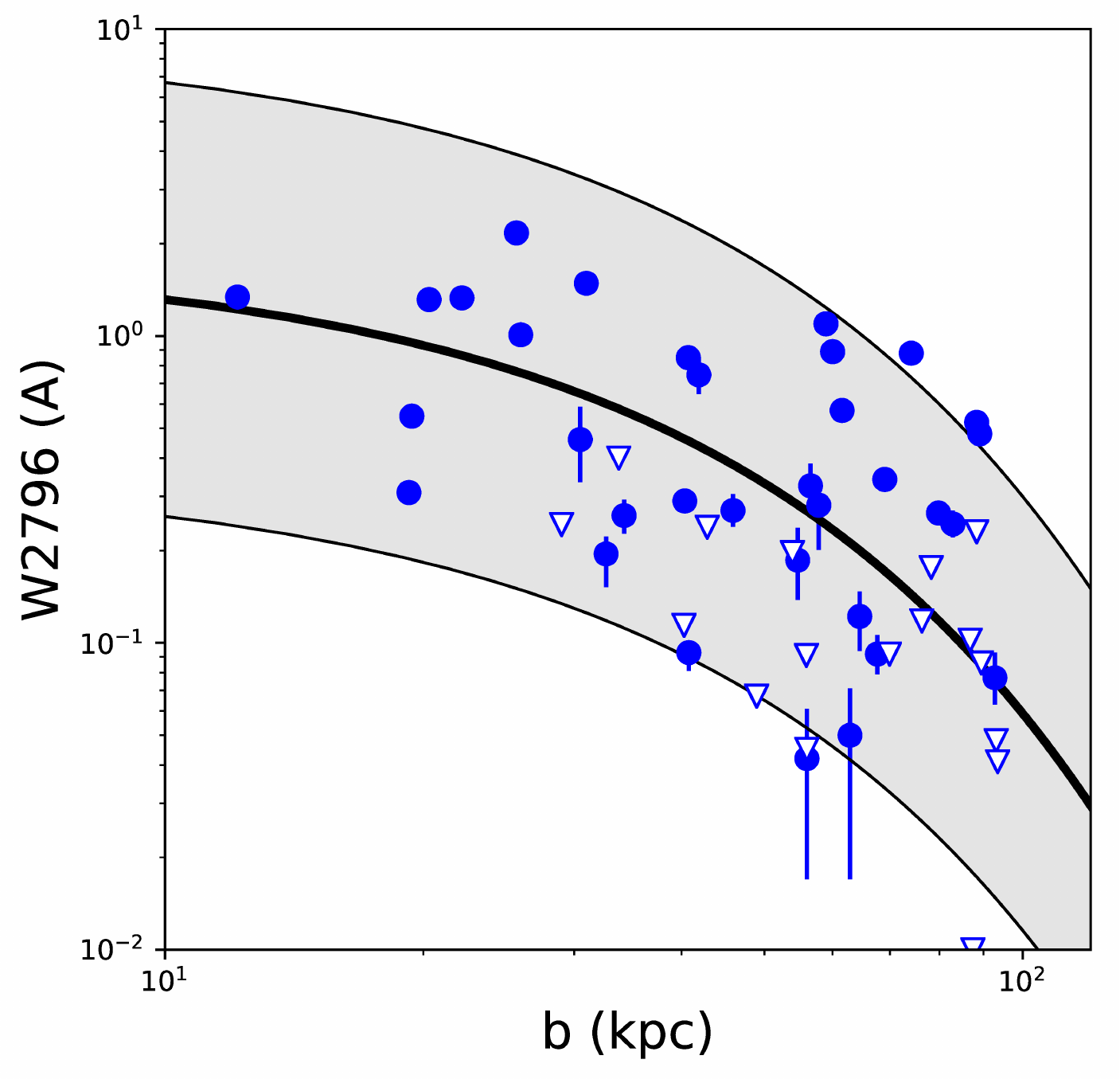}
   \end{center}
    \caption{Rest-frame equivalent width of \mgII\ 2796 versus impact
      parameter. Absorption strength declines with the distance of
      the sightline from the galaxy, consistent with previous studies.
      For comparison, the black line shows the maximum likelihood fit
      from \cite{Nielsen2013}; the shaded region shows the root-mean-square
      variation between this fit and their sample.
      }
      \label{fig:ew_b} 
       \end{figure}

In Figure~\ref{fig:ew_b} we plot the radial distribution of the rest-frame
equivalent width of the \mgII\ $\lambda 2796$ line as a function of impact
parameter.  The absorption strength declines as the separation of the sightline 
from the target galaxy increases, conforming to the previously established correlation
between these quantities \citep{HWChen2010,Nielsen2013}. 
Scaling the impact parameter to the virial radius \citep{Churchill2013a,Churchill2013b} 
or halo scale radius produced similar results. This $W - b$ relation defines
the expected value of  \wavg\  in each sightline, and we define the excess
equivalent width by $\Delta W(\lambda 2796) \equiv W_{r} / \wavg$.

The 2-sigma upper limits for the non-detections have median value of 0.1 \AA. 
Most of the quasar  spectra are therefore sensitive enough to detect a \mgII\ 
absorber of average strength  out to $b = 85$~kpc, as indicated by the solid line
in Figure~\ref{fig:ew_b}.
The spectra without \mgII\ detections are generally sensitive enough to exclude
typical absorption strengths; the upper limits are below $\wavg$.  Figure~\ref{fig:ew_b} 
shows four exceptions at large impact parameter: \j081521+082623, \j135522+303324, 
\j171220+291806, and \j114927+142002. Most of the non-detections do not exclude 
a \mgII\ system a standard deviation weaker than the median strength.

The transition from the ground state of \mgI\ to the 3s3p $^1P_0$ level
has a large oscillator strength, $f = 1.83$ \citep{Morton2003}. Neutral
magnesium produces absorption at $\lambda 2852.9642 (1 + z_{abs})$, a wavelength 
covered by our spectra for 48 of 50 sightlines.\footnote{The \mgI\ line is
     not covered by our \j142816+585432 and \j123049+071036 spectra.}
We detect \mgI\ absorption in 12 sightlines. 
We find that the
\mgI\ Doppler shifts are consistent with the \mgII\ Doppler shifts.
Table~\ref{tab:mgi} lists
absorber equivalent widths. The line centers are optically thick, and
these measurements place lower limits on the column density 
$N(MgI) \ge\ 7.59 \times 10^{12}$\col $W({\rm \AA})$.

The ionization potential of \mgI\ is 7.65~eV, substantially lower than
that of hydrogen.  The detection of \mgI\ does not require the system 
to have a large  neutral hydrogen column. The $N(MgI)/N(MgII)$ ratio 
is sensitive to the total hydrogen density.  The  calculations in Figure 6 of 
\cite{Ellison2003} illustrate this point for a $\log N(HI) = 19$\col\ slab of 
CGM photoionized by the UV background. The $N(MgI)/N(MgII) $ ratio will be
very low, less than a part per million for clouds at the average density of the CGM,
which is between $n_H \approx 1.5-4.0 \times 10^{-4}$\cm3\ in the inner CGM \citep{Werk2014}.
High gas density, roughly 10\cm3\ for $\log N(HI) = 19$\col, raises the 
$N(MgI)/N(MgII)$ ratio to 1\%.

Sightlines with \mgI\ detections are not expected to select the large
neutral hydrogen columns associated with galactic disks. Indeed
damping wings on the hydrogen absorption-line profiles identify
systems with $N(\hI) \ge\ 2 \times 10^{20}$\col, and these damped \lya\ 
absorbers (DLAs) are not typically detected in \mgI\ absorption \citep{Rao2000}. 
Under the assumption that all our sightlines probe typical halo hydrogen columns, 
$\ge\ 10^{14}$\col \citep{Chen2005}, then the \mgI\ detections identify clouds of
higher than average density and low ionization parameter.

We find the strongest \mgI\ system, $W_{r,2853} = 1.64$~\AA,  
in the \j104151+174603 sightline.  The \j104151+174558 \mgI\ absorption 
is about four times stronger than the second ranked system, 
suggesting this sightline has some unique attribute.  
The target galaxy is a confirmed group member, but it is not clear whether
the environment explains the strong absorption. This galaxy also
has the lowest stellar mass  in our sample. The quasar sightline intersects 
the disk plane of at a radius of just $R = 25.5$~kpc, and we note that 
six of the 12 sightlines with \mgI\ detections intersect the disk plane 
at $R \le\ 40$~kpc. Extended disks with hydrogen columns lower than
DLAs could also produce some, but not all,  of the \mgI\ systems. 

In Section~\ref{sec:environment}, we identified \j091954+291345 as 
interacting with another galaxy. The widely separated \mgII\ components 
may come from separate galaxies. The \j091954+291508 spectrum detects both 
velocity components in \mgI.  We will argue in Section~\ref{sec:multi} that
dense gas in galactic outflows provides another source of \mgI\ absorption.

%%%%%%% TABLE 3
%\input tables/quasar_table.tex
\input table3.tex

%LBRT

%%%%%%% TABLE 4 - MgI
%\input tables/mgi_table.tex
\input table4.tex

%%%%%%%%%%%%%%%%%%%%%%%%%%%%%%%%%%%%%%%%%%%%%%%%%%%%%%%%%%%%%%%%%%%%%%%%%%%%%%%%
\section{Kinematics of Low-Ionization Circumgalactic Gas} \label{sec:results}

Figure~\ref{fig:mgII_profiles} shows the \mgII\ line profiles in the reference 
frames of the target galaxies. Asymmetric profiles are common, indicating that
multiple velocity components contribute to the absorption systems. Blending of 
unresolved components can produce larger equivalent widths than a single saturated 
component. Furthermore, because the resolution does not resolve saturated components,
line strengths only place lower limits on the Mg$^+$ column density. For these reasons,
we describe each absorption system by its equivalent-width-weighted mean velocity, \vavg, 
and the rest-frame equivalent width of \mgII\ $\lambda 2796$, $W_r$. 

The components contributing to an absorption system may include a warped disk, 
streams produced by the tidal disruption of satellites, or a galactic outflow.
Cold flow accretion may also produce absorption, although calculations suggest these
streams are more common at higher redshift. Numerous recent studies address the 
origin of cool circumgalactic gas \citep{Voit2015,McCourt2015,McCourt2016,Voit2017,
GronkeOh2018,Schneider2018}, but no consensus regarding its formation and destruction 
has been reached. Observations of the circumgalactic gas kinematics offer a new
perspective on these problems.

A unique aspect of our study is the large number of measured galactic rotation curves.
We use them to illustrate the line-of-sight velocities produced by circular orbits in 
the plane of the galactic disk. These centrifugally supported disk components are 
indicated in Figure~\ref{fig:mgII_profiles}. Sec.~\ref{sec:discussion} will discusses 
individual velocity components further. In this section, we describe the
absorption systems.

\subsection{Corotation with Galactic Disks}

We detected \mgII\ absorption in 33 sightlines and measured galactic rotation curves
for 31 host galaxies. We compare \vavg\ to the projected disk rotation speed.
If the sign of the \mgII\ Doppler shift  matches the sign of the disk rotation,  
then we say the system {\it corotates} with the disk.  When the equivalent-width-weighted
absorption cannot be distinguished from the systemic velocity of the galaxy, i.e., the
Doppler shift is less than our measurement uncertainties, we label the system
{\it systemic absorption}. {\it Counter-rotating} absorbers have a net Doppler shift,
but the sign of the Doppler shift is opposite that of the galactic disk.

\begin{figure}  
 \begin{center}
  \includegraphics[scale=0.6,angle=0,trim = 0 0 0 15]{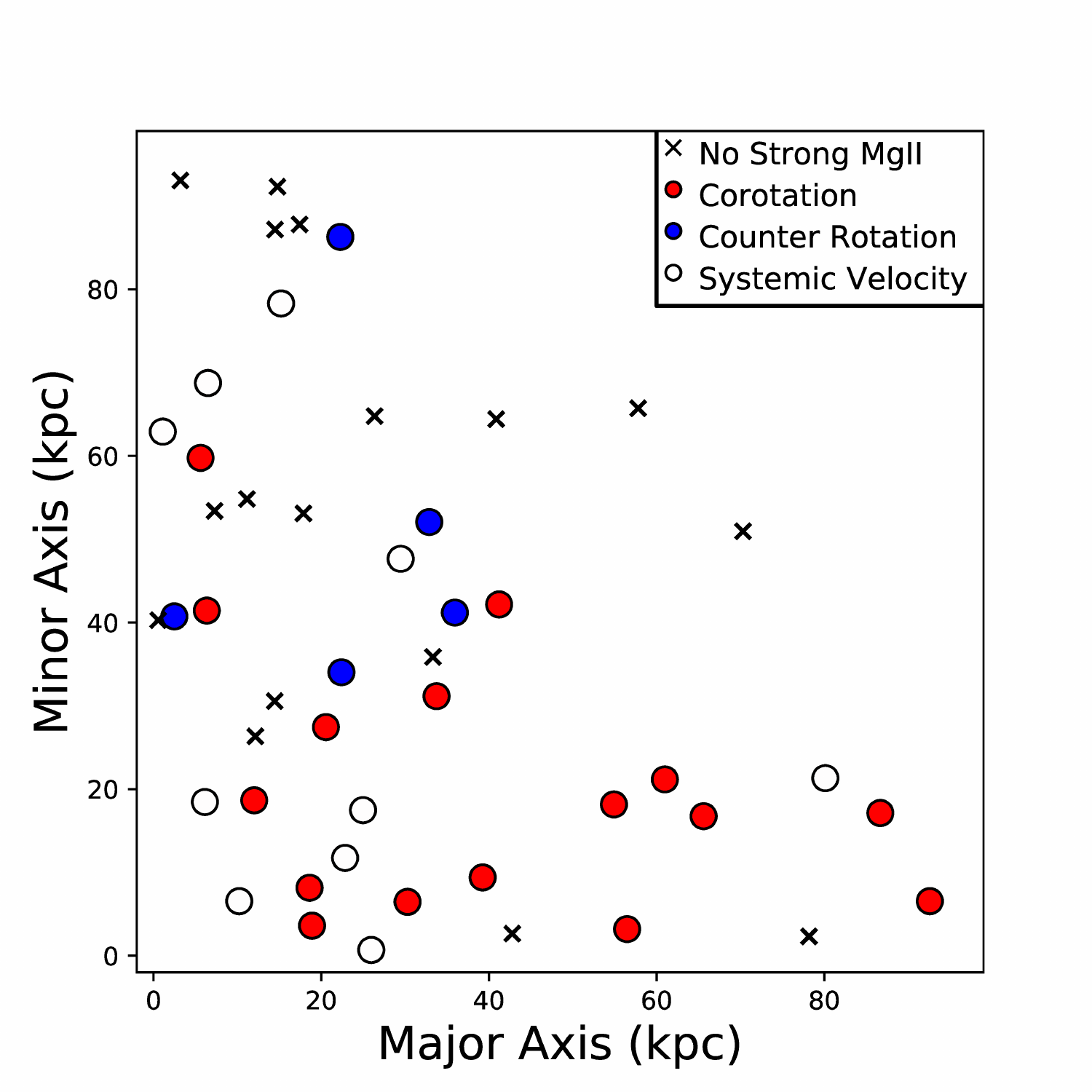}
   \end{center}
    \caption{{\it Top:}
      Doppler shifts of \mgII\ systems in quasar sightlines. The location of
      each sightline is plotted relative to the projected plane of the galactic
      disk on the sky. The redshifted  side of the major axis has been aligned
      with the positive x-axis of the coordinate system. The colorbar describes
      corotation (red), counter rotation (blue), and systemic absorption (white).  
      Sightlines without \mgII\ detections are marked with an {\it X}.
      }
      \label{fig:ab_corot}
       \end{figure}

Figure~\ref{fig:ab_corot} shows the locations of the quasar sightlines 
relative to the projection of the galactic disks on the sky. 
Each galaxy image has been rotated to align
the semi-major axis with the abscissa.  The location
of the sightline on the sky is given by the polar coordinates $(b,\alpha)$.

%LBRT

At azimuthal angles less than 46\deg, we find 12 corotating systems. Ten
of these are within 30\deg\ of the major axis, and we described them previously
in \cite{Ho2017}. We add \j122411+495511 ($ 45.6\deg$) and \j144727+403206
($ 41.5\deg$). Over the same range of azimuthal angle, the \vavg\ measurements 
are indistinguishable from the systemic velocity along 5 additional sightlines.
We find no counter-rotating absorbers. The absence of counter-rotating gas is 
highly significant. 

To appreciate the significance of these statistics,
consider tossing two coins. There are four possible outcomes, and two of
these result in the coins showing the same face.  This thought experiment
is analogous to measuring whether the CGM and the galactic disk have an
angular momentum component in the same direction.  Out of 12 successful
coin tosses (or net Doppler shift measurements), the chance of getting
alignment all 12 times is $P = 2.4 \times 10^{-4}$. This result comes from
a binomial distribution with a mean of 6 and standard deviation of 1.73.

If we consider only those sightlines within 45\deg\ of the minor axis, then
the number of  counter-rotating \mgII\ systems is similar to the number of 
corotating systems.  We find no correlation between the sign of the 
Doppler shift of the CGM and the rotation of galactic disks near the minor axis.

%LBRT
\begin{figure}  
 \begin{center}
  \includegraphics[scale=0.30,angle=-90,trim = 10 -40 40 0]{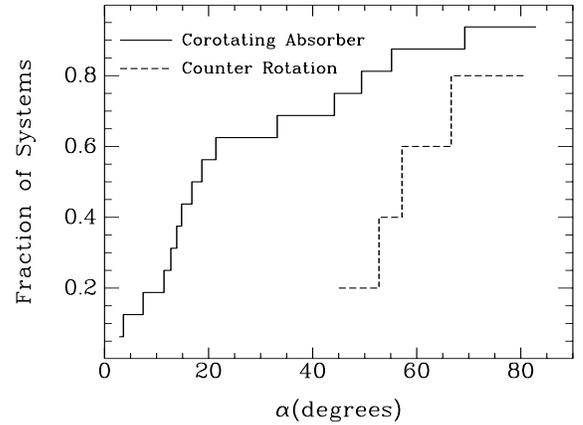}
   \end{center}
    \caption{
      Cumulative azimuthal distributions of corotating and counter-rotating \mgII\ systems.
      To compare these distributions we use Kuiper's statistic, a KS-test defined on a circle 
      to eliminate sensitivity to the endpoints. The null hypothesis states that the 
      sightlines with corotating systems are drawn from the same parent distribution 
      as the sightlines with counter-rotating systems. The probability of 
      the null hypothesis is 3.8\%, so these distributions are significantly different. The 
      systemic absorbers in contrast are randomly distributed in azimuthal angle as shown
      in Figure~\ref{fig:ab_corot}. 
      }
      \label{fig:ks_alpha}
       \end{figure}

Figure~\ref{fig:ks_alpha} compares the azimuthal distribution of corotating
and counter-rotating \mgII\ systems.  A KS test quantifies the chance
of finding such a large separation between two samples drawn from the same
underlying distribution. We used Kuiper's statistic to provide uniform sensitivity 
across azimuthal angle \citep{numrec} and found the probability of this null 
hypothesis is just 3.8\%. The corotators and counter-rotators do not have the
same azimuthal distribution.

%LBRT
\begin{figure} 
 \begin{center}
  \includegraphics[scale=0.3,angle=-90,trim = 40 -30 40 0]{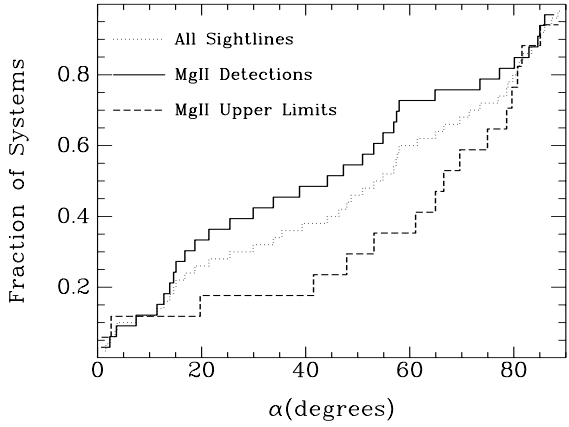}
   \end{center}
    \caption{Cumulative azimuthal distribution of \mgII\ detections and non-detections.
      The probability that these samples were drawn from the same parent distribution
      is 7.3\%, where we have again used the Kuiper statistic to perform the KS-test.
      Comparison to the distribution of all sightlines indicates that the difference is 
      driven in large part by the small fraction of non-detections near the major axis.
      }
     \label{fig:ks_hits} 
      \end{figure}

Figure~\ref{fig:ks_hits} shows how the cumulative distribution of \mgII\
detections and non-detections differ in azimuthal angle. The paucity of 
non-detections at low $\alpha$ means the covering factor of strong \mgII\
absorption is high there.  Three of the non-detections correspond to sightlines 
that probe the large open area in the upper right quadrant of Figure~\ref{fig:ab_corot}. 
The sensitivity limits for each sightline are provided in Table~\ref{tab:quasar} 
and must be taken into account when using these data to measure covering fractions.

\subsection{Velocity Range \& Absorption Strength}

\begin{figure} 
 \begin{center}
  \includegraphics[scale=0.58,angle=0,trim = 0 0 0 0]{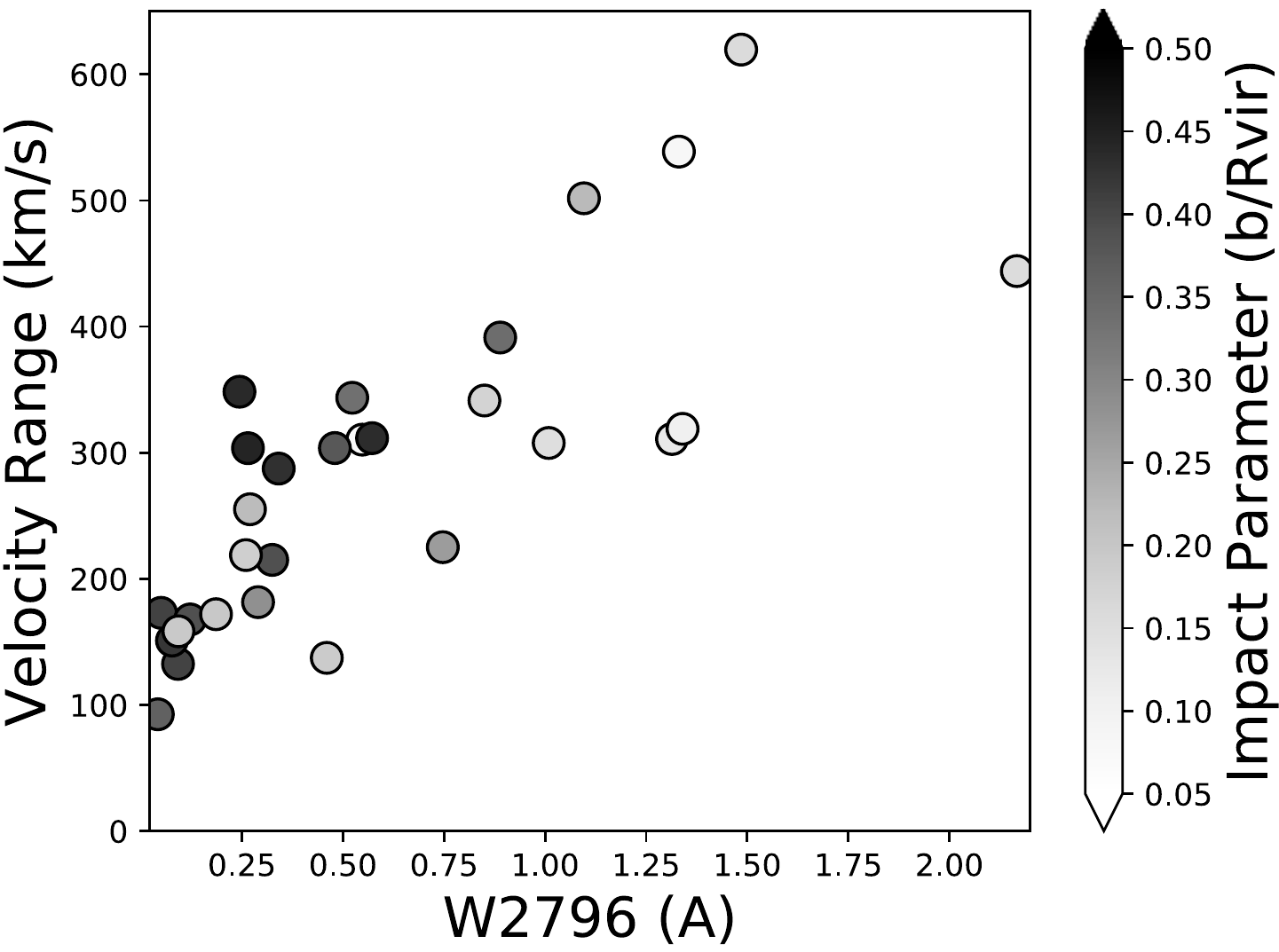}
   \end{center}
    \caption{Velocity range of \mgII\ absorption versus
      rest-frame equivalent width of \mgII\ 2796. The grayscale 
      identifies the impact parameter of each sightline.  The Spearman
      rank-order correlation coefficient is $r_s = 0.82 ~(4.3\sigma)$, 
      indicating a significant correlation. 
      }
      \label{fig:vavg_ew} 
       \end{figure}

We expect the line strength to be sensitive to the gas kinematics because
of optical depth effects. In Figure~\ref{fig:vavg_ew} we show that the
equivalent width correlates strongly with the velocity
range of the \mgII\ absorption. A large
velocity range for the absorption indicates a kinematic disturbance.
Turbulent gas motion may broaden individual absorption components.  
The largest velocity ranges reach many hundred\kms, however, and
are more likely produced by multiple velocity components.

\begin{figure} 
 \begin{center}
 \vbox{
   \includegraphics[scale=0.34,angle=-90,trim = 0 0 60 0]{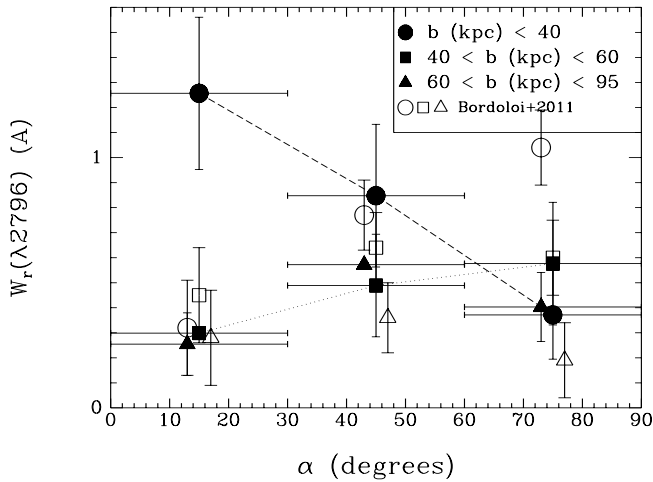}
   }
    \end{center}
     \caption{Absorption strength vs. azimuthal angle, binned by impact
       parameter. The minor-axis excess seen at
       $b > 40$~kpc agrees with \cite{Bordoloi2011} and \cite{Lan2018}
       who analyzed large numbers of sightlines.  
       The absorption strength at low $b$, however, increases toward the
       major axis; previous studies did not sample the low $b$ sightlines well.
      }
      \label{fig:w_alpha} 
       \end{figure}

For a given impact parameter then, absorption tends to be stronger than average 
along sightlines within 45\deg\ of the minor axis. 
We applied a KS test to the sightlines with an excess and a deficit of absorption.
Kuiper's statistic yields a 20\% probability that the 
azimuthal distributions of excess and deficit absorbers were drawn
from the same parent population.\footnote{
      This analysis includes \j142816+585432 and \j154956+070044,
      the two galaxies without rotation curve measurements, even
      though they cannot be shown in Figure~\ref{fig:w_avg}.} 
The significance of the azimuthal variation of the excess absorption is therefore
marginal in our sample. Other studies, however, robustly show that equivalent
width increases toward the minor axis \citep{Bordoloi2011,Lan2018}. Two factors explain the 
difference. First, our sample includes a larger fraction of low $b$, low $\alpha$
sightlines; these major axis sightlines also show excess equivalent width, diluting the trend
measured in the opposite direction at larger $b$. Second, studies like the one
of \cite{Lan2018} have many, many more sightlines. We conclude that there is
strong azimuthal variation in excess equivalent width, which increases with azimuthal
angle at $b > 40$~kpc.  For the first time, however, we demonstrate that
equivalent widths actually increase toward the major axis at smaller impact parameters.
Figure~\ref{fig:w_alpha} bins our data by impact parameter and further illustrates
these azimuthal variations.

\begin{figure} 
 \begin{center}
 \vbox{
  \includegraphics[scale=0.6,angle=0,trim = 0 0 0 0]{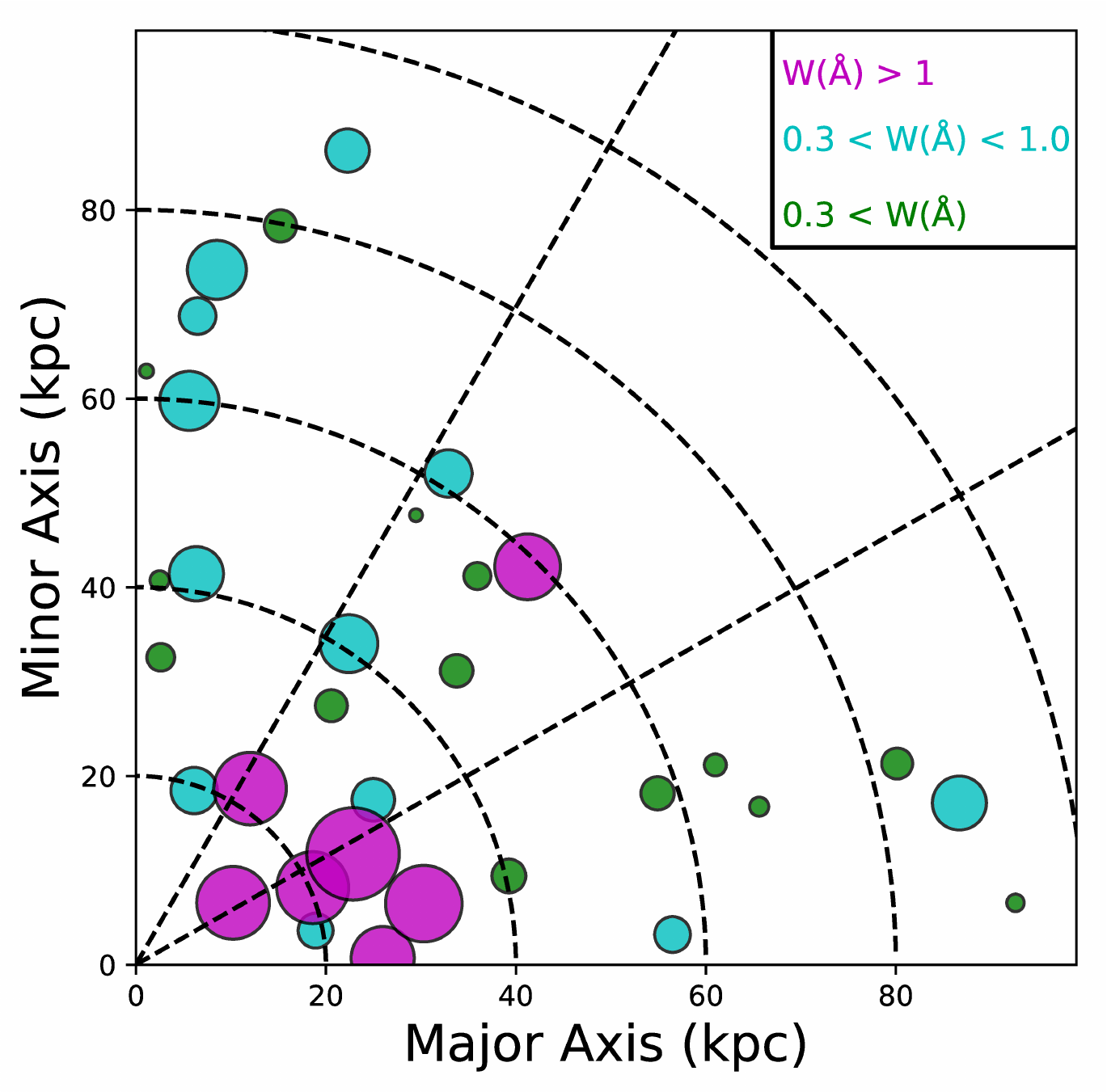}
   }
    \end{center}
     \caption{Absorption strength vs. azimuthal angle.  
       We have scaled the symbol size by the equivalent width, 
       illustrating the very strong absorbers ($W_r \ge\ 1.0$ \AA),
       strong absorbers ($0.3 \le W_r < 1.0$ \AA), and
       optically thin absorbers  ($W_r < 0.3$ \AA).  Many of the
       very strong absorbers were found in low impact parameter 
       sightlines near the major axis, so they may intersect
       extended disks.  The strong absorbers confirm the established 
       trend that absorption strength increases near the minor axis. 
      }
      \label{fig:clm_ab} 
       \end{figure}

\begin{figure}
 \begin{center}
  \includegraphics[scale=0.6,angle=0,trim = 0 0 0 10]{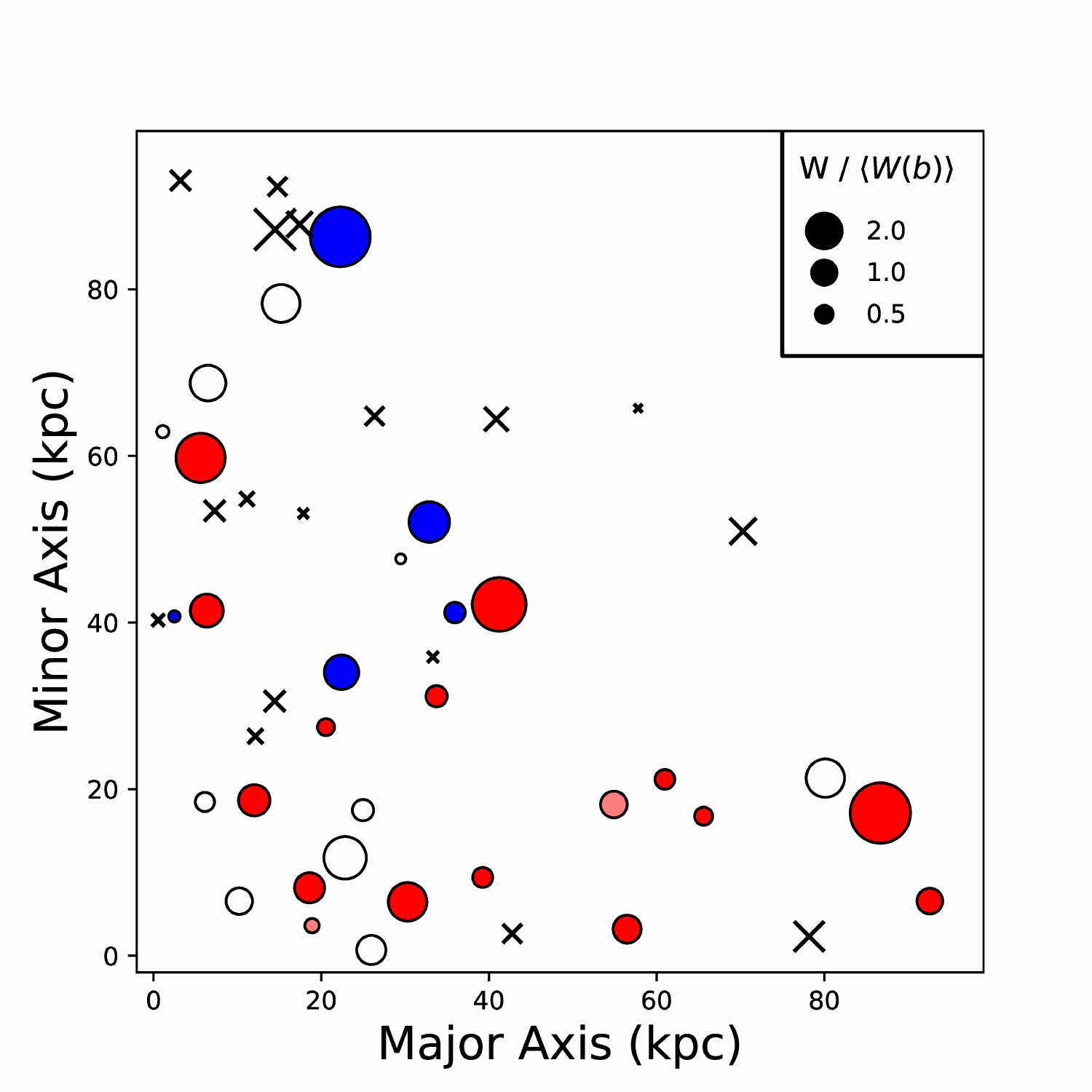}
   \end{center}
    \caption{Relative \mgII\ absorption strength. The coordinate
      system is the same as in Figure~\ref{fig:ab_corot}.  Large
      and small symbols indicate an excess or a deficit of absorption,
      respectively, compared to the average equivalent width at that 
      impact parameter.  The solid line in Figure~\ref{fig:ew_b} illustrates
      the factor $\langle W(b) \rangle$ that we have used to remove
      the strong dependence of absorption strength on impact parameter.
      The largest  excess occurs in \j091954+291408. Typically, however,
      large excess equivalent width is found most often along sightlines
      with large azimuthal angles, $\alpha > 45\deg$, and larger
      impact parameters, $b > 40$~kpc.} 
      \label{fig:w_avg}
       \end{figure}

To gain insight into the nature of the extra component, we examined
how line strength depends on azimuthal angle. Figure~\ref{fig:clm_ab}
separates the sightlines into very strong, strong, and optically thin
\mgII\ absorption.  The strong systems confirm the previously established
trend \citep{Bordoloi2011,Bouche2012,Kacprzak2012,Lan2014,Lan2018}, 
namely that the equivalent width increases toward the minor
axis. Previously missed, however, is the very strong absorption near the
major axis at small impact parameter. The very strong absorption at
$ b < 40$~kpc was missed by previous surveys due their small number of 
close pairs. One possible explanation for this very strong component 
is absorption by the galactic disks.

We detect optically thin systems at random locations in the $(b,\alpha)$ projection.
Their strength is not correlated with azimuthal angle. These weak
lines directly measure the column density $N({\rm Mg}^+)$. We find no variation in 
$N({\rm Mg}^+)$ column with azimuthal angle.

%LBRT 10

To gain more insight into which sightlines show excess absorption,
we divide the rest-frame equivalent widths by the average value
of the \cite{Nielsen2013} fit. The resulting ratio is used to scale 
the size of the points in Figure~\ref{fig:w_avg}, which shows the
sightline location relative to a coordinate system determined by 
the projected geometry of each galaxy. We find 14 systems with
$W_r / \wavg  > 1.5$,  and 10 of these have azimuthal angles $\alpha > 45\deg$.
The excess absorption, $W_r/\wavg$, has a weak positive correlation with specific SFR.
Note that the absorption strength, $W_r$, is not correlated with the galactic SFRs 
in our sample. 

%LBRT 11

%LBRT 12

Do all the excess equivalent width systems have blended velocity 
components?
Our sample includes five systems with $W_r \ge\ 3 \wavg$.
Three of these systems --  \j091954+291408, \j095424+093648, and \j142816+585432 
lie above the shaded region in Figure~\ref{fig:ew_b}, and the other two systems 
\j122411+495511 and \j165931+373528 lie nearby. 
As shown above, two components produce the large excess equivalent width in 
\j091954+291408. Line wings are prominent on the \j095424+093648, \j122411+495511,
 and \j165931+373528 \mgII\ profiles. Strong excess absorption is therefore produced
by blends of multiple components.

We previously showed
that the stronger component in the \j165931+373528 spectra can
be easily described by a galactic outflow \citep{Kacprzak2014}.
Like the \j165931+373528 sightline, the \j122411+495511 and \j095424+093648 
sightlines probe large azimuthal angles.  
We will discuss whether galactic outflow create their equivalent width excess.

%LBRT 13

\section{Discussion} \label{sec:discussion}
%\section{Models for \mgII\ Kinematics} \label{sec:models}

We find that a high fraction of \mgII\ systems corotate with galactic
disks, particularly in sightlines near the galactic major axis. 
This result rules out the simplest model for the CGM kinematics; 
the motion of the low-ionization gas is not random.  Matched signs
for the absorbers and disk Doppler shifts indicate the inner CGM
has a component of angular momentum in the direction of the disk's
angular momentum vector.  The high angular momentum of cold flow gas 
in simulations is related to the fact that it tends to be accreted 
along filaments \citep{Stewart2011b}. Those filaments also direct
satellites toward the central galaxy, and the gas stripped from 
the satellite winds provides another important source of material for building
galactic disks \citep{Angles-Alcazar2017,Hafen2018}. The detection of
high angular momentum gas in the CGM raises the question of whether
the warped extended disks that form in the simulations might be common
even at low redshift. Direct 21-cm observations of disks on 
circumgalactic scales are rare \citep{Pisano2014} because they
require detecting \ion{H}{1} column densities well below the critical 
value where the neutral gas is vulnerable to the cosmic ionizing 
background \citep{Bland-Hawthorn2017}.

We also confirm the presence of a kinematic disturbance concentrated
around the minor axis where no net corotation is measured. This result
is consistent with the classical picture of a bipolar wind. Yet simulated
outflows from disks in a cosmological setting, where the disk is free
to wobble and move, sometimes produce more spherical outflows even at  
low redshift. The short lifetime of accelerated clouds raises concerns
about interpreting the kinematic disturbance as a continuous outflow.
What we observe might be better described as condensation triggered
by a wind, a fine distinction that has important implications for
modeling feedback.

To gain further insight into these outstanding issues, we compare 
the circumgalactic gas kinematics to two axisymmetric models in this
section. Doing so requires an assumption about the location of the 
absorbing gas along each sightline, introducing potential systemic errors
that do not affect the results of the previous section. In Section~\ref{sec:thin_disk} 
we compare the \vavg\ measurements to an extended disk model in order
to further evaluate the possibility that corotating gas feeds galactic disks. 
Minor-axis sightlines may miss extended disk-like structures because they 
intersect the disk plane at larger galactocentric radii than do major-axis
sightlines at similar impact parameters on the sky. We resolve the orientations 
of galactic disks in Section~\ref{sec:winds} and test the conjecture that 
bipolar outflows produce the excess absorption at high azimuthal angles.
After identifying outflow components and describing their relationship to
galaxy properties, we return to our discussion of the disk plane
in Section~\ref{sec:ovi}, where we place limits on the spatial extent of 
the corotating gas.

\subsection{Thin Disk Component} \label{sec:thin_disk}

In this section we describe the velocity component added 
to an absorption system by a centrifugally supported disk.  Thin disk
models will not explain the velocity range of an entire \mgII\ system.
A sightline intersects a thin disk over a short pathlength and samples
a small range in line-of-sight velocity, producing a narrow line profile.
The broader range observed may be attributed to spatially distinct
components along a sightline, gas inflow within a disk \citep{Ho2017},
and/or a cylindrical model for the corotation \citep{Steidel2002}. 
Higher resolution spectroscopy would resolve distinct velocity 
components.\footnote{We detected two components from 
       the CGM of \j091954.11+291345.3, and in this section we consider 
       only the stronger component.}
We evaluate which systems might include a component produced by  a
centrifugally supported gas near the disk plane.

For our purposes here, an extended disk is an axisymmetric structure with an angular
momentum vector roughly parallel to that of the galactic gas disk. An extended disk has a 
finite radius. Sightlines that intersect the disk plane at small radii are more 
likely to detect an extended disk. All the major axis sightlines intercept the disk plane 
at $R < 100$~kpc. A few minor-axis sightlines also pass through the disk plane at 
relatively small radii, $R \sles\ 40 $~kpc, but other high minor-axis sightlines intercept 
that plane at many hundred kpc. Disk inclination has a large effect on whether minor-axis 
sightlines intersect an extended disk. We project all the sightlines onto the disk plane 
for purposes of illustration because we do not know the size of these structure a priori.

Consider an extended thin disk whose
angular momentum vector is parallel to that of the galactic disk. The disk
lies in the xy-plane and has an angular momentum vector in the z-direction.
We calculate the coordinates $(x_0,y_0,0)$ where each  sightline intersects the disk plane.
In cylindrical polar coordinates the sightline intersects the
disk at a point (R,$\phi$,0), where the angle $\phi$ is measured from the x-axis
to the point $(x_0,y_0)$.
The inclination \inc\ describes the angle between the z-axis and the sightline,
and we adopt the convention that \inc\ is positive (negative) for quasars at
positive (negative) z-coordinates.

We choose the x-axis so that it is aligned with the receeding semi-major axis of the disk on the sky. 
Then we have  $x_0 = b \cos \alpha$, where the azimuthal angle $\alpha$ is defined on the interval
$(-180\deg,180\deg]$. The value of $y_0$ depends on the disk inclination as well as $b \sin \alpha$.
We adopt the convention where the quasar is at some very large value of $y$, and the observer is
at a very negative value of $y$. A positive inclination then produces 
$y_0 = - b \sin (\alpha) / \cos(\inc)$, and the sign is reversed for a negative disk inclination.
The quasar sightline intersects the disk plane at radius
\begin{equation}
R = b \sqrt{1 + \sin^2 \alpha \tan^2 \inc}.
\end{equation}

We project the rotation of the extended disk onto the plane of the sky.
Disks produce absorption at a maximum Doppler shift, $V_c(R) \sin \inc$, in major axis sightlines.
The same disks produce absorption at the systemic velocity in minor-axis sightlines because the 
velocities of the circular orbits are perpendicular to the sightline. Smaller Doppler shifts 
will be detected at intermediate azimuthal angles. The orbital motion produces a line-of-sight velocity 
\begin{equation}
\vlos = \frac{V_c(R) \sin \inc \cos \alpha}{\sqrt{1 + \sin^2 \alpha \tan^2 \inc}}.
\label{eqn:disk_vlos} \end{equation}
in the quasar sightline. The major-axis sightlines therefore
have a more favorable geometry for identifying a disk kinematically. 
Ambiguity about which side of the disk is nearer will not affect the predicted line-of-sight velocities
for circular orbits. 

\begin{figure} 
 \begin{center}
  \includegraphics[scale=0.58,angle=0,trim = 0 0 0 0]{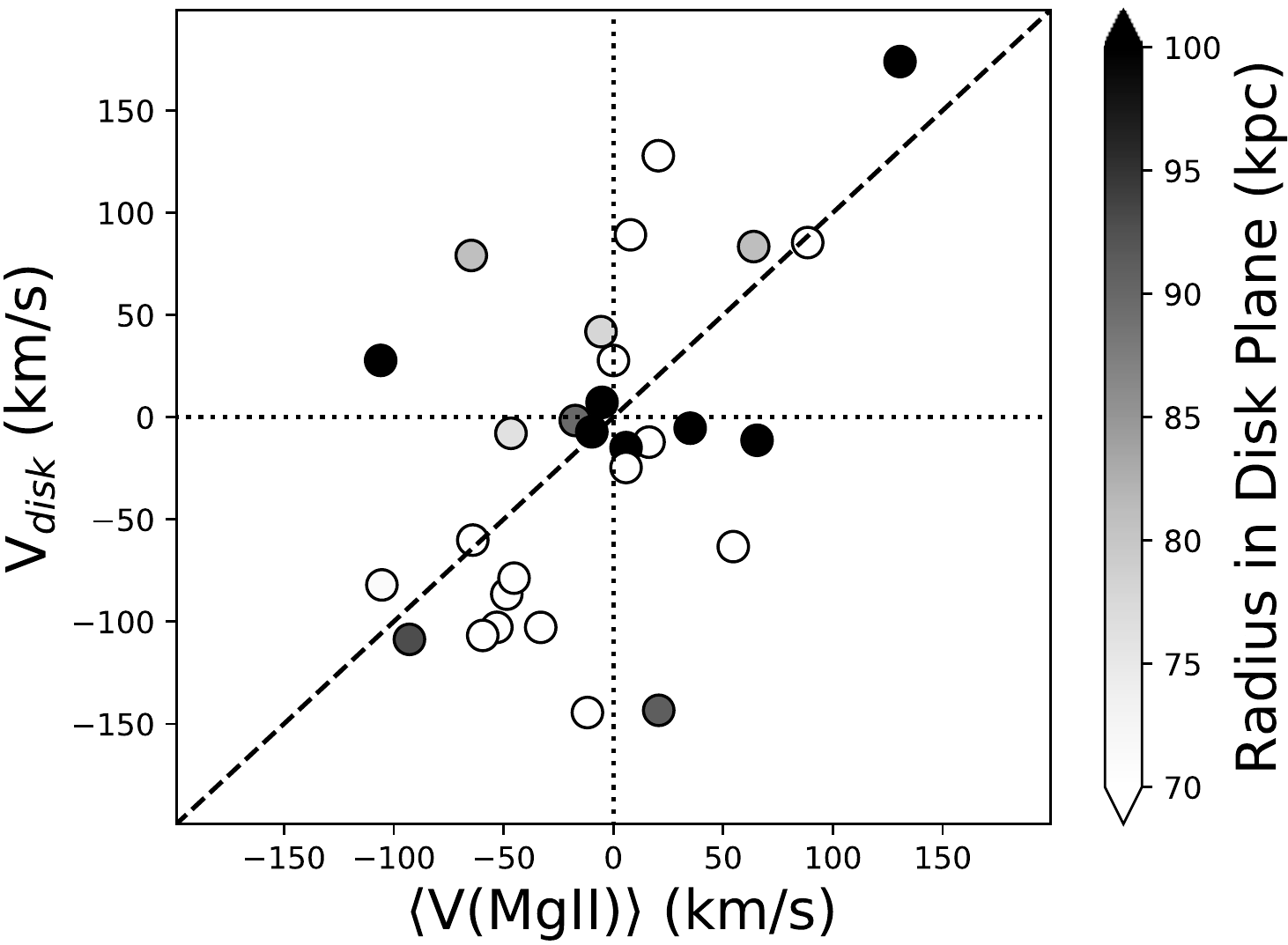}
   \end{center}
    \caption{Disk model versus \mgII\ Doppler shifts. 
      The gas in the model lies in the plane defined by 
      the galactic disk and moves on circular orbits.
      This projected velocities of this model are positively
      correlated with the \vavg\ measurements. The correlation
      is strongest among the sightlines that intersect the disk 
      plane at radii $R < 70$~kpc.  The Spearman rank order correlation coefficent
      is $r_S  = 0.64$ for this subset of the data, which rules out the
      the null hypothesis (no correlation) at the $2.4 \sigma$ level.
      }
      \label{fig:vv} 
       \end{figure}

Motivated by the correlation in the signs of the Doppler shifts of the
\mgII\ systems and the disk model, we examined the relationship between
the magnitude of \vavg\ and the projected speed. Starting with the subset 
of sightlines at $R \sgreat\ 30 $~kpc, we computed the Spearman rank order
correlation coefficient, finding a positive correlation. We then 
increased the sample size by one sightline iteratively until all sightlines
were included. The Spearman rank-order correlation coefficient increased up to $R
\le\ 70$~kpc and then decreased sharply as we continued to add sightlines at 
$R \ge\ 90$~kpc. This result suggests the extent of the corotating gas that is 
{\it disky} is roughly $R_{max} \approx\ 80 \pm 10  $~kpc. 
The lighter colored points in Figure~\ref{fig:vv} illustrate this correlation 
between the \mgII\ Doppler shifts and the disk model.\footnote{ 
      Four of the 33 galaxy -- quasar pairs with \mgII\ detections are
      excluded from this section. Following \cite{Ho2017}, we exclude the 
      \j124601.81+173156.4 and \j103640.74+565125.9 sightlines because the 
      absorbers may not be related to the CGM of the target galaxy.
      Our NIRC2 imaging resolved \j142815.41+585442.1 and \j154956.73+070056.0 
      into multiple galaxies, so we could not measure their rotation curves 
      and therefore exclude them as well.} 

This result strongly suggests that
centrifugal support of the low-ionization CGM is significant out to at least $R_{max}$. 
We observed very few major-axis sightlines at larger radii, so our results do not exclude 
corotating gas at $R > R_{max}$. Inspection of Figure~\ref{fig:vv} also shows that the
correlated points at $R < R_{max}$ generally have lower speeds than would be expected 
for circular orbits. Discrepancies introduced by the warping of large disks would scatter
the residual velocities in both directions. The lower observed speeds indicate that 
the angular momentum of the low-ionization gas is not entirely sufficient to support 
it against gravity, so the gas spirals in toward the galaxy.  We have simply extrapolated
a flat rotation curve to make this comparison because it makes our results easier to
reproduce. Halo circular velocities, see Section~4.1 and Figures 6 and 7 of \cite{Ho2017}, 
are lower than this extrapolation predicts but still higher than the deprojected \vavg\
values.

\begin{figure}
 \begin{center}
  \includegraphics[scale=0.6,angle=0,trim = 0 0 0 10]{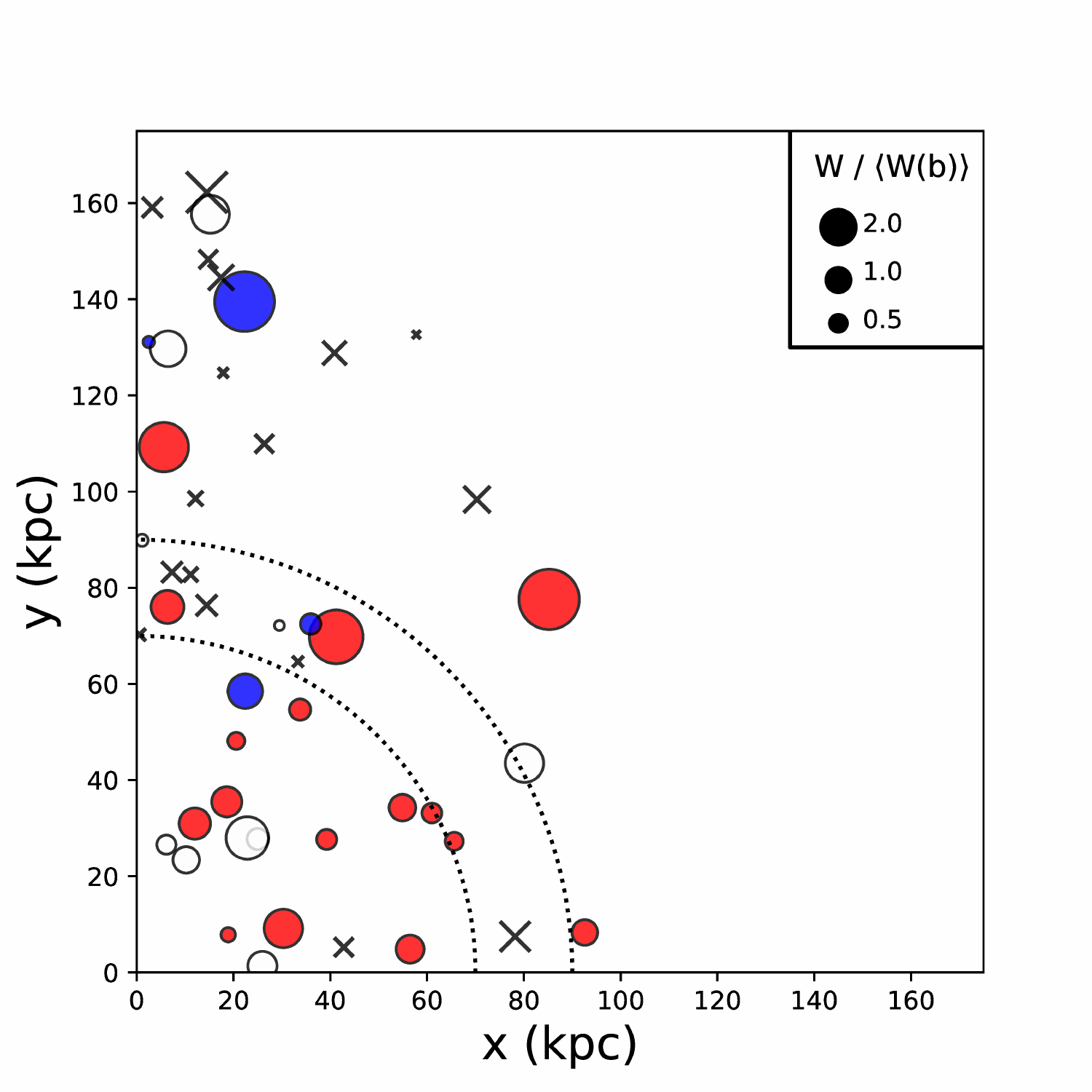}
   \end{center}
    \caption{Relative \mgII\ absorption strength in the disk plane.  Symbol
      color represents corotation (red), counter rotation (blue), and systemic
      velocity (white) with respect to circular orbits in the disk plane. 
      We scale the symbol size by $W_r / \langle W_r(b) \rangle$, where the
      black line in Figure~\ref{fig:ew_b} defines the expected absorption 
      strength $\langle W_r(b) \rangle$ as measured by \cite{Nielsen2013}.
      The lines show disk radii of $R_{max} = 70$ and 90~kpc.  The sightlines
      with the largest excess equivalent width intersect the disk plane at
      $R \ge\ R_{max}$. We conclude that absorption from an extended disk is 
      unlikely to be the primary source of the excess absorption.
      }
      \label{fig:R_phi_excess}
       \end{figure}

Sightlines that do not detect \mgII\ place limits on the shape of the corotating 
structure.  Extended disks, for example, are axisymmetric by definition, and sightlines
intersecting the disk plane at any coordinates $(R,\phi)$ will detect absorption so
long as $R$ is less than the maximum extent of the disk. In Figure~\ref{fig:R_phi_excess} 
the covering fraction of \mgII\ detections is near unity out to $R \approx 70$~kpc,
a radius consistent with the value of $R_{max}$ estimated above from the velocity
correlation. We estimate that the typical galaxy in our sample may be surrounded
by a gaseous disk-like structure which extends to $\approx 70- 80$~kpc.

%LBRT 14

Absorption strengths provide further insight about the relationship of the gas kinematics
to the disk plane.  If extended disks produced all the \mgII\ absorption, then we would 
expect the equivalent widths to correlate better with disk radius $R$ than they do with 
impact parameter. We confirm that \wavg\ decreases with increasing  $R$ ($r_S = -0.47, 2.5\sigma$),
as expected based on the well-established correlation between \wavg\ and $b$ ($r_S = -0.56, 3.0\sigma$). 
Projection onto the disk plane does not increase the correlation coefficient, however, so the \mgII\
kinematics do not support a pure disk description, a result that was already clear from the velocity
widths of \mgII\ systems.  The open question is which absorption components are consistent
with absorption near the disk plane.

We again use the ratio of the \mgII\ equivalent width to $\wavg$, the absorption
excess, to takes out the well-measured decline in absorption strength
with impact parameter. Excess absorption appears as large symbols in
Figure~\ref{fig:R_phi_excess}.  The sightlines with the largest absorption excess intersect 
the disk plane at $R > 80$~kpc, where the \vavg\ velocities are not very  correlated with the
projected disk rotation. We therefore conclude that most the systems with excess absorption
include velocity components {\it not} associated with gas in the disk plane. 
Outflowing gas far above the disk plane, for example, may contribute velocity components to these systems.

%LBRT  15                                                                                                   

We find the largest absorption excess in the \j091954+ 291408 sightline, which is 
near the major axis ($\alpha =15.3\deg$).
The \j091954+291508 sightline is unique within our sample, as
it shows two well separated \mgII\ components. We have interpreted
the stronger component, which is redshifted with respect to the
systemic velocity, as the one associated with the target galaxy. This stronger
component corotates with the galactic disk,  following the trend for 
low azimuthal angle sightlines. The symbol size in Figure~\ref{fig:R_phi_excess}
represents only this stronger component. This sightline impacts the disk plane at 
$R = 115$~kpc.  
Previous work has shown that the CGM extends to larger radii around group galaxies 
\citep{Bordoloi2011,Johnson2015,Nielsen2018}, and the environment of \j091954+291345
may contribute to the excess absorption.\footnote{ 
     We find evidence that our target galaxy \j091954+291345 is 
     involed in an interaction with other group members.
     The NIRC2 image shows a prominent 2-arm spiral that may be two tidal tails.
     The candidate group includes two luminous galaxies with red colors in addition to our
     target.  Spectroscopic redshifts are needed to confirm these potential group members:
     (1) J091954+291345 is 4\farcs4 from our target and has a similar impact parameter, and
     (2) J091954+291336 is 8\farcs7 from our target and further than our target from the 
     quasar sightline.  }
The second component may be associated with a different galaxy or a group.
Tidal streams generated by a galaxy -- galaxy interaction may contribute to
one or both of these absorption components.

Galactic outflows produce most of the equivalent width, however, in  
the three systems with the next highest absorption excesses. In the 
next section, we present a strategy for recognizing galactic outflows
and then show that these absorption-line profiles are inconsistent
with a disk component but easily described by a simple outflow model.
We return to the disk-like component in Section~\ref{sec:ovi} where we 
use non-detections to further constrain their properties.

\subsection{Outflow Component}  \label{sec:winds}

Galactic winds blow out of disks along the path of least resistance, forming
bipolar structures whose axis is perpendicular to the disk plane \citep{DeYoung1994}. 
Cool, dense outflows within a few kpc of galaxies have been directly imaged in optical 
emission lines \citep{Lehnert1995,Martin1998}. Mapping the Doppler shifts across these
nebulae reveals multiple velocity components whose separation and relative strength are
well described by the surface of a conical structure \citep{HAM1990}. The term {\it galactic wind}
was originally introduced to describe the much hotter fluid inside this cone \citep{Chevalier1985};
yet x-ray emission from hot winds has only been detected from a few nearby starburst galaxies
 \citep{Strickland2000}.  The cooler component, which we will call the {\it outflow}
to distinguish it from the hot wind, has been detected via blueshifted absorption
lines in galaxy spectra across much of cosmic time \citep{Heckman2000,Martin2005,
Martin2012,Erb2012,Kornei2012,Rubin2014}. The broad absorption component from an 
outflow typically overlaps the absorption component from interstellar gas \citep{Martin2012}.
This line blending complicates measurements of outflow properties.  We chose quasar sightlines
that avoid the interstellar medium, eliminating this confusion.

%LBRT
\begin{figure}
 \begin{center}
 \includegraphics[scale=0.35,angle=0,trim = 80 0 0 0]{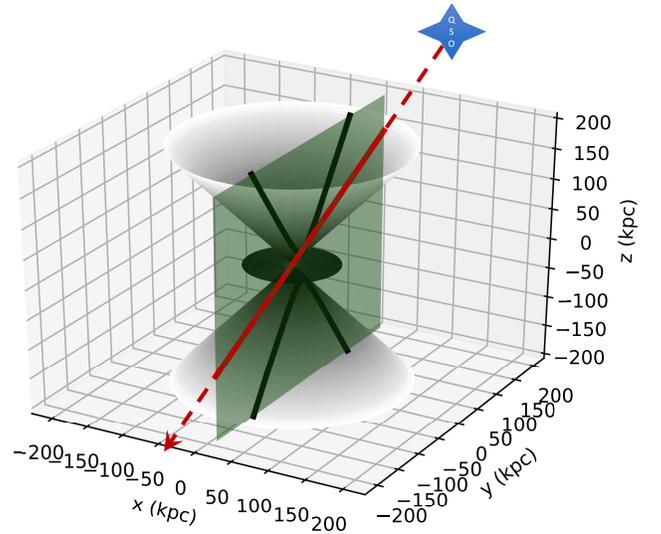}
   \caption{Coordinate system describing the three-dimensional geometry of a 
     galactic disk and bipolar outflow. The outflow is radial in a
     cone perpendicular to the disk. Clouds of low-ionization gas could 
     populate the interior of the cone or be concentrated along its surface.
     The angular momentum vector 
     of the disk defines the direction of the z-axis. The angle between the 
     z-axis and the cone defines the {\it outflow opening angle}, $\theta_{max}$.
     The  positive x-axis corresponds to the receding side of the major axis.
     A quasar sightline lies in a plane at $x = x_0$. The intersection
     of this plane with the outflow cone defines a parabola. The observer
     lies at a negative $y$-coordinate approaching infinity. 
    }
    \label{fig:3Dschema}
     \end{center}
      \end{figure}

The schematic drawing in Figure~\ref{fig:3Dschema} illustrates the intersection
of a quasar sightline and an outflow. Outflow opening angles range from 
$\theta_{max} = 30\deg$ to 45\deg\ at intermediate redshift based on the 
fraction of galaxy spectra with outflow detections \citep{Martin2012}.
Moving the quasar closer
to the minor axis on the sky produces a wider parabola in the $x = x_0$ plane, reaching 
the opening angle of the cone for a minor-axis sightline. Sightlines close to the minor 
axis therefore intercept outflowing gas over the largest pathlengths. Assuming clouds
populate the outflow cone, the longer pathlengths increase the range of observed
Doppler shifts. We attribute the observed increase in  \mgII\ equivalent widths with 
increasing azimuthal angle to this geometrical effect.

\begin{figure}
 \begin{center}
   \includegraphics[scale=0.35,angle=0,trim = 50 00 0 0]{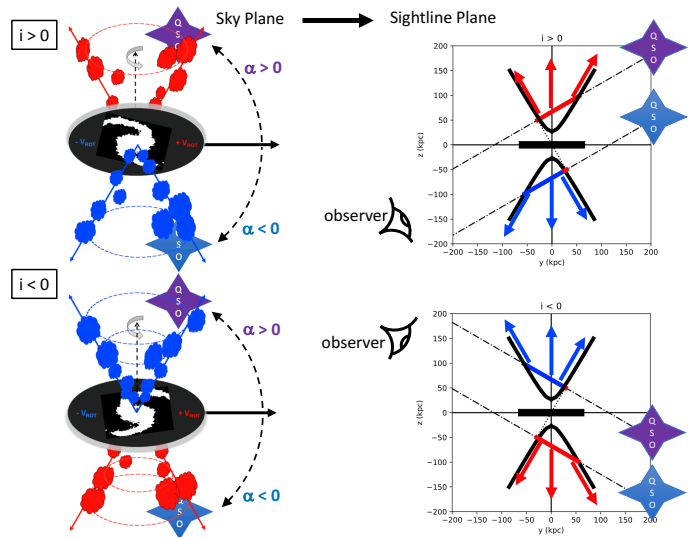}
   \caption{Geometry connects the sky plane to the sightline plane. 
     {\it Left:} The image on the sky is rotated to align the rotation axis
     of the disk with the z-axis of the coordinate system; then the redshifted
     side of the rotation curve is on the right. The disk tilts shown in 
     the top and bottom panels can be distinguished by the winding direction
     of the spiral arms.  Determining the sign of the disk inclination this
     way makes a unique prediction for the sign of the Doppler shift from
     a conical outflow.
     {\it Right:}
     Schematic view of the outflow cone in the plane of the sightline.
     Positive and negative values of the azimuthal angle are shown. 
     Along each sightline (dot-dash lines), the sign of the projected radial
     outflow velocity is denoted as redshifted (red) or blueshifted (blue).
    }
    \label{fig:tip}
     \end{center}
      \end{figure}

A degeneracy regarding which side of the minor axis is tipped toward the observer 
presents a problem, allowing outflowing gas to be either redshifted or blueshifted
in a quasar sightline. The Doppler shift of the \mgII\ system does not uniquely
identify outflowing gas the way it identifies corotating gas. 
The left-hand panels of Figure~\ref{fig:tip} illustrate the 
projection of outflow cone on the sky.

Determining which side of the disk is tipped toward the observer eliminates 
the ambiguity. Figure~\ref{fig:tip} illustrates a strategy for measuring the
{\it 3D orientation} of the disk. In a self-gravitating, collisionless 
system, only trailing spiral patterns are long lived \citep{Carlberg1985}. Most
spiral patterns therefore lag further behind the direction of rotation with increasing 
radius. A rotation curve and the winding direction of the arms determine the disk tilt, 
leading to a complete description of the disk orientation in three-dimensional space.

The F390W images resolve spiral patterns. The winding direction of the arms
is apparent in roughly 20\% of the {\it Ks} images. Figures~\ref{fig:nirc2_hst}, 
\ref{fig:nirc2}, and \ref{fig:color} show examples.
We define the sign of the disk inclination as shown in Figure~\ref{fig:tip}.
Images and measurements for galaxies probed with major-axis sightlines are presented 
in Ho  \et (2019, in prep). Table~\ref{tab:galaxy} lists the signed inclinations 
for galaxies observed with sightlines at large azimuthal angles. We use 
the 3D orientations of these disks to predict the velocity
range of an outflow component and then discuss whether an outflow is detected below.

We will use the terms blueshifted-lobe and redshifted-lobe to describe
the Doppler shift of gas moving radially outward along the symmetry axis of the cone.
The lobe seen in front of the disk plane will have a net blueshift, while the lobe behind the 
disk plane will have a net redshift. The disk plane behind the blueshifted lobe is tipped away 
from the observer, while the near side of the disk plane will be seen in front of the
redshifted lobe. This red/blue nomenclature identifies each outflow lobe but is a bit misleading. 
A single lobe may produce both redshifted and blueshifted absorption in a sightline.  
In Figure~\ref{fig:tip} for example, the near and far sides of the 
cone would acquire opposite  Doppler shifts if the opening angle and/or inclination were increased.
A quasar sightline may also intersect opposite lobes of a cone, but this geometry is not
relevant in this paper because we selected disks viewed at high inclination.

We assume clouds populate the interior of the cone as well as its surface
in order to illustrate the full range of velocity components that an outflow
could produce. The outflow model has a constant radial velocity $v_r$ which 
can be increased (decreased) to broaden (shrink) the predicted velocity range.
We constrain the opening angle $\theta_{max}$ independently. In principle the line profiles 
contain additional information related to the radial density gradient, which we do not
model here because our spectra do not resolve the absorption components.

%LBRT

\subsubsection{Outflow Dominated \mgII\ Systems}

\begin{figure}
 \begin{center}
 \includegraphics[scale=0.55,angle=0,trim = 50 20 0 10]{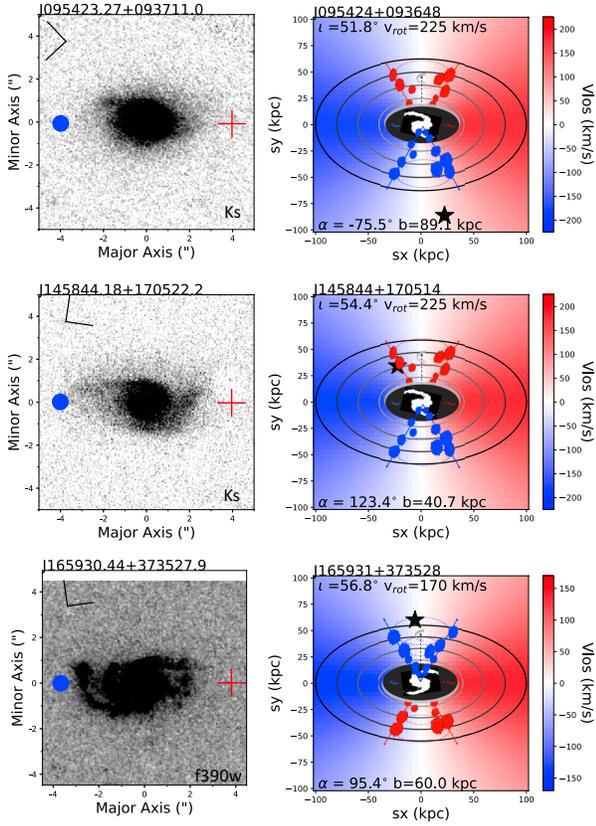}
   \caption{Examples of galaxies oriented in ways that produce outflow
     and disk plane components which can be easily distinguished. 
{\it Left:}
Detection of spiral arms.
For ease of comparison, we orient the galactic 
disks so that the receding side  of the major axis (red plus) is on the right; the compass rose
indicates north and  east on the sky. If the spiral arms are trailing, then the disks 
are tipped as shown in the schematic to the right.
{\it Right:} Schematic of outflow cone projected onto the velocity field
in the plane of an extended disk. For a positive disk inclination, the bottom lobe of the outflow cone is
the  closer, blueshifted side. The black star marks the quasar sightline. In two of these
examples, the net absorption from the outflow has the opposite sign of gas rotating in the disk plane.
In the third example, a disk component has a projected speed close to the systemic velocity.
}
    \label{fig:outflow_geometry}
     \end{center}
      \end{figure}

We selected \mgII\ systems  likely to be outflow dominated from the
subsample for which we resolved spiral arms. These systems
have excess equivalent width and are found in minor-axis sightlines.
Our high-resolution images resolve spiral arms in seven galaxies meeting these
criteria. Their equivalent widths range from 2 to 6 times the average at
their impact parameters.  Among them, our test is most sensitive for three
that are orientated in ways that cleanly separate outflow and disk components.
Their images are shown in Figure~\ref{fig:outflow_geometry}, and we discuss them
in turn. We have applied unsharp masking techniques to some of the {\it Ks} images
to enhance visibility of the arms (Ho \et 2019, in prep).

%LBRT

\begin{figure}
 \begin{center}
  \includegraphics[scale=0.70,angle=-90,trim = 20 40 150 400]{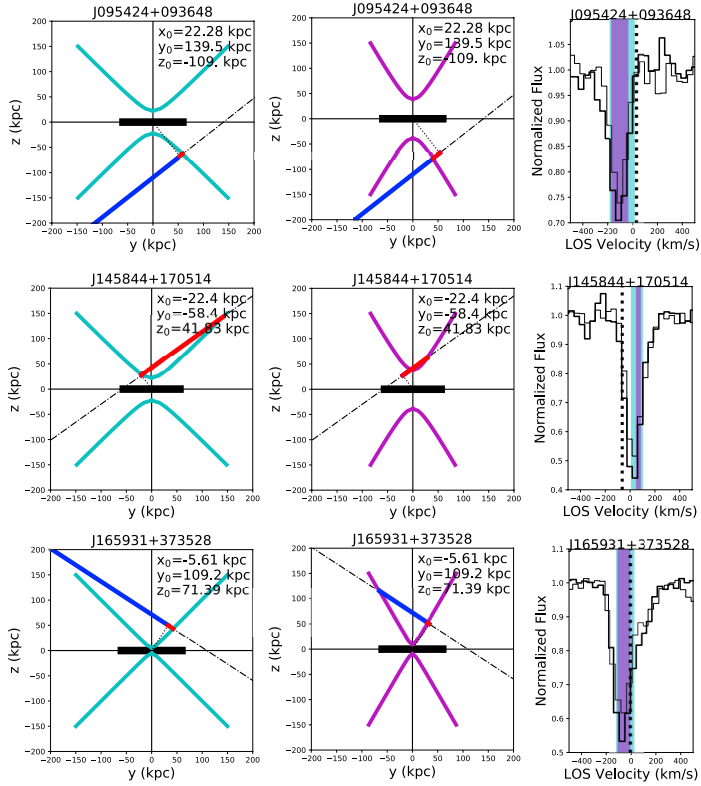}
   \caption{Intersection of outflow cones with the plane of the quasar sightline. For two
     values of the cone's opening-angle, a radial outflow within the boundaries of each 
     cone is projected onto the quasar sightline. The color along each sightline denotes 
     the sign of the observed Doppler shift. The right most panels show the \mgII\ $\lambda 2796$
     (bold line) and $\lambda 2803$ (thin line) line profiles relative to the target galaxy systemic
     velocity.
     We overlay the line-of-sight velocity range for outflow speeds $v_r = 180, 100, {\rm ~and~} 
     120$\kms\
     for \j095424+093648, \j145844+170514, and \j165931+373528, respectively.
     The wider cone with $\theta_{max} = 45\deg$  
     predicts absorption velocities within the cyan band, and the purple band shows
     the velocity range across the $\theta_{max} = 30\deg$ cone.
    }
    \label{fig:best_outflows}
     \end{center}
      \end{figure}

The \j095424+093648 sightline is near the minor axis of our target,
\j095423+093711. The projected rotation speed where the sightline intersects the
disk plane is $+28$\kms, barely distinguishable from the systemic velocity.
If the spiral arms seen in Figure~\ref{fig:outflow_geometry} are trailing, then 
the inclination of the disk is positive, and our sightline would pass through the 
blueshifted lobe of an outflow before intersecting the disk plane at radius $R = 141$~kpc.
We detect a blueshifted \mgII\ system at $\vavg = -106$\kms. A radial outflow speed of 180\kms\ 
combined with an outflow opening angle of 30\deg\ describes the observed velocity 
range of \mgII\ absorption in Figure~\ref{fig:best_outflows}. No absorption is detected
at the velocity of the disk component. The line profile shows a weaker, second component 
blueward of the strongest component. The sightline geometry, see the left panel
of Figure~\ref{fig:best_outflows}, shows that the outflow absorption comes from
increasingly larger radii as the blueshift increases, qualitatively explaining the line profile
shape.  Increasing the opening angle much beyond 
$\theta_{max} = 30\deg$ would generate a redshifted component from the  outflow that 
is not observed in the \mgII\ line profile.

The disk of \j145844+170522 has a positive inclination. Figure~\ref{fig:outflow_geometry} 
shows that the \j145844+170514 sightline intersects the near side of the disk plane at 
$R=63$~kpc before encountering the redshifted outflow lobe.  The azimuthal angle, 
$\alpha = 123.4\deg$, places the quasar 33.4\deg\ from the minor axis. In this geometry, the 
Doppler shift of an outflow component would distinguish  it from a disk component.
The projected rotation speed is -63\kms. We measure a net redshift for the \mgII\ system, 
which is one of just a few counter-rotating systems at $R < R_{max}$.
Figure~\ref{fig:best_outflows} shows that the projected velocity range  of a $v_r = 
100$\kms\ outflow overlaps the strongest absorption component. A weaker disk component
may produce the blended, blueshifted component, a prediction we hope to test with 
higher resolution spectroscopy. The outflow component clearly produces the 
equivalent width excess.

Our high resolution images draw attention to the disturbed morphology of \j165930+373527;
a large arm extends southward wrapping towards the east. 
The wrapping direction of the arms and the rotation curve require a negative disk inclination
if the arms are trailing. In this orientation the quasar sightline intersects the blueshifted outflow lobe. 
The main absorption component is much more blueshifted than expected for
a galactic disk component in this sightline. We infer an outflow speed of roughly 100\kms,
about 20\kms\ higher than \cite{Kacprzak2014} fit because we measure a correspondingly higher galaxy
redshift. Figure~\ref{fig:best_outflows}
shows that outflow angles approaching $\theta_{max} \sgreat\  45\deg$ produce
some redshifted absorption; but the blue component limits the fitted outflow speed
such that the outflow cone does not describe the prominent red wing on the \mgII\
line profile even if $\theta_{max}$ approaches $90\deg$. It is possible that
the redshifted absorption wing is produced by streams associated with the
interaction driving the tidal arm. The target galaxy is isolated in the sense that
there are no brighter galaxies with a consistent (photometric) redshift within 100~kpc.

%LBRT

\subsubsection{Multi-component Sightlines} \label{sec:multi}

\begin{figure*}
 \begin{center}
  \includegraphics[scale=0.70,angle=0,trim = 60 20 20 20]{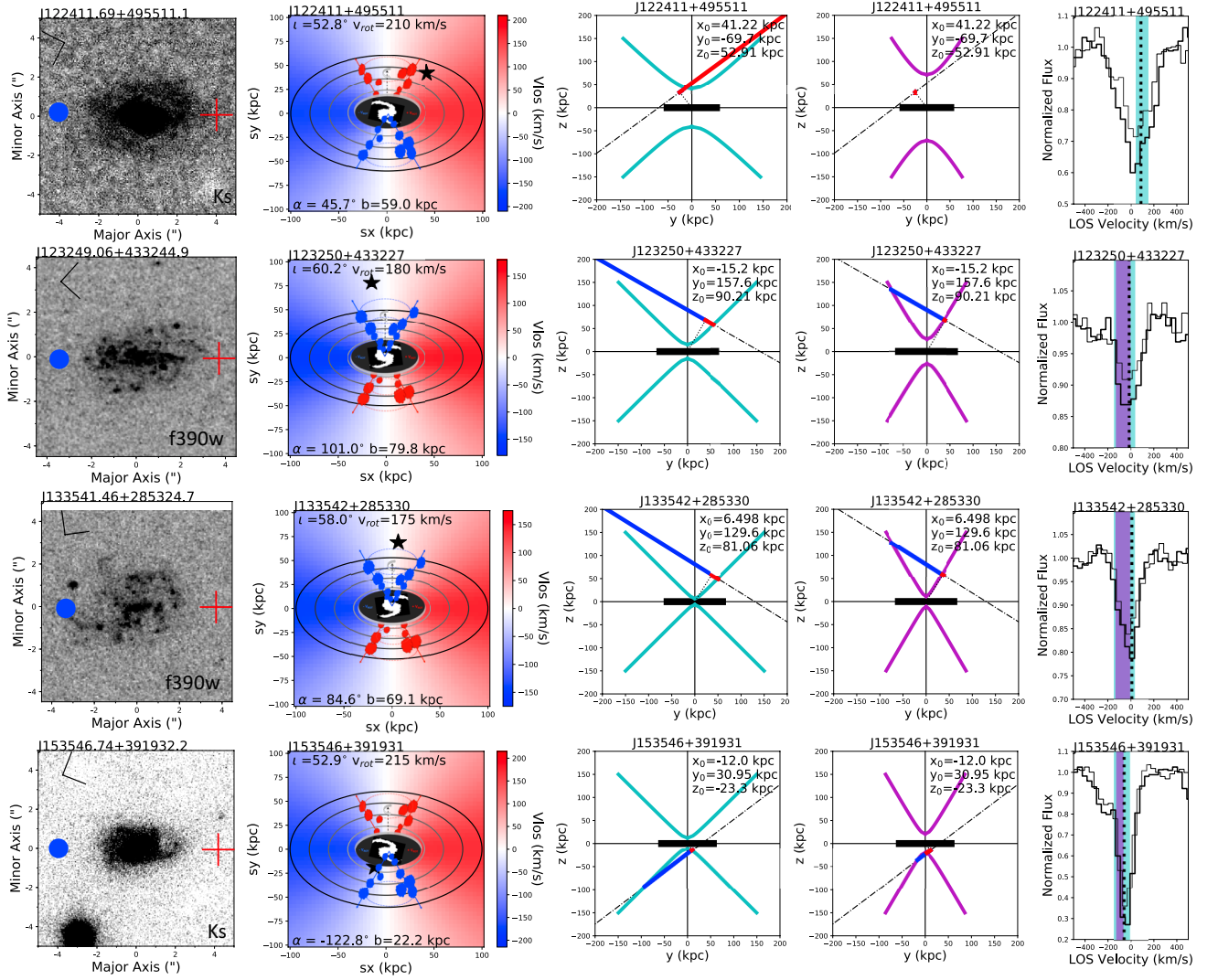}
   \caption{Examples of \mgII\ absorption systems with blended velocity components 
     {\it and} a measured 3D disk orientation that constrain the geometry. We compare
     the line profile to the projected velocity of rotation in the disk plane (vertical, dotted 
     line) and outflow cones with opening angles of 45\deg\ (cyan band) and 30\deg (purple band).
     With no tweaking of any parameter, a fiducial outflow speed $v_r =150$\kms\ provides an explanation 
     for the redshifted component in the \j122411+495511 sightline and the blue wing in the
     \j153546+391931 sightline.
   }
    \label{fig:disk_outflow}
     \end{center}
      \end{figure*}

Figure~\ref{fig:disk_outflow} shows more minor-axis sightlines near resolved
galactic disks. These systems also have stronger than average \mgII\ absorption.
Their geometry is less favorable for distinguishing outflow and disk components,
which have similar line-of-sight velocities, so we do not
fit an outflow speed in these cases. We adopt an outflow 
speed of $v_r = 150$\kms\ and illustrate the velocity range where 
this outflow would contribute to the line profile in Figure~\ref{fig:disk_outflow}. 

The \j122413+495515 disk has a positive inclination. The \j122411+495511 sightline 
intersects the redshifted half of the disk plane at $R=81$~kpc. The azimuthal angle
is not close enough to the minor axis to intersect a $\theta_{max} \approx 30\deg$ 
outflow cone.  A large opening angle is required to produce an outflow component. 
The narrow absorption component produced by the fiducial 150\kms\ outflow speed and 
opening angle $\theta_{max} = 45$\deg\ is coincident with a disk component.
Increasing the outflow speed to $v_r \approx 200$\kms\ will spread the outflow
absorption across the redshifted line wing visible in the last column of 
Figure~\ref{fig:disk_outflow}. A very large  opening angle, roughly $\theta_{max}
> 52\deg$, would also produce the blueshifted component. The line profile appears to 
be a blend of multiple, unresolved components. A disk model can only explain a
single, narrow component.  An outflow with a large opening angle can explain the 
broad range of absorption velocities.

The absorption near the galaxies \j123249+433244 and \j133541+285324 likely
includes an outflow component for several reasons. The \j123249+433244 and 
\j133542+285330 sightlines
are not only close to the minor axis;  they also intersect the disk plane at very large radii. The
\mgII\ systems are fairly weak in absolute terms, but their equivalent widths
are twice the predicted value for their large impact parameter. Both sightlines 
intersect outflow cones with a wide range of opening angles. 
Outflowing gas would be blueshifted along most of each sightline. However we predict
substantial absorption from these outflows near the systemic velocity. The
radial velocity vector is nearly perpendicular to our sightline on the far side of
the cone in Figure~\ref{fig:disk_outflow}, which is where the radius is small and the
outflow density highest.

The absorption system associated with \j153546+291932 has a narrow, asymmetric
component which is blueshifted and a weaker  blueshifted line wing at -200\kms.
Figure~\ref{fig:disk_outflow} shows that
our sightline is nearly parallel to the surface of an outflow cone
$\theta_{max} \approx 45\deg$,  generating both a long pathlength 
through the outflow and very little deprojection of the outflow velocity. 
This geometry may provide better than average sensitivty to
the fastest, lowest column regions of the outflow. 
Fitting the broad blue wing requires 
low-ionization outflow moving at $v_r \sgreat\ 200$\kms. At our fiducial
outflow speed of 150\kms, outflowing gas would contribute to the narrow,
blueshifted component,  which may be blended with an  absorption component
produced nearer the disk plane. The sightline intersects the disk plane at
a relatively small radius $R=33$~kpc.

%LBRT

\subsubsection{Properties of Outflows \& Their Host Galaxies}

We have presented the most secure outflow detections. Six of these seven galaxies have
SFRs slightly above the SFR main sequence in Figure~\ref{fig:color_mass}, the exception
being \j123249+433244 with specific $SFR = 0.068$~Gyr~$^{-1}$. The highest estimated specfic SFR 
is 0.81~Gyr$^{-1}$ in \j165930+373527. We identified four starburst galaxies in 
Section~\ref{sec:interactions} with higher specific SFRs. The minor-axis sightline probing
\j133740+055211 intersects the disk plane where the line-of-sight velocity is just
-8\kms, so this pair has a good geometry for distinguishing outflow and disk components.
The equivalent width excess of $W_r / \wavg = 1.7$ also favors an outflow detection.
The detected \mgII\ absorption is Doppler shifted -47\kms, which is consistent with 
an outflow if (and only if) the disk inclination is positive. High-resolution blue imaging
with HST would resolve the spiral arms in the disk, allowing us to measure its 3D orientation
and test this outflow prediction.\footnote{Our sightlines do not constrain the outflow 
        properties of the other three starburst galaxies.  The NIRC2 imaging resolves
        spiral arms in only one of the starburst galaxies, \j123049+071050, which is 
        paired with a major axis sightline ($\alpha = 4.0\deg$) that does not intersect
        a conical outflow. The sightline paired with \j150150+553227 does not detect \mgII,
        and the NIRC2 image resolved \j142815+585442 into two galaxies.}

We identified spiral arms and measured the disk inclinations for five other galaxies
whose CGM was detected in a sightline near the minor axis ($\alpha > 45\deg$). Three
of the associated sightlines address the question of whether weaker than average \mgII\ 
systems also detect outflows.  The strongest case for an outflow among these is in the
\j155504+362847 spectrum which detects absorption $34$~kpc from \j155505+382848
at 0.46 of the average equivalent width. The {\it Ks} image marginally resolves a
tightly wrapped two-arm spiral pattern that would require a negative disk inclination, but
the wrapping direction is not as certain as it is for the other spiral patterns discussed
in this paper. Higher signal-to-noise ratio blue imaging needs to be obtained; but if
the preliminary 3D orientation is correct, an outflow cone will produce blueshifted 
absorption in the quasar sightline between roughly -145 and -14\kms.
This sightline intersects the disk plane at
$R = 52$~kpc where circular orbits project to a line-of-sight velocity of -78\kms.
We measure a \mgII\ Doppler shift of -45\kms,  consistent with either disk plane or outflow absorption.
It seems likely that both components are detected because the line profile shows two blended,
blueshifted components. 

We measure positive inclinations for the disks of \j143603+375138 and \j154956+070056. 
The \mgII\ equivalent widths in  the \j143603+375131 and \j154956+ 070056 sightlines 
are, respectively, 0.20 and 0.32 times the average value for the impact parameters.
Our analysis for these pairs does not rule out outflows, but it does not offer 
compelling evidence for outflows either. For example, 
the \j143603+375131 sightline selects the blueshifted outflow lobe but also intersects
redshifted outflowing gas on the far side of this lobe. The \mgII\ absorption is near
the systemic velocity, $\vavg = +35^{+4}_{-10}$\kms. Similarly,
a 150\kms\ outflow from \j154956+070056\
would produce absorption at Doppler shifts from -142\kms\ to +46\kms\ relative to 
the target galaxy systemic velocity in the \j154956+070056 spectrum. This sightline
intersects the disk plane at $R=48$~kpc where
the projected rotation speed is positive but near the systemic velocity. We detect
\mgII\ absorption at the systemic velocity.

Our analysis rules out outflow detections in some sightlines.
We have already seen in Figure~\ref{fig:ab_corot} that some minor-axis sightlines do 
not detect a \mgII\ system.  In addition, we resolve spiral patterns in the NIRC2
imaging that require a positive disk inclination for \j135051+250506 and a negative
disk inclination for \j152027+421519. The blueshifted \mgII\ system in the \j1520228
sightline is not consistent with the projected outflow speeds of 25 to 245\kms\ (for
$\theta_{max} = 45\deg$ and $v_r = 150$\kms). The \j135051+250506 sightline
only intercepts
outflows with a large opening angle, $\theta_{max} \sgreat\ 45\deg$, and the implied Doppler
shifts are redshifted 78 - 148\kms, which is inconsistent with the \mgII\ system
detected at the systemic velocity. Whether these non-detections indicate
the absence of outflows from the target galaxies is not immediately clear. Alternatively, 
the covering fraction of low-ionization gas entrained in the winds may be less than unity.

The \mgI\ detections include some of these outflow systems. 
The velocity of the main \mgI\ absorption in the \j165931+373528 spectrum
corresponds to the main \mgII\ component,  which we have attributed to absorption from
a galactic outflow. The \mgI\ absorption velocity is coincident with the outflow component 
in \j145844+170514.  Our weakest \mgI\ detection is in the \j122411+495511 spectrum, and the
low SNR does not constrain the \mgI\ velocity very well. We explained in Section~\ref{sec:quasar_als} 
why \mgI\ detections may indicate the presence of dense gas. Detection of \mgI\ absorption 
in two of our three strongest outflow systems may indicate the presence of dense clumps of 
gas in the outflows.

We found both outflow systems and corotating systems associated with galaxies having
estimated halo masses exceeding $10^{12}$\msun.  Galaxy formation theory has increasingly 
emphasized the importance of a critical halo mass of roughly $10^{12}$\msun\ \citep{Keres2005,Dekel2006}, 
above which gas accretion fed by cold flows is greatly suppressed. In Table~\ref{tab:galaxy} we list
11 galaxies with halo masses in the range $12.01 < \log M_h < 12.63$. However abundance matching introduces  
factors of two'ish (at the one-sigma level) uncertainties when applied to an individual galaxy 
\citep{Behroozi2010}.   Even our targets with the highest stellar masses have blue colors.
This selection may bias them toward the lower end of the halo mass range for their stellar mass,
perhaps explaining why their CGM absorption properties do not stand out. The more significant
statement is that the median halo mass of our sample is a factor of 2.5 lower than the theoretical 
critical mass, offering an explanation for why cold flow disks would still be forming at low redshift.

\subsection{The Maximum Extent of  Corotating Gas} \label{sec:ovi}

Our Quasars Probing Galaxies program does not find an edge, or
maximum extent, for the corotating CGM. The results of Section~\ref{sec:thin_disk}
require the corotating structure be non-axisymmetric at radii
$R \sgreat\ R_{max} \approx 70-80$~kpc. Gas streams are one such example.
Streams produced by cold flow gas \citep{Stewart2011a,Stewart2011b} 
or satellite winds \citep{Angles-Alcazar2017,Hafen2018} have both been shown
to feed the growth of extended disks in numerical simulations. These streams
have smaller covering fractions than do the warped disks \citep{Stewart2013}
and offer a plausiable explanation for the corotation we observe at $R > 70$~kpc.

This result motivates an examination of corotation at larger impact parameters. 
Our sightlines probe the inner CGM. Observations of absorption-line-systems
already exist at larger impact parameters, but the rotation curves and orientations
of the host galaxies have not typically been measured. One complication 
will be that the strength of low-ionization metal lines decreases rapidly toward larger
impact parameters.

It has recently been argued that low-ionization absorbers 
lie at much smaller radii than the high-ionization gas producing \ovi\ 
absorption \citep{Stern2018}. Combining the COS-Halos and \cite{Johnson2015,Johnson2017}
measurements, the equivalent width -- impact parameter relation for \ovi\ is less steep than 
that for the lower ionization metal lines \citep{Stern2018}.  \cite{Heckman2017},
however, showed the strength of the metal lines falls off more slowly (like the
\ovi) in the CGM of starburst galaxies.  It is not clear whether detections of
low-ionization lines at $b \ge\ 0.5 \rvir$ require a starburst. Indeed, the
duration of starburst activity (by definition almost) is limited to the dynamical 
timescale of the galaxy, which is short compared to that of the halo on spatial
scales 10 to 100 times larger.  

High ionization lines including \neviii\ \citep{Burchett2018} as well as \ovi\ 
will probably be the tool of choice to investigate CGM kinematics on larger scales.
While neutral hydrogen is also detected at large impact parameters, the metal lines
are more biased towards gas clouds that lie within the virial radius.  Hydrogen gas
beyond the virial radius can produce absorption near the systemic velocity because
of well-known degeneracies between Doppler shifts induced by peculiar velocities
and cosmological redshift.

Comparing the angular momenta of the gas traced by \ovi\ and \mgII\ systems could 
yield new insight the origin of \ovi\ systems. We note that inner CGM sightlines 
with \ovi\ detections often show some kinematic correspondence
with the low-ionization components \citep{Werk2014}.  The high ionization systems
are often offset to one side of the systemic velocity, so some \ovi\ components
may corotate with the galaxy.  High-ionization gas behind  virial shocks has recently
been shown to have significant rotation in cosmological hydrodynamical simulations 
\citep{Oppenheimer2018}.  In an alternative picture, where the \ovi\ lies at much larger radii
\citep{Stern2018}, the high-ionization gas feeds a reservoir of low-ionization gas as it cools,
so the \ovi\ absorption would be Doppler shifted with the same sign as the \mgII\ systems
but have smaller absolute velocity shifts than the low ions.

\begin{figure} 
 \begin{center}
  \includegraphics[scale=0.7,angle=-90,trim = 0 35 60 0]{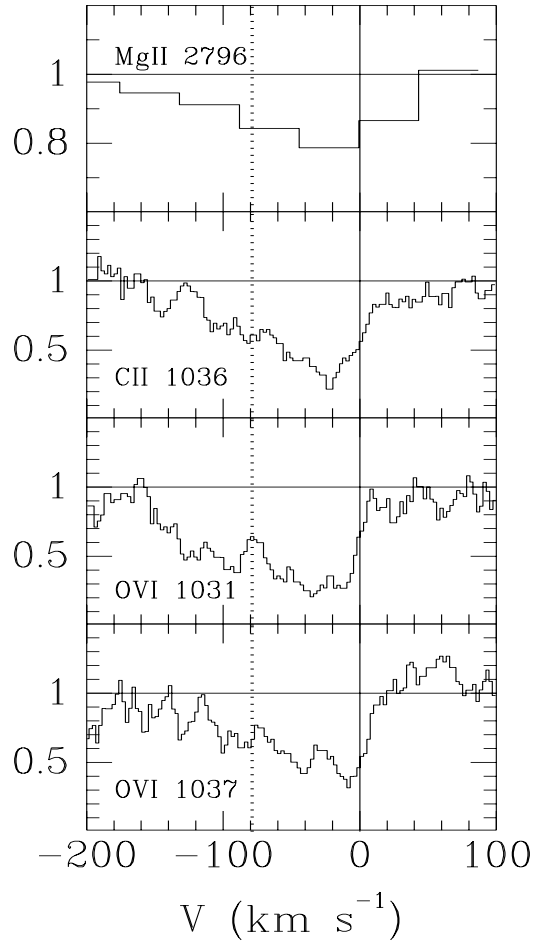}
   \end{center}
    \caption{Absorption by the CGM of \j155505.27+362848.4 in the
      spectrum of \j155504.39+362847.9.  The blue wing on the \mgII\
      profile indicates multiple components, and these are better resolved
      by the \cII\ detection in the HST/COS spectrum (GO 11598, PI Tumlinson). 
      {\clm
      Absorption from \ovi\ is detected over the same velocity range as the
      low-ion absorption. The asymptotic rotation speed of the galaxy, 255\kms,
      projects to -105\kms\ along the quasar sightline ($\alpha = 44.6\deg$, 
      $i=55.2\deg$).} Both the high-ionization and low-ioniation CGM corotate
      with this galactic disk.
      }
      \label{fig:ovi} 
       \end{figure}

We searched the HST archives and found COS observations for two of our targets.
The \j155504+362847 sightline has an intermediate azimuthal angle ($\alpha = 44.6$);
it intersects the disk at $R = 55$~kpc. Figure~\ref{fig:ovi} compares the projected
line profiles to the projected rotation speed in the disk plane. The strongest 
component in the low-ionization lines is consistent with rotation in the disk plane,
and this component is also the strongest \ovi\ component. Corotating gas is detected 
in this \ovi\ system.  Several of the \ovi\ absorption components, however, are
more blueshifted than a disk component.  Our ability to determine the velocity offset
of an outflow component is limited by the marginal constraints on the sign of the 
disk inclination. The contrast of the spiral arms in our infrared image is poor,
but our image processing suggests a negative disk inclination.  An outflow component in the
quasar sightline would then be blueshifted for this disk orientation,  so the system
may include one of more components from an outflow and/or a disk component.

The outflow dominated \mgII\ system, \j165931+373528, is only marginally detected
in \ovi. The sightline is near the minor axis ($\alpha = 84.6$). Apparently not
all outflows produce strong \ovi\ absorption.  We do not draw any general 
conclusions from these two sightlines.

However \cite{Kacprzak2018}  recently presented measurements of galactic rotation
for the hosts of \ovi\ absorption-line systems. They found no significant excess
of corotating \ovi.
Clearly corotating \ovi\ absorption is not ubiquitous.  More comparisons of galaxy 
rotation curves to the gas kinematics in higher ionization lines would clearly provide
additional insight into the nature of corotating circumgalactic gas.

%LBRT

\section{Conclusions}

Circumgalactic gas flows that feed the disk fuel for star formation
and those driven by feedback can be identified via the Doppler shifts and
linewidths of metal-line absorption systems detected in spectra of background
beacons. In this paper we investigated how the orientation of galactic
disks can be used to distinguish these components from other sources of
low-ionization halo gas.  We used LGSAO at Keck Observatory and HST
to establish the orientation of the galactic disks on the sky. When 
the winding direction of spiral arms is measured, our rotation curve
measurements determine the 3D orientation of the disk because long-lived
spiral patterns trail the rotation direction.

Circumgalactic gas kinematics were measured  relative to these
disks. We observed 50 quasars located behind galaxies drawn at random
from a well-defined population:  star-forming galaxies with $\mstar \approx\ 
10^{10}\msun$ at redshift $z \approx\ 0.2$.  The Keck/LRIS spectra
provide sensitivity to an absorber of average strength over the full
range of impact parameters, $b \approx\ 12 - 98$~kpc. We found the
following results.

\begin{enumerate}

\item
The equivalent-width weighted Doppler shifts of \mgII\ absorbers often
show a net Doppler shift with the same sign as the galactic rotation. This 
comparison was made as a function of azimuthal angle at impact parameters 
$ b < 0.5 r_{vir}$.  We found no examples of net counter-rotating absorption
within 45\deg\ of the major axis. Corotation describes the velocity offsets
of \mgII\ absorption over a larger range of azimuthal angles than we
reported in \cite{Ho2017}. Differences between the azimuthal distribution 
of corotators and counter-rotators depend on galaxy inclination, and our 
galaxy selection criteria favored edge-on disk orientations

\item
Absorber strengths at fixed impact parameter generally increase with
azimuthal angle. We found the opposite trend, however, at very small
impact parameters; sightlines at $b < 40$~kpc detect stronger \mgII\
systems near the major axis. The strong absorption requires  multiple 
velocity components rather than exceptional column densities, so these
trends identify kinematically disturbed circumgalactic gas.

\item
We computed the Doppler shift for a velocity component generated by gas orbiting 
in the disk plane. When we could determine the sign of the disk inclination, 
we also predicted the line-of-sight velocities for an outflow component. Outflows
uniquely fitted the observed \mgII\ velocity range for the minor-axis sightlines
with the largest absorption excess, demonstrating that outflow components 
increase equivalent width near the minor axis at redshifts as low as $z \approx\ 0.2$.
We found that the measured \vavg\ values were positively correlated with the 
projected disk velocities ($r_S = 0.64, 2.4\sigma$) for the subset of sightlines 
that intersect the disk plane at $R \le\ 70$~kpc. This correlation draws attention
to the important role of angular momentum in supporting the low-ionization CGM.
Blending of multiple components in our spectra may weaken the significance of 
this correlation, something we plan to test with higher resolution spectroscopy.

\end{enumerate}

Motivated by the result that a significant portion of the inner CGM
is rotating in a manner related to the galactic disk, we make the following observations.
Among all the sightlines that intersected the disk plane at $R < 70$~kpc, only one failed
to detect \mgII\ absorption, and only one detected counter-rotating absorption.  
We conclude that on scales $R < 70$~kpc, the corotating CGM is likely axisymmetric,
reminiscent of warped extended disks \citep{Stewart2011a,Stewart2011b} but also
consistent with corotating structures that are less flattened \citep{ElBadry2018}.

We emphasize that centrifugally supported disks fail to produce the full range of absorption 
velocities \citep{Steidel2002,Kacprzak2010}. One possible resolution is that
the corotating gas is only partially supported by centrifugal forces, and it 
gradually spirals into the galactic disk, thereby broadening the line profile
\citep{Ho2017}.  Ho \et (2019) measured inflow rates from
a sample of simulated EAGLE galaxies with masses and redshifts matched to the work
presented in this paper, and they found inflow rates in the disk plane sufficient
maintain current SFRs.  It is therefore plausible that the corotating gas identifies
material that feeds the growth of galactic disks at low redshift.

We did not detect a maximum impact parameter for corotation among $\alpha\ < 45\deg$
sightlines.  The covering fraction of \mgII\ detections dropped, however, and their
velocity offsets were no longer correlated with the galactic rotation speeds
beyond $R \sgreat\ 70$~kpc. Inspection of numerical simulation paints a scenario in which 
streams feeding the disk produce corotation at these distances. Such streams extend 
well beyond the inner CGM in some simulations, so it will be important to directly 
measure velocity offsets at larger impact parameters than we considered in this work.

Radial outflow at 120-180\kms\ within a cone easily described the velocity components
producing the minor-axis excess equivalent width.  This result would appear to conflict
with numerical results indicating that individual clouds are destroyed on timescales
much shorter than the time it would take to accelerate the cold gas
\citep{Scannapieco2015,Bruggen2016,Schneider2017}. Yet recent numerical work has
followed up on the cloud shattering argument of \cite{McCourt2016} and identified
regimes where the cold gas not only survives but entrains many times its original
mass through mixing layers with the hot wind \citep{GronkeOh2018}. Hence it seems
plausible that the interaction of a hot wind with the CGM triggers cooling that 
creates a low-ionization outflow, a scenario supported by 
high-resolution simulations of galactic winds \citep{Schneider2018,Sparre2019}.  
We therefore wondered whether the low-ionization clouds might form beyond the galaxies
at radii closer to the large impact parameters where we observe them.

We calculated the minimum radius where each of the seven sightlines intersects the 
surface of the outflow cone, finding values from $r = 22$~kpc (\j153546+391931) to
89~kpc (\j095424+093648). 
A hot wind moving radially outward at $v_r \ge\ 1000$\kms\
would disturb the ambient CGM at these radii on timescale of 22 to 87 Myr, short enough
that the UV continuum would still be enhanced from the starburst. The Doppler shifts
of the cooler outflows measured down-the-barrel in galaxy spectra are much slower \citep{Martin2012}.
Taking $v_r \approx\ 100$\kms\ for illustration, the timescale required to reach our
sightlines grows to 220 to 870 Myr, and we might not identify the stellar population
as a starburst. We therefore expect a stronger correlation between the outflow systems
and the galactic SFRs in the scenario where a hot wind triggers the formation of the
cold, low-ionization outflow that we detect in \mgII\ absorption.

Our sightlines through the halos of the four starburst galaxies happened to not
be optimal for detecting outflows.  Other work, however, has shown that equivalent
widths are generally large in sightlines intersecting the CGM of starburst galaxies \citep{Heckman2017}We have confirmed outflow components in seven galaxies with large absorption excesses,
but they all have fairly normal star formation activity. The specific SFRs of the associated host galaxies 
range from 0.07 to 0.81~Gyr$^{-1}$. It is possible that the detected outflows are relics of past 
starburst activity. Alternatively the feedback may be stronger than expected owing to 
additional acceleration mechanisms such as cosmic rays \citep{Everett2008,Ruszkowski2017,Jacob2018}.

\acknowledgements
The authors thank Sara Ellison, Andrey Kravtsov, and Jonathan Stern for 
useful discussions about this work.
Some of the data presented herein were obtained at the 
W. M. Keck Observatory, which is operated as a scientific 
partnership among the California Institute of Technology, 
the University of California and the National Aeronautics and 
Space Administration. The Observatory was made possible by the 
generous financial support of the W. M. Keck Foundation.
The authors wish to recognize and acknowledge the very significant cultural role
and reverence that the summit of Maunakea has always had within the indigenous 
Hawaiian community. We are most fortunate to have the opportunity to conduct
observations from this mountain. 
The analysis presented in this paper was supported by 
the National Science Foundation under AST-1817125 (CLM).
We gratefully acknowledge support for CWC from NSF AST-1517816. 
NASA supported the HST imaging analysis through grant HST-GO-14754.001-A
from the Space Telescope Science Institute. GGK acknowledges the support of the 
Australian Research Council through the Discovery Project DP170103470.
The two ESI rotation curve observations were obtained by Swinburne Keck program 
2016A\_W056E.
A portion of this work was performed at the Aspen Center for Physics, 
which is supported by National Science Foundation grant PHY-1607611

\facility{Keck:I (LRIS)}
\facility{Keck:II (NIRC2)}
\facility{Keck:II (LGSAO)}
\facility{APO: (DIS)}
\facility{HST: (COS)}
\facility{HST: (WFC3)}

% Appendix
%\appendix

\clearpage

\vspace{15truecm}

\begin{center}
{\bf Figure Sets for Online Journal}
\end{center}

\setcounter{figure}{2}

%LBRT
\begin{figure*}
 \begin{center}
  \includegraphics[scale=0.21,angle=0,trim = 0 50 0 0]{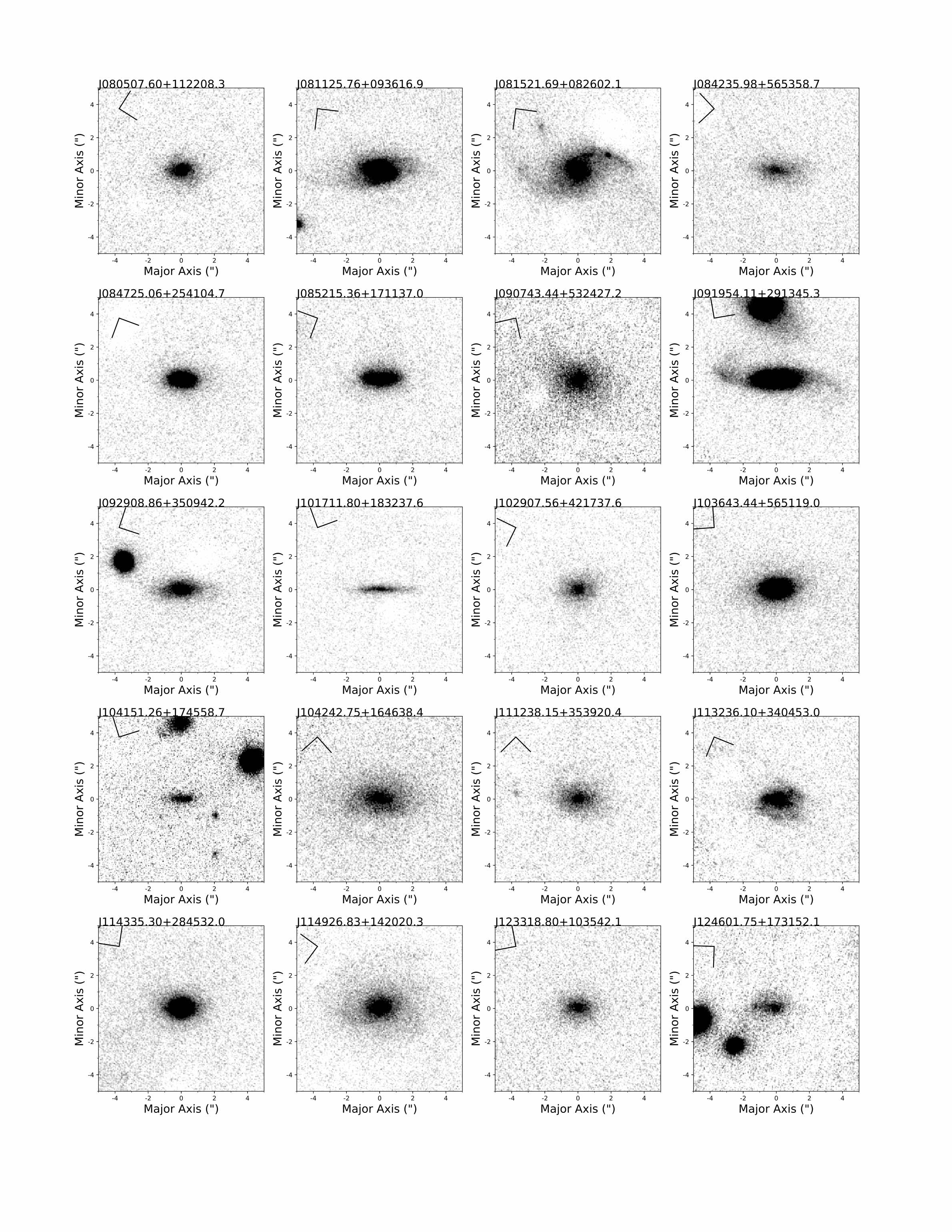}
   \end{center}
 \caption{NIRC2/LGSAO {\it Ks}-band images.
 {\it (b)} - The NIRC2 {\it Ks} band images of target galaxies that were not shown in the
      main text.} 
    \end{figure*}

\setcounter{figure}{2}

%LBRT
\begin{figure*}
 \begin{center}
  \includegraphics[scale=0.21,angle=0,trim = 0 50 0 0]{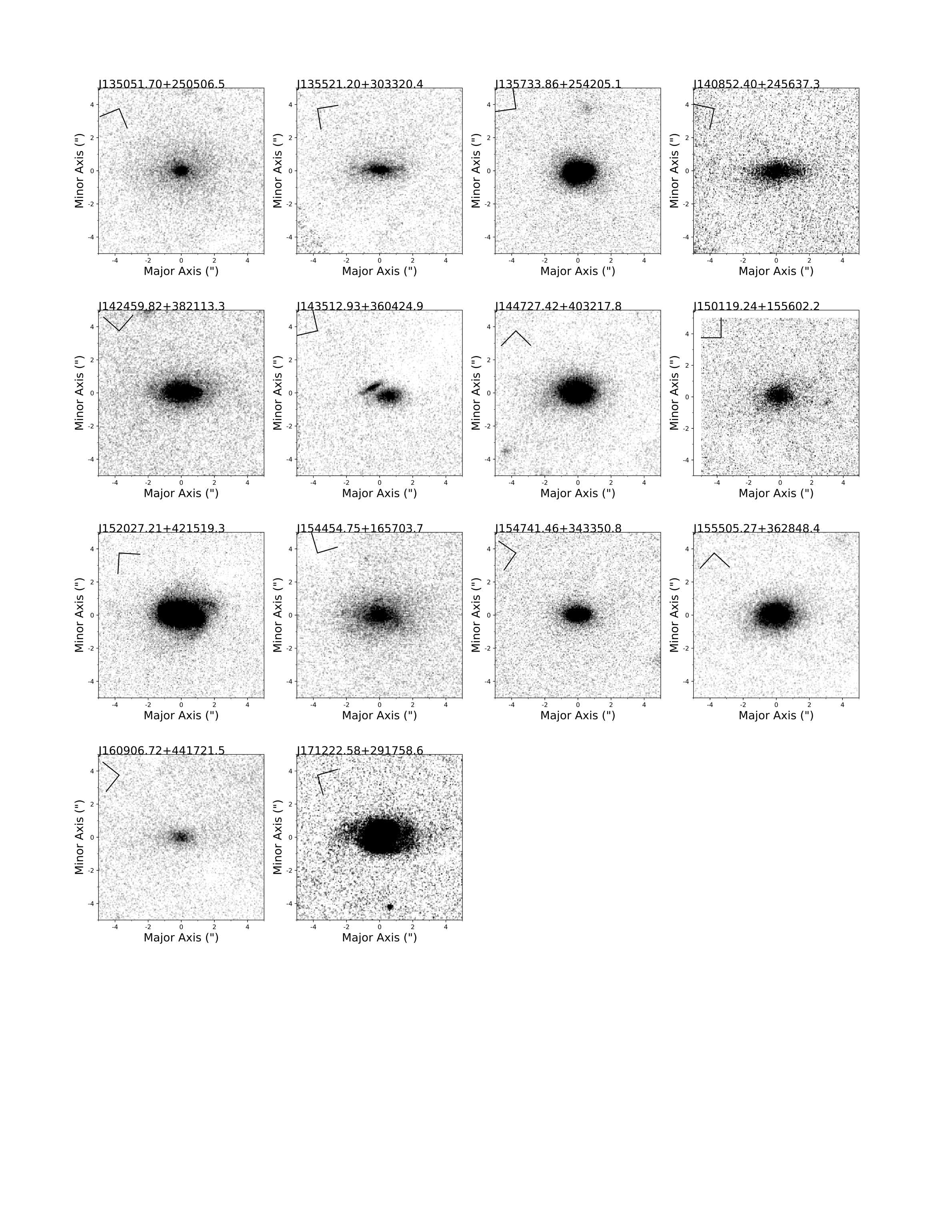}
   \end{center}
    \caption{ NIRC2/LGSAO {\it Ks}-band images.
      {\it c} - Continued from Figure 3b.
      Note that we did not obtain images of \j142910+324840 and \j150150+553227. } 
    \end{figure*}

\clearpage

\bibliographystyle{apj}
\bibliography{library}

\end{document}

%% file: mymac.tex
\newcommand{\be}{\begin{equation}} \newcommand{\ba}{\begin{eqnarray}}
\newcommand{\ee}{\end{equation}} \newcommand{\ea}{\end{eqnarray}}

\def\-{{\em{---}}}

\def \h         {\hbox{$\, h$} }

\def\H7{\mbox {$h_{0.7}$}}

\def\hI{\mbox {\ion{H}{1}}}
\def\mgI{\mbox {\ion{Mg}{1}}}

\def\ovi{\mbox {\ion{O}{6}}}
\def\neviii{\mbox {\ion{Ne}{8}}}

\def\mgII{\mbox {\ion{Mg}{2}}}

\def\feII{\mbox {\ion{Fe}{2}}}

\def\cII{\mbox {\ion{C}{2}}}

\def\mstar{\mbox {$M_*$}}
\def\mhalo{\mbox {$M_h$}}
\def\rvir{\mbox {$r_v$}}
\def\rrc{\mbox {$R_{RC}$}}
\def\j{J}
\def\inc{\mbox {$\iota$}}
\def\zabs{\mbox {$z_{abs}$}}
\def\vinf{\mbox {$v_{\infty}$}}
\def\vlos{\mbox {$v_{los}$}}
\def\vavg{\mbox {$\langle V(MgII) \rangle$}}
\def\wavg{\mbox {$\langle W(b) \rangle$}}
\def\vphi{\mbox {$v_{\phi}$}}
\def\vrot{\mbox {$V_{rot}$}}
\def\vc{\mbox {$V_{c}$}}

\def\IZw18{I~Zw~18}
\def\m82{M82}

\def\h{\mbox {\rm H}}

\def\deg{\mbox {$^{\circ}$}}
\def\msun{\mbox {${\rm ~M_\odot}$}}

\def\lya{\mbox {Ly$\alpha$}}

\def\Ha{\mbox {H$\alpha$}}

\def\lstar{\mbox {$L^*$}}

\def\h0{\mbox {~H$_0$}}
\def\q0{\mbox {~q$_0$}}
\def\o3hb{[OIII]$\lambda5007$~/~H$\beta$~}
\def\O1ha{[OI]$\lambda6300$~/~H$\alpha$~}

\def\s2ha{[SII]$\lambda\lambda6717,31$~/~H$\alpha$~}
\def\2z2{HeII~$\lambda4686$~}
\def\z7{[NII]~$\lambda6583$ }
\def\N2{[NII]~$\lambda6583$~/~H$\alpha$~}
\def\16z2{[SII]~$\lambda\lambda6717, 6731$ }

\def\asec{\ifmmode {'' }\else $''~$\fi}  % arc sec
\def\amin{\ifmmode {' }\else $'~$\fi}    % arc min
\def\arcsper{\ifmmode \rlap.{'' }\else $\rlap{.}'' $\fi} % '' %Arcsec period
\def\arcmper{\ifmmode \rlap.{' }\else $\rlap{.}' $\fi} % '  %Arcmin period
\def\sles{\lower2pt\hbox{$\buildrel {\scriptstyle <}
   \over {\scriptstyle\sim}$}} % approximately less than
\def\sgreat{\lower2pt\hbox{$\buildrel {\scriptstyle >}
    \over {\scriptstyle\sim}$}} % approximately greater than
\def\kms{\mbox {~km~s$^{-1}$}}

\def\cm3{~cm$^{-3}$}
\def\col{\mbox {~cm$^{-2}$}}
\def\mpc3{~Mpc$^{3}$}
\def\mpc-3{~Mpc$^{-3}$}

\def\um{~${\mu}$m}
\def\sec{{Sec.}~}

\def\et{{\rm et\thinspace al.}\ }   % et al.
\def\apjs{ApJS}

%% file: table1.tex
% use long table only if the table spans more than 1 page
%\startlongtable
%
\begin{deluxetable*}{llll}[htb]
\tablecaption{Observations\label{tb:obs}}
%\tabletypesize{\scriptsize}
\tabletypesize{\footnotesize}
%\tabletypesize{\small}
%\tabletypesize{\normalsize}
%\rotate
\tablewidth{0pt}
\tablehead{
\colhead{Target} &
\colhead{Keck/LRIS} &
\colhead{APO/DIS or Keck/ESI\tablenotemark{a}} &
\colhead{Keck/NIRC2}
\\
\colhead{} &
\colhead{Date (Exposure (s))} &
\colhead{Date (Exposure (s))} &
\colhead{Date (Exposure (s))}
}
\startdata
% GAL/QSO                       Keck/LRIS                   DIS/ESI                   Keck/NIRC2
J080507+112208/J080508+112157 & 2014 Feb 1  (4710/4500)     & 2017 Mar 22 (7200)     & 2017 Apr 12 (600)\\
J081125+093616/J081125+093626 & 2013 Apr 11 (1840/1800)     & \nodata                & 2017 Jan 26 (600)\\
J081521+082602/J081521+082623 & 2014 Feb 1  (2835/2250)     & \nodata                & 2017 Jan 26 (600)\\
J081708+091750/J081708+091751 & 2014 Feb 1  (2700/2700)     & 2017 Jan 7  (3400)     & 2017 Jan 26 (600)\\
J084235+565358/J084234+565350 & 2015 Mar 21 (2700/3520)     & \nodata                & 2017 Jan 26 (600)\\
J084725+254104/J084723+254105 & 2015 Mar 22 (3600/3520)     & \nodata                & 2017 Jan 26 (600)\\
J085215+171137/J085215+171143 & 2014 May 2  (2700/2640)     & \nodata                & 2017 Jan 26 (600)\\
J090743+532427/J090745+532421 & 2015 Mar 21 (3600/3520)     & \nodata                & 2017 Apr 12 (1200)\\
                              &                             &                        & 2017 Jan 26 (600)\\
J091954+291345/J091954+291408 & 2015 Mar 22 (1800/1760)     & \nodata                & 2017 Jan 26 (600)\\
J092908+350942/J092907+350942 & 2014 Feb 1  (3810/3600)     & \nodata                & 2017 Jan 26 (600)\\
J095423+093711/J095424+093648 & 2013 Apr 11 (1840/1800)     & 2017 Mar 22 (3000)     & 2017 Jan 26 (600)\\
J101711+183237/J101713+183232 & 2015 Mar 22 (2700/2640)     & 2017 Mar 21 (7200)     & 2015 May 6  (300)\\
J102907+421737/J102907+421752 & 2015 Mar 22 (2700/2640)     & 2016 Apr 10 (7200)     & 2017 Jan 26 (600)\\
J103643+565119/J103640+565125 & 2015 Mar 22 (1800/1760)     & \nodata                & 2017 Jan 26 (600)\\
J104151+174558/J104151+174603 & 2014 May 2  (2400/2340)     & 2016 Apr 10 (7000)     & 2017 Apr 12 (1200)\\
                              &                             &                        & 2015 May 6  (300)\\
J104242+164638/J104244+164656 & 2013 Apr 11 (1240/1200)     & \nodata                & 2015 May 6  (600)\\
J111238+353920/J111239+353928 & 2014 Feb 1  (2500/3600)     & \nodata                & 2015 May 6  (600)\\
J113236+340453/J113235+340428 & 2013 Apr 11 (1840/1800)     & \nodata                & 2015 May 6  (600)\\
J114335+284532/J114334+284511 & 2013 Apr 11 (2760/2700)     & \nodata                & 2017 Jan 26 (600)\\
J114926+142020/J114927+142002 & 2013 Apr 11 (1240/1200)     & \nodata                & 2017 Jan 26 (600)\\
J122413+495515/J122411+495511 & 2014 May 3  (1800/1760)     & \nodata                & 2017 Jan 26 (600)\\
J123049+071050/J123049+071036 & 2015 Mar 21 (3600/3520)     & \nodata                & 2015 May 6  (600)\\
J123249+433244/J123250+433227 & 2014 Feb 1  (3720/3600)     & 2016 May 13 (4200)     & 2015 May 6  (600)\\
J123318+103542/J123317+103538 & 2014 May 3  (1800/1760)     & \nodata                & 2015 May 6  (600)\\
J124601+173152/J124601+173156 & 2014 May 3  (2700/2640)     & \nodata                & 2017 Apr 12 (1200)\\
                              &                             &                        & 2017 Jan 26 (600)\\
J133541+285324/J133542+285330 & 2014 Feb 1  (2340/2250)     & 2016 May 30 (5200)     & 2015 May 6  (600)\\
J133740+055211/J133741+055216 & 2013 Apr 11 (3660/3600)     & 2017 Feb 22 (5100)     & 2017 Jan 26 (600)\\
J135051+250506/J135051+250454 & 2015 Mar 22 (3600/1800)     & 2017 Mar 22 (2700)     & 2017 Apr 12 (600)\\
J135521+303320/J135522+303324 & 2014 May 3  (3600/3520)     & \nodata                & 2015 May 6  (600)\\
J135733+254205/J135734+254204 & 2014 May 2  (2400/2340)     & 2015 Mar 25 (3600)     & 2017 Apr 12 (600)\\
J140852+245637/J140852+245631 & 2013 Apr 11 (2760/2700)     & \nodata                & 2017 Apr 12 (600)\\
J142459+382113/J142501+382100 & 2014 May 2--3 (1800/1760)   & 2016 Apr 2  (5100)     & 2015 May 6  (600)\\
J142815+585442/J142816+585432 & 2014 Feb 1  (1320/1320)     & \nodata                & 2017 Apr 12 (600)\\
J142910+324840/J142911+324824 & 2013 Apr 11 (1540/1500)     & \nodata                & \nodata          \\
%J142910+324840/J142911+324824 & 2013 Apr 11 (1540/1500)     & 2017 Apr 21 (3600)     & \nodata          \\
J143512+360424/J143511+360437 & 2015 Mar 22 (1800/1760)     & \nodata                & 2015 May 6  (600)\\
J143603+375138/J143603+375131 & 2015 Mar 22 (1800/1760)     & 2016 Jun 6  (2100) (E) & 2015 May 6  (600)\\
J144727+403217/J144727+403206 & 2014 May 3  (3600/3520)     & 2016 May 13 (4800)     & 2017 Apr 12 (600)\\
J145844+170522/J145844+170514 & 2015 Mar 22 (1800/1760)     & 2016 May 30 (4200)     & 2017 Jan 26 (600)\\
J150119+155602/J150118+155623 & 2014 May 3  (3600/3520)     & \nodata                & 2015 May 6  (600)\\
J150150+553227/J150150+553241 & 2014 Feb 1  (3400/3000)     & \nodata                & \nodata          \\
J152027+421519/J152028+421530 & 2015 Mar 21 (2700/1740)     & 2016 Apr 2 (4500)      & 2017 Apr 12 (600)\\
J153546+391932/J153546+391931 & 2015 Mar 21 (1800/1740)     & 2016 Jun 6 (1000) (E)  & 2015 May 6  (600)\\
J154454+165703/J154453+165710 & 2015 Mar 22 (2700/2640)     & \nodata                & 2015 May 6  (600)\\
J154741+343350/J154741+343357 & 2014 May 3  (1800/1760)     & \nodata                & 2015 May 6  (600)\\
J154956+070056/J154956+070044 & 2013 Apr 11 (1840/1800)     & \nodata                & 2015 May 6  (600)\\
J155505+362848/J155504+362847 & 2015 Mar 21 (1800/1760)     & 2016 Jul 4 (3200)      & 2017 Apr 12 (600)\\
J160906+441721/J160907+441734 & 2015 Mar 21--22 (5400/5280) & \nodata                & 2017 Apr 12 (600)\\
J160951+353838/J160951+353843 & 2014 May 2  (1800/1760)     & \nodata                & 2015 May 6  (600)\\
J165930+373527/J165931+373528 & 2013 Apr 11 (2240/2200)     & 2013 Jul 12 (4400)     & 2015 May 6  (600)\\
J171222+291758/J171220+291806 & 2014 May 2--3 (3600/3520)   & \nodata                & 2015 May 6  (1200)
\enddata
\tablenotetext{a}{The label (E) indicates the galaxy spectra were obtained from Keck/ESI instead of APO/DIS.
                  The Keck/ESI observations were conducted through the Keck observing time allocated 
				  to the Swinburne University.
				  }
\end{deluxetable*}

%% file: table2.tex
% .../sample_all/galaxies.txt
%

%\movetabledown=1.25in

\begin{longrotatetable}
\begin{deluxetable*}{llllllllllllllll}
%\tabletypesize{\small}
\tabletypesize{\scriptsize}
\tablecaption{Properties of Target Galaxies \label{tab:galaxy}}                                
\tablehead{                                                                                                  
\colhead{Galaxy} & 
\colhead{RA} & 
\colhead{DEC} &
\colhead{z} &
\colhead{PA} &
\colhead{\inc}  &
\colhead{\vinf} &
\colhead{\rrc} &
\colhead{Ks} &
\colhead{NUV} &
\colhead{FUV} &
\colhead{\mstar}  &
\colhead{$M_u$} &
\colhead{$M_r$} &
\colhead{\mhalo} &
\colhead{\rvir} 
\\                                           
\colhead{} & 
\colhead{(J2000)} & 
\colhead{(J2000)} &
\colhead{} &
\colhead{(\deg)} &
\colhead{(\deg)} &
\colhead{(\kms)} & 
\colhead{(kpc)} &
\colhead{(mag)} &
\colhead{(mag)} &
\colhead{(mag)} &
\colhead{(\msun)} &
\colhead{} &
\colhead{} &
\colhead{(\msun)} &
\colhead{(kpc)} 
\\                                           
}                                                                                                            
\colnumbers                                                                                                  
\startdata                                                                                                   
J080507.60+112208.3 &    08:05:07.60&  +11:22:08.36& 0.21811&  32.5   &45.6 &190     &3.28    &$18.71 \pm 0.03$ & \nodata          & \nodata           &9.74    & $-19.45$ & $-20.70$ &11.47 &154  \\
J081125.76+093616.9 &    08:11:25.76&  +09:36:16.93& 0.20118&  96.8   &66.4 &\nodata &\nodata &$17.58 \pm 0.02$ & \nodata          & \nodata           &10.46   & $-19.57$ & $-21.27$ &12.05 &243  \\
J081521.69+082602.1 &    08:15:21.69&  +08:26:02.11& 0.25055&  97.7   &57.5 &\nodata &\nodata &$17.53 \pm 0.08$ & \nodata          & \nodata           &10.62   & $-20.60$ & $-22.03$ &12.31 &288  \\
J081708.29+091750.3 &    08:17:08.29&  +09:17:50.31& 0.17395&  -28.8  &46.0 &80      &2.44    &$16.82 \pm 0.01$ & \nodata          & \nodata           &10.56   & $-19.44$ & $-21.42$ &12.18 &273  \\
J084235.98+565358.7 &    08:42:35.98&  +56:53:58.78& 0.21824&  -133.0 &59.6 &\nodata &\nodata &$19.22 \pm 0.05$ & $ \geq\ 22.24$   & $ \geq\ 21.68$    &9.64    & $-18.86$ & $-20.36$ &11.42 &148  \\
J084725.06+254104.7 &    08:47:25.06&  +25:41:04.70& 0.19591&  110.2  &52.1 &115     &2.35    &$18.24 \pm 0.04$ & $ \geq\ 23.96$   & $ \geq\ 23.76$    &9.90    & $-18.12$ & $-20.24$ &11.56 &166  \\
J085215.36+171137.0 &    08:52:15.36&  +17:11:37.07& 0.16921&  -159.7 &76.7 &200     &2.68    &$18.36 \pm 0.02$ & \nodata          & \nodata           &9.81    & $-18.89$ & $-20.26$ &11.50 &162  \\
J090743.44+532427.2 &    09:07:43.44&  +53:24:27.20& 0.18587&  167.3  &48.7 &150     &2.25    &$18.77 \pm 0.20$ & $21.61 \pm 0.10$ & $22.51 \pm 0.23$  &9.71    & $-18.84$ & $-20.37$ &11.45 &154  \\
J091954.11+291345.3 &    09:19:54.11&  +29:13:45.34& 0.23288&  -9.7   &72.5 &250     &3.44    &$17.14 \pm 0.16$ & $21.05 \pm 0.03$ & $ \geq\ 21.54$    &10.54   & $-19.82$ & $-21.47$ &12.18 &263  \\
J092908.86+350942.2 &    09:29:08.86&  +35:09:42.21& 0.18585&  18.1   &64.8 &\nodata &\nodata &$18.77 \pm 0.08$ & \nodata          & \nodata           &9.78    & $-18.71$ & $-20.30$ &11.49 &159  \\
J095423.27+093711.0 &    09:54:23.27&  +09:37:11.09& 0.20253&  -133.0 &+51.8 &225     &1.03    &$16.99 \pm 0.02$ & $20.91 \pm 0.07$ & $21.97 \pm 0.26$  &10.43   & $-19.90$ & $-21.69$ &12.01 &236  \\
J101711.80+183237.6 &    10:17:11.80&  +18:32:37.60& 0.18966&  -20.1  &83.7 &165     &3.59    &$19.36 \pm 0.08$ & $ \geq\ 22.16$   & \nodata           &9.47    & $-17.79$ & $-19.39$ &11.34 &141  \\
J102907.56+421737.6 &    10:29:07.56&  +42:17:37.64& 0.26238&  -153.9 &50.3 &155     &2.09    &$19.00 \pm 0.14$ & $ \geq\ 22.41$   & $ \geq\ 21.81$    &9.93    & $-19.63$ & $-21.14$ &11.59 &165  \\
J103643.44+565119.0 &    10:36:43.44&  +56:51:19.00& 0.13629&  -93.9  &58.0 &210     &1.74    &$17.68 \pm 0.04$ & $20.93 \pm 0.02$ & $ 21.37 \pm 0.05$ &9.94    & $-18.94$ & $-20.61$ &11.56 &173  \\
J104151.26+174558.7 &    10:41:51.26&  +17:45:58.79& 0.14232&  -17.4  &73.7 &75      &1.55    &$20.44 \pm 0.09$ & \nodata          & \nodata           &8.72    & $-17.81$ & $-18.70$ &11.01 &113  \\
J104242.75+164638.4 &    10:42:42.75&  +16:46:38.44& 0.19610&  138.4  &54.2 &\nodata &\nodata &$17.95 \pm 0.06$ & $20.57 \pm 0.19$ & $ \geq\ 21.30$    &9.99    & $-19.80$ & $-21.28$ &11.61 &174  \\
J111238.15+353920.4 &    11:12:38.15&  +35:39:20.49& 0.24671&  135.0  &48.5 &\nodata &\nodata &$18.50 \pm 0.09$ & $ \geq\ 22.31$   & $ \geq\ 21.62$    &9.91    & $-19.43$ & $-21.00$ &11.58 &164  \\
J113236.10+340453.0 &    11:32:36.10&  +34:04:53.03& 0.22210&  112.1  &51.5 &\nodata &\nodata &$18.13 \pm 0.04$ & $20.90 \pm 0.16$ & $ \geq\ 21.26$    &9.90    & $-19.69$ & $-21.03$ &11.56 &165  \\
J114335.30+284532.0 &    11:43:35.30&  +28:45:32.00& 0.14013&  -81.5  &50.1 &\nodata &\nodata &$17.97 \pm 0.07$ & $20.68 \pm 0.30$ & $20.97 \pm 0.41$  &9.95    & $-18.94$ & $-20.59$ &11.57 &174  \\
J114926.83+142020.3 &    11:49:26.83&  +14:20:20.35& 0.21887&  -143.8 &52.6 &\nodata &\nodata &$17.80 \pm 0.10$ & \nodata          & \nodata           &10.36   & $-20.14$ & $-21.73$ &11.94 &221  \\
J122413.10+495515.4 &    12:24:13.10&  +49:55:15.43& 0.26014&  -153.0 &+52.8 &210     &2.49    &$17.26 \pm 0.03$ & $21.20 \pm 0.07$ & $21.79 \pm 0.14$  &10.55   & $-20.32$ & $-22.05$ &12.20 &264  \\
J123049.01+071050.5 &    12:30:49.00&  +07:10:51.56& 0.39946&  -40.0  &37.8 &190     &1.66    &$18.39 \pm 0.03$ & $22.66 \pm 0.16$ & $ \geq\ 23.62$    &10.43   & $-20.41$ & $-21.94$ &12.09 &224  \\
J123249.06+433244.9 &    12:32:49.06&  +43:32:44.96& 0.21479&  42.0   &-60.2 &180     &1.80    &$17.94 \pm 0.05$ & $ \geq\ 21.30$   & $ \geq\ 21.33$    &10.06   & $-19.42$ & $-21.12$ &11.67 &179  \\
J123318.80+103542.1 &    12:33:18.80&  +10:35:42.13& 0.21040&  -100.6 &48.7 &175     &2.13    &$18.96 \pm 0.07$ & $21.67 \pm 0.11$ & $ \geq\ 21.24$    &9.56    & $-19.06$ & $-20.32$ &11.38 &144  \\
J124601.75+173152.1 &    12:46:01.75&  +17:31:52.15& 0.26897&  -178.1 &62.6 &60      &2.55    &$19.83 \pm 0.13$ & \nodata          & \nodata           &9.25    & $-19.40$ & $-20.15$ &11.25 &127  \\
J133541.46+285324.7 &    13:35:41.46&  +28:53:24.78& 0.28301&  -7.7   &-58.0 &175     &4.85    &$18.74 \pm 0.05$ & $21.17 \pm 0.07$ & $21.83 \pm 0.15$  &9.89    & $-20.04$ & $-21.20$ &11.57 &160  \\
J133740.48+055211.9 &    13:37:40.48&  +05:52:11.96& 0.16178&  168.7  &57.0 &115     &2.30    &$18.34 \pm 0.08$ & $20.76 \pm 0.06$ & $ \geq\ 21.12$    &9.67    & $-19.30$ & $-20.62$ &11.43 &154  \\
J135051.70+250506.5 &    13:50:51.70&  +25:05:06.58& 0.13709&  157.6  &50.9 &180     &2.75    &$17.99 \pm 0.18$ & $21.31 \pm 0.19$ & $21.33 \pm 0.45$  &9.72    & $-18.91$ & $-20.54$ &11.44 &158  \\
J135521.20+303320.4 &    13:55:21.20&  +30:33:20.41& 0.20690&  80.6   &+71.9 &\nodata &\nodata &$18.59 \pm 0.16$ & $ \geq\ 21.32$   & $ \geq\ 21.30$    &9.76    & $-19.20$ & $-20.64$ &11.48 &156  \\
J135733.86+254205.1 &    13:57:33.86&  +25:42:05.14& 0.25995&  -98.1  &44.9 &160     &2.49    &$17.74 \pm 0.07$ & \nodata          & \nodata           &10.22   & $-19.76$ & $-21.39$ &11.81 &196  \\
J140852.40+245637.3 &    14:08:52.40&  +24:56:37.38& 0.16686&  -167.8 &74.5 &\nodata &\nodata &$19.31 \pm 0.06$ & \nodata          & \nodata           &9.55    & $-18.49$ & $-20.11$ &11.37 &147  \\
J142459.82+382113.3 &    14:24:59.82&  +38:21:13.37& 0.21295&  -48.6  &60.7 &190     &2.15    &$17.90 \pm 0.07$ & $21.05 \pm 0.10$ & $ \geq\ 21.47$    &10.15   & $-19.58$ & $-21.20$ &11.73 &189  \\
J142815.41+585442.1 &    14:28:15.46&  +58:54:42.23& 0.37731&  -171.7 &66.4 &\nodata &\nodata &$19.01 \pm 0.17$ & $21.38 \pm 0.19$ & $ \geq\ 21.91$    &10.10   & $-20.60$ & $-21.68$ &11.75 &175  \\
J142910.91+324840.2 &    14:29:10.91&  +32:48:40.20& 0.26130&  96.6   &53.9 &\nodata &\nodata &\nodata          & $21.56 \pm 0.02$ & $22.24 \pm 0.19$  &10.79   & $-20.12$ & $-21.99$ &12.63 &366  \\
J143512.93+360424.9 &    14:35:12.93&  +36:04:24.98& 0.26225&  -102.8 &60.3 &\nodata &\nodata &$19.50 \pm 0.16$ & $ \geq\ 22.06$   & $ \geq\ 21.70$    &9.73    & $-18.89$ & $-20.45$ &11.48 &151  \\
J143603.16+375138.5 &    14:36:03.16&  +37:51:38.55& 0.31462&  -120.0 &71.9 &260     &3.80    &$18.28 \pm 0.06$ & \nodata          & \nodata           &10.32   & $-20.20$ & $-21.82$ &11.93 &208  \\
J144727.42+403217.8 &    14:47:27.42&  +40:32:17.89& 0.24497&  134.4  &55.2 &205     &3.58    &$17.63 \pm 0.07$ & $22.45 \pm 0.16$ & $ \geq\ 23.03$    &10.30   & $-19.47$ & $-21.37$ &11.88 &209  \\
J145844.18+170522.2 &    14:58:44.18&  +17:05:22.26& 0.21965&  8.3    &+54.4 &225     &3.30    &$17.32 \pm 0.03$ & $21.26 \pm 0.35$ & $ \geq\ 20.89$    &10.42   & $-20.01$ & $-21.79$ &12.01 &233  \\
J150119.24+155602.2 &    15:01:19.24&  +15:56:02.25& 0.17779&  -90.0  &60.0 &\nodata &\nodata &$18.44 \pm 0.09$ & $21.70 \pm 0.13$ & $ \geq\ 21.31$    &9.75    & $-18.80$ & $-20.39$ &11.47 &157  \\
J150150.28+553227.5 &    15:01:50.28&  +55:32:27.58& 0.15139&  -68.9  &55.0 &\nodata &\nodata &\nodata          & $ \geq\ 21.75$   & $ \geq\ 22.46$    &9.17    & $-17.56$ & $-18.99$ &11.20 &130  \\
J152027.21+421519.3 &    15:20:27.21&  +42:15:19.37& 0.21099&  93.5   &-55.4 &220     &2.13    &$16.69 \pm 0.04$ & $ \geq\ 21.99$   & $ \geq\ 21.83$    &10.58   & $-19.91$ & $-21.90$ &12.23 &277  \\
J153546.74+391932.2 &    15:35:46.74&  +39:19:32.27& 0.25587&  21.4   &+52.9 &215     &1.64    &$17.34 \pm 0.08$ & \nodata          & \nodata           &10.57   & $-20.03$ & $-21.81$ &12.23 &271  \\
J154454.75+165703.7 &    15:44:54.75&  +16:57:03.78& 0.20215&  -16.5  &56.3 &\nodata &\nodata &$17.91 \pm 0.11$ & \nodata          & \nodata           &10.12   & $-19.88$ & $-21.19$ &11.71 &186  \\
J154741.46+343350.8 &    15:47:41.46&  +34:33:50.86& 0.18392&  -145.3 &59.3 &175     &1.59    &$18.10 \pm 0.07$ & \nodata          & \nodata           &9.96    & $-18.86$ & $-20.53$ &11.59 &172  \\
J154956.73+070056.0 &    15:49:56.73&  +07:00:56.09& 0.16235&  -104.8 &+63.0 &\nodata &\nodata &$19.58 \pm 0.07$ & \nodata          & \nodata           &9.28    & $-18.75$ & $-19.90$ &11.25 &135  \\
J155505.27+362848.4 &    15:55:05.27&  +36:28:48.40& 0.18926&  132.6  &55.2 &255     &3.59    &$17.86 \pm 0.09$ & \nodata          & \nodata           &10.12   & $-19.72$ & $-21.10$ &11.70 &187  \\
J160906.72+441721.5 &    16:09:06.72&  +44:17:21.50& 0.14732&  -141.5 &70.1 &145     &3.19    &$19.48 \pm 0.33$ & \nodata          & \nodata           &9.41    & $-18.18$ & $-19.79$ &11.30 &141  \\
J160951.62+353838.5 &    16:09:51.62&  +35:38:38.58& 0.28940&  178.7  &65.2 &45      &2.69    &$18.98 \pm 0.06$ & \nodata          & \nodata           &9.93    & $-19.95$ & $-21.00$ &11.60 &163  \\
J165930.44+373527.9 &    16:59:30.44&  +37:35:27.90& 0.19984&  -8.2   &-56.8 &170     &2.72    &$17.62 \pm 0.04$ & $20.21 \pm 0.17$ & $20.70 \pm 0.31$  &10.01   & $-20.38$ & $-21.52$ &11.63 &175  \\
J171222.58+291758.6 &    17:12:22.58&  +29:17:58.67& 0.21020&  74.0   &58.8 &\nodata &\nodata &$17.42 \pm 0.03$ & \nodata          & \nodata           &10.18   & $-19.68$ & $-21.26$ &11.76 &193  \\
\enddata                                                                                                     
\tablecomments{
Column Descriptions:
(1) Target galaxy.
(2,3)  Coordinates of target galaxy.
(4) Emission-line redshift of target galaxy. 
The redshift corrections were most
substantial for J123049.01+071050.5 ($0.167 \rightarrow  0.39946$),
J142815.41+585442.1  ($0.243 \rightarrow  0.37731 $), and
J143603.16+375138.5  ($0.20011 \rightarrow  0.31462 $).
(5) Position angle of galaxy major axis.
(6) Inclination of galactic disk estimated as $\inc = \arccos(b/a)$,
    where $b$ and $a$ are the lengths of the semi-minor and semi-major 
   axes, respectively. If the 3D orientation of the disk has been established, 
   then we provide the sign of the inclination.
%% IF YOU USE J155505, THEN I = -55.2; AND I =+50.9 FOR J135051
(7) Asymptotic rotation speed of galactic disk. An arctangent
model, $\vphi (R)  = 2/\pi \vinf \arctan(R/\rrc)$, for the disk
was projected onto the sky, convolved with a Gaussian model for the seeing 
disk, and fit to the Doppler shifts measured along the spectroscopic slit.
(8) Scale radius for the rotation curve model.
(9) AB magnitude of target galaxy from NIRC2 photometry. 
(10,11) AB magnitudes of target galaxy from GALEX NUV and FUV photometry.
(12) Stellar mass fit to SDSS ugriz, NIRC2 Ks, and GALEX NUV/FUV photometry, except as noted. The stellar 
population synthesis models have a Chabrier initial mass function and an 
exponentially decaying SFR. Fits were computed using FAST \citep{Kriek2009}. 
(13,14) Absolute magnitudes in $u$ and $r$ bands from SED fit. 
(15) Halo mass estimate from abundance matching \citep{Behroozi2010}.
{\clm
We adopt the halo mass that contains an average stellar mass matched
to our measured stellar mass. We use the measured spectroscopic redshift for each galaxy.
We propagate uncertainties in the measured stellar mass but do not account
for the scatter in the stellar mass -- halo mass relation.
}
(16) Halo virial radius, defined by the overdensity
$\Delta_{vir}(z) = 18 \pi^2 + 82 x -39x^2$ \citep{BryanNorman1998} with respect
to the critical density and evaluated at the redshift shown in column~4. 
}
% \tablenotetext{a}{Stellar mass fitting did not include GALEX NUV photometry.}
% \tablenotetext{b}{Stellar mass fitting did not include GALEX FUV photometry.}
% \tablenotetext{c}{Stellar mass fitting included an upper limit in GALEX NUV photometry.}
% \tablenotetext{d}{Stellar mass fitting included an upper limit in GALEX FUV photometry.}
% %\tablenotetext{e}{Stellar mass fitting included GALEX NUV photometry.}
% %\tablenotetext{f}{Stellar mass fitting included GALEX FUV photometry.}
% \tablenotetext{e}{Stellar mass fitting did not include NIRC2 Ks photometry.}

%\tablecomments{}                                                                      
\end{deluxetable*}
\end{longrotatetable}

%% file: table3.tex
% .../sample_all/quasars.txt
%
%\startlongtable
\begin{longrotatetable}
\begin{deluxetable*}{llllllllllll}                                                                                    
%\tabletypesize{\small}
%\tabletypesize{\scriptsize}
\tabletypesize{\footnotesize}
\tablecaption{Summary of Sightlines \label{tab:quasar}}                                
\tablehead{                                                                                                  
\colhead{Quasar} & 
\colhead{RA} & 
\colhead{DEC} &
\colhead{Target} &
\colhead{$\alpha$} &
\colhead{b} &
\colhead{$W_{r}^{2796}$} &
\colhead{$W_{r}^{2803}$} &
\colhead{$\langle W(b) \rangle$} &
\colhead{$\langle V \rangle _W^{2796}$} &
\colhead{$[V_{min},V_{max}]$} &
\colhead{$Resel$} 
\\                                           
\colhead{} & 
\colhead{(J2000)} & 
\colhead{(J2000)} &
\colhead{Galaxy} &
\colhead{(\deg)} &
\colhead{(kpc)} &
\colhead{(\AA)} &
\colhead{(\AA)} &
\colhead{(\AA)} &
\colhead{(\kms)} &
\colhead{(\kms)} &
\colhead{(\kms)} 
\\                                           
}                                                                                                            
\colnumbers                                                                                                  
\startdata                                                                                                   
J080508.50+112157.2 & 08:05:08.50  & +11:21:57.21  & J080507.60+112208.3   &82.3 (97.7)   &  62.9 &  $0.050_{-0.033}^{+0.021}$  & $0.046_{-0.012}^{+0.015}$ & 0.212  & $-17_{-30}^{+31}$  & $[-102 , 71]$  &  165    \\
J081125.55+093626.2 & 08:11:25.55  & +09:36:26.27  & J081125.76+093616.9   &64.7          & 33.81 &  $ \leq\ 0.400$             & $ \leq\ 0.089 $           &  0.579  & \nodata           &    \nodata     &  160    \\
J081521.64+082623.9 & 08:15:21.64  & +08:26:23.95  & J081521.69+082602.1   &80.6          & 88.37 &  $ \leq\ 0.230$             & $ \leq\ 0.742 $           &  0.0880 & \nodata           &    \nodata     &  165    \\
J081708.71+091751.4 & 08:17:08.71  & +09:17:51.49  & J081708.29+091750.3   &71.6 (108.4)  &  19.4 &  $0.548_{-0.024}^{+0.026}$  & $0.444_{-0.023}^{+0.025}$ & 0.953  & $16_{-6}^{+5}$     & $[-134 , 176]$ &  165    \\
J084235.00+565350.2 & 08:42:35.00  & +56:53:50.24  & J084235.98+565358.7   & 3.5          & 42.87 &  $ \leq\ 0.238$             & $ \leq\ 0.197 $           &  0.424  & \nodata           &    \nodata     &  105    \\
J084723.56+254105.4 & 08:47:23.56  & +25:41:05.40  & J084725.06+254104.7   &18.2 (161.8)  &  67.7 &  $0.092_{-0.013}^{+0.014}$  & $0.111_{-0.020}^{+0.022}$ & 0.180  & $-105_{-9}^{+9}$   & $[-176 , -44]$ &  105    \\
J085215.34+171143.8 & 08:52:15.34  & +17:11:43.85  & J085215.36+171137.0   &22.8 (157.2)  & 20.32 &  $1.314_{-0.043}^{+0.047}$  & $1.190_{-0.046}^{+0.054}$ & 0.923  & $-49_{-5}^{+5}$    & $[-225 , 86]$  &  165    \\
J090745.28+532421.4 & 09:07:45.28  & +53:24:21.47  & J090743.44+532427.2   &58.2 (-58.2)  & 56.03 &  $0.042_{-0.025}^{+0.019}$  & $0.044_{-0.025}^{+0.020}$ & 0.269  & $-6_{-15}^{+20}$   & $[-50 , 42]$   &  105    \\
J091954.28+291408.3 & 09:19:54.28  & +29:14:08.36  & J091954.11+291345.3   &15.3 (15.3)   & 88.36 &  $0.216_{-0.013}^{+0.014}$  & $0.159_{-0.029}^{+0.019}$ & 0.088  & $-147_{-7}^{+5}$   & $[-266 , -8]$  &  105    \\
                    &              &               &                       &15.3 (15.3)   & 88.36 &  $0.523_{-0.002}^{+0.018}$  & $0.345_{-0.029}^{+0.018}$ & 0.088  & $131_{-4}^{+5}$    & $[-51 , 292]$  &         \\
J092907.44+350942.0 & 09:29:07.44  & +35:09:42.06  & J092908.86+350942.2   &71.4          & 56.02 &  $ \leq\ 0.045$             & $ \leq\ 0.082 $           &  0.269 & \nodata            &    \nodata     &  165    \\
J095424.10+093648.3 & 09:54:24.10  & +09:36:48.33  & J095423.27+093711.0   &75.5 (-75.5)  & 89.12 &  $0.480_{-0.025}^{+0.021}$  & $0.333_{-0.040}^{+0.030}$ & 0.086  & $-106_{-4}^{+4}$   & $[-284 , 20]$  &  160    \\
J101713.07+183232.2 & 10:17:13.07  & +18:32:32.24  & J101711.80+183237.6   &53.4 (126.6)  & 61.58 &  $0.572_{-0.033}^{+0.036}$  & $0.401_{-0.035}^{+0.037}$ & 0.222  & $65_{-8}^{+6}$     & $[-93 , 219]$  &  105    \\
J102907.73+421752.9 & 10:29:07.73  & +42:17:52.93  & J102907.56+421737.6   &19.1 (160.9)  & 64.56 &  $0.122_{-0.028}^{+0.025}$  & $0.123_{-0.017}^{+0.019}$ & 0.200  & $-53_{-20}^{+17}$  & $[-134 , 34]$  &  105    \\
J103640.74+565125.9 & 10:36:40.74  & +56:51:25.98  & J103643.44+565119.0   &21.4 (21.4)   & 57.84 &  $0.281_{-0.08}^{+0.06}$    & $0.088_{-0.043}^{+0.048}$ & 0.253  & $392_{-22}^{+16}$  & $[284 , 516]$  &  105    \\
J104151.33+174603.4 & 10:41:51.33  & +17:46:03.40  & J104151.26+174558.7   &29.8 (29.8)   & 12.15 &  $1.340_{-0.035}^{+0.036}$  & $1.268_{-0.031}^{+0.046}$ & 1.224  & $0_{-3}^{+3}$      & $[-174 , 145]$ &  165    \\
J104244.23+164656.1 & 10:42:44.23  & +16:46:56.14  & J104242.75+164638.4   &88.0          & 93.12 &  $ \leq\ 0.048$             & $ \leq\ 0.019 $           & 0.075  & \nodata            &    \nodata     &  160    \\
J111239.11+353928.2 & 11:12:39.11  & +35:39:28.20  & J111238.15+353920.4   &78.5          & 55.95 &  $ \leq\ 0.091$             & $ \leq\ 0.019 $           & 0.270  & \nodata            &    \nodata     &  165    \\
J113235.64+340428.3 & 11:32:35.64  & +34:04:28.36  & J113236.10+340453.0   &80.9          & 93.49 &  $ \leq\ 0.041$             & $ \leq\ 0.075 $           & 0.0737 & \nodata            &    \nodata     &  160    \\
J114334.84+284511.6 & 11:43:34.84  & +28:45:11.67  & J114335.30+284532.0   &82.2          & 53.89 &  $ \leq\ 0.197$             & $ \leq\ 0.224 $           & 0.290  & \nodata            &    \nodata     &  160    \\
J114927.97+142002.2 & 11:49:27.97  & +14:20:02.27  & J114926.83+142020.3   &78.8          & 89.51 &  $ \leq\ 0.086$             & $ \leq\ 0.021 $           & 0.0846 & \nodata            &    \nodata     &  160    \\
J122411.69+495511.1 & 12:24:11.69  & +49:55:11.15  & J122413.10+495515.4   &45.6 (45.6)   & 58.97 &  $1.096_{-0.038}^{+0.034}$  & $0.721_{-0.032}^{+0.031}$ & 0.243  & $64_{-6}^{+6}$     & $[-196 , 306]$ &  165    \\
J123049.67+071036.9 & 12:30:49.67  & +07:10:36.97  & J123049.01+071050.5   & 4.0 (-176.0) & 97.73 &  $0.077_{-0.014}^{+0.016}$  & $0.047_{-0.012}^{+0.012}$ & 0.075  & $-93_{-17}^{+12}$  & $[-172 , -21]$ &  105    \\
J123250.30+433227.3 & 12:32:50.30  & +43:32:27.38  & J123249.06+433244.9   &79.0 (101.0)  & 79.77 &  $0.265_{-0.019}^{+0.02}$   & $0.180_{-0.025}^{+0.033}$ & 0.118  & $6_{-8}^{+8}$      & $[-152 , 151]$ &  165    \\
J123317.75+103538.1 & 12:33:17.75  & +10:35:38.19  & J123318.80+103542.1   & 3.9 (-3.9)   & 56.58 &  $0.325_{-0.022}^{+0.059}$  & $0.302_{-0.039}^{+0.047}$ & 0.264  & $20_{-29}^{+8}$    & $[-86 , 129]$  &  165    \\
J124601.81+173156.4 & 12:46:01.81  & +17:31:56.45  & J124601.75+173152.1   &10.8 (-169.2) & 19.25 &  $0.309_{-0.027}^{+0.026}$  & $0.270_{-0.029}^{+0.031}$ & 0.958  & $-293_{-8}^{+9}$   & $[-404 , -200]$&  165    \\
J133542.57+285330.4 & 13:35:42.57  & +28:53:30.48  & J133541.46+285324.7   &76.3 (76.3)   & 69.05 &  $0.341_{-0.018}^{+0.018}$  & $0.268_{-0.019}^{+0.020}$ & 0.171  & $-5_{-6}^{+7}$     & $[-155 , 133]$ &  165    \\
J133741.39+055216.9 & 13:37:41.39  & +05:52:16.90  & J133740.48+055211.9   &81.3 (-98.7)  & 41.92 &  $0.747_{-0.1}^{+0.107}$    & $0.510_{-0.112}^{+0.097}$ & 0.438  & $-47_{-15}^{+10}$  & $[-155 , 70]$  &  160    \\
J135051.46+250454.8 & 13:50:51.46  & +25:04:54.85  & J135051.70+250506.5   &37.5 (37.5)   & 30.49 &  $0.460_{-0.126}^{+0.129}$  & $0.501_{-0.199}^{+0.177}$ & 0.650  & $8_{-20}^{+17}$    & $[-78 , 60]$   &  105    \\
J135522.89+303324.7 & 13:55:22.89  & +30:33:24.76  & J135521.20+303320.4   & 1.6          & 78.22 &  $ \leq\ 0.176$             & $ \leq\ 0.108 $           &  0.125 & \nodata            &    \nodata     &  165    \\
J135734.41+254204.6 & 13:57:34.41  & +25:42:04.62  & J135733.86+254205.1   &12.1 (-167.9) & 30.99 &  $1.485_{-0.018}^{+0.019}$  & $1.168_{-0.013}^{+0.013}$ & 0.638  & $-33_{-3}^{+3}$    & $[-423 , 197]$ &  165    \\
J140852.98+245631.4 & 14:08:52.98  & +24:56:31.45  & J140852.40+245637.3   &65.2          & 29.00 &  $ \leq\ 0.243$             & $ \leq\ 0.392 $           &  0.684 & \nodata            &    \nodata     &  160    \\
J142501.46+382100.5 & 14:25:01.46  & +38:21:00.52  & J142459.82+382113.3   & 7.8 (172.2)  & 82.92 &  $0.244_{-0.023}^{+0.026}$  & $0.117_{-0.01}^{+0.009}$  & 0.106  & $21_{-14}^{+12}$   & $[-128 , 221]$ &  165    \\
J142816.69+585432.6 & 14:28:16.69  & +58:54:32.62  & J142815.41+585442.1   &53.3          & 72.39 &  $0.879_{-0.029}^{+0.034}$  & \nodata                   & 0.144  & $-180_{-5}^{+4}$   & $[-316, -15]$  &  165    \\
J142911.26+324824.0 & 14:29:11.26  & +32:48:24.03  & J142910.91+324840.2   &67.8          & 69.94 &  $ \leq\ 0.092$             & $ \leq\ 0.138 $           & 0.166  & \nodata            &    \nodata     &  160    \\
J143511.53+360437.2 & 14:35:11.53  & +36:04:37.22  & J143512.93+360424.9   &48.7          & 87.22 &  $ \leq\ 0.010$             & $ \leq\ 0.058 $           & 0.0906 & \nodata            &    \nodata     &  105    \\
J143603.56+375131.3 & 14:36:03.56  & +37:51:31.37  & J143603.16+375138.5   &86.5 (-93.5)  & 40.81 &  $0.093_{-0.012}^{+0.009}$  & $0.066_{-0.010}^{+0.010}$ & 0.455  & $35_{-10}^{+4}$    & $[-46 , 113]$  &  105    \\
J144727.50+403206.3 & 14:47:27.50  & +40:32:06.34  & J144727.42+403217.8   &41.5 (41.5)   & 45.94 &  $0.270_{-0.031}^{+0.035}$  & $0.168_{-0.026}^{+0.036}$ & 0.381  & $89_{-12}^{+12}$   & $[-34 , 221]$  &  165    \\
J145844.76+170514.7 & 14:58:44.76  & +17:05:14.79  & J145844.18+170522.2   &56.6 (123.4)  & 40.75 &  $0.850_{-0.029}^{+0.032}$  & $0.722_{-0.028}^{+0.030}$ & 0.456  & $55_{-5}^{+5}$     & $[-89 , 253]$  &  105    \\
J150118.32+155623.0 & 15:01:18.32  & +15:56:23.04  & J150119.24+155602.2   &57.6          & 76.29 &  $ \leq\ 0.118$             & $ \leq\ 0.260 $           & 0.134  & \nodata            &    \nodata     &  165    \\
J150150.93+553241.1 & 15:01:50.93  & +55:32:41.19  & J150150.28+553227.5   &89.2          & 40.29 &  $ \leq\ 0.114$             & $ \leq\ 0.118 $           & 0.463  & \nodata            &    \nodata     &  165    \\
J152028.18+421530.3 & 15:20:28.18  & +42:15:30.32  & J152027.21+421519.3   &48.9  (-48.9) & 54.67 &  $0.186_{-0.048}^{+0.051}$  & $0.136_{-0.035}^{+0.036}$ & 0.282  & $-65_{-23}^{+29}$  & $[-146 , 26]$  &  105    \\
J153546.29+391931.1 & 15:35:46.29  & +39:19:31.18  & J153546.74+391932.2   &57.2 (-122.8) & 22.19 &  $1.331_{-0.022}^{+0.023}$  & $1.074_{-0.040}^{+0.040}$ & 0.865  & $-64_{-2}^{+3}$    & $[-403 , 136]$ &  105    \\
J154453.86+165710.1 & 15:44:53.86  & +16:57:10.11  & J154454.75+165703.7   &47.1          & 48.96 &  $ \leq\ 0.067$             & $ \leq\ 0.057 $           & 0.343  & \nodata            &    \nodata     &  105    \\
J154741.88+343357.3 & 15:47:41.88  & +34:33:57.32  & J154741.46+343350.8   & 4.0 (-176.0) & 25.99 &  $1.009_{-0.025}^{+0.023}$  & $0.898_{-0.030}^{+0.028}$ & 0.759  & $-12_{-3}^{+3}$    & $[-152 , 156]$ &  165    \\
J154956.87+070044.9 & 15:49:56.87  & +07:00:44.93  & J154956.73+070056.0   &85.4 (-85.4)  & 32.68 &  $0.195_{-0.043}^{+0.027}$  & $0.086_{-0.032}^{+0.035}$ & 0.602  & $5_{-15}^{+8}$     & $[-76 , 103]$  &  160    \\
J155504.39+362847.9 & 15:55:04.39  & +36:28:47.96  & J155505.27+362848.4   &44.6 (135.4)  & 34.30 &  $0.260_{-0.033}^{+0.033}$  & $0.151_{-0.047}^{+0.050}$ & 0.570  & $-45_{-12}^{+11}$  & $[-176 , 43]$  &  105    \\
J160907.45+441734.4 & 16:09:07.45  & +44:17:34.43  & J160906.72+441721.5   &7.3  (172.7)  & 40.37 &  $0.290_{-0.02}^{+0.023}$   & $0.299_{-0.023}^{+0.026}$ & 0.462  & $-59_{-6}^{+5}$    & $[-151 , 30]$  &  105    \\
J160951.81+353843.7 & 16:09:51.81  & +35:38:43.78  & J160951.62+353838.5   &25.5 (-154.5) & 25.69 &  $2.167_{-0.033}^{+0.035}$  & $2.065_{-0.040}^{+0.053}$ & 0.767  & $6_{-3}^{+2}$      & $[-213 , 231]$ &  165    \\
J165931.92+373528.8 & 16:59:31.92  & +37:35:28.80  & J165930.44+373527.9   &84.6 (95.37)  & 60.03 &  $0.889_{-0.024}^{+0.028}$  & $0.769_{-0.023}^{+0.024}$ & 0.234  & $-10_{-5}^{+5}$    & $[-158 , 233]$ &  160    \\
J171220.81+291806.9 & 17:12:20.81  & +29:18:06.96  & J171222.58+291758.6   &35.9          & 86.83 &  $ \leq\ 0.102$             & $ \leq\ 0.010 $           & 0.0928 & \nodata            &    \nodata     &  165    \\
\enddata                                                                                                     
\tablecomments{
Column Descriptions:
(1) Quasar name.
(2,3) Quasar coordinates.
(4) Target Galaxy.
(5) Azimuthual angle between the galaxy -- quasar position angle and the galaxy major axis. 
We  define $\alpha$ on the interval $[0\deg,90\deg]$ when  we stack  the sightlines on a single quadrant
of the sky coordinates. When we need to distinguish among the four quadrants, 
we define azimuthal angle relative to the receding side of the major axis on the 
interval $-180\deg\ < \alpha\ \le 180\deg$ such that  positive values are consistent
with positive z-values as defined in Section~\ref{sec:thin_disk}.
(6) Projected distance of the quasar sightline from the target galaxy.
(7) Rest-frame equivalent width of \mgII\ $\lambda 2796$; the $2\sigma$ upper limit is listed for non-detections.
(8) Rest-frame equivalent width of \mgII\ $\lambda 2803$; the $2\sigma$ upper limit is listed for non-detections.
(9) Predicted equivalent width. We adopt the fitted $W_r -- b$ relation from \citep{Nielsen2013}. 
(10) Equivalent-width-weighted Doppler shift of the \mgII\ $\lambda 2796$ line, measured relative to the target galaxy redshift.
(11) Velocity range of the absorption system, not corrected for the instrumental profile.
(12) Velocity resolution (full width at half maximum intensity).
}
\end{deluxetable*}
\end{longrotatetable}

%% file: table4.tex
\startlongtable
\begin{deluxetable}{lll}
\tablecaption{Mg~I Detections \label{tab:mgi}}
\tablehead{
\colhead{Quasar} &
\colhead{$W_{r}(MgI)$}  &
\colhead{$W_{r}(MgI)/$}
\\
\colhead{} &
\colhead{(\AA)} &
\colhead{$W_{r}(MgII)$}
}
%\colnumbers                                                                                          
\startdata
J104151.33+174603.428 & $1.64 \pm 0.05$ & 1.2 \\
J160951.81+353843.728 & $0.38 \pm 0.03$ & 0.18 \\
J085215.34+171143.828 & $0.29 \pm 0.03$ & 0.22 \\
J153546.29+391931.128 & $0.26 \pm  0.01$ &  0.20 \\
J135734.41+254204.628 & $0.25 \pm 0.01$ & 0.17 \\
J154741.88+343357.328 & $0.15 \pm  0.02$ & 0.15 \\
J165931.92+373528.828 & $0.12 \pm 0.02$ & 0.13 \\
J160907.45+441734.428 & $0.11 \pm  0.02$ & 0.38 \\
J145844.76+170514.728 & $0.10 \pm  0.02$ & 0.12 \\
J091954.28+291408.328 & $0.093 \pm  0.012$ & 0.18 \\
J124601.81+173156.428 & $0.070 \pm 0.016$ & 0.23 \\
J122411.69+495511.128 & $0.049 \pm 0.015$ & 0.045 \\
\enddata
\tablecomments{
{\it Col 1.} Quasar sightline.
{\it Col 2.} Rest-frame Equivalent width of \mgI\
$\lambda 2853$ absorption near the redshift of the target galaxy.
{\it Col 3.} Equivalent width ratio of rest-frame \mgI\
$\lambda 2853$ to  \mgII\ $\lambda 2796$.
}
\end{deluxetable}